\documentclass[12pt,a4]{iopart}
\usepackage{graphicx}

\begin{document}

\title{\bf Correlations of Heavy Quarks Produced at Large Hadron Collider}

\title[Correlations of......]{}

\author{Mohammed Younus$^\dagger$\footnote{E-mail: mdyounus@vecc.gov.in},
 Umme Jamil$^\ddagger$
\footnote{Present Addrees: Department of Physics,
D. R. College, Golaghat, Assam 785621, India}, 
Dinesh K. Srivastava$^\dagger$\footnote{E-mail: dinesh@vecc.gov.in}}

\address{$^\dagger$Variable Energy Cyclotron Center, 1/AF, Bidhan Nagar, Kolkata 700 064, India\\
$^\ddagger$Saha Institute of Nuclear Physics, 1/AF, Bidhan Nagar, Kolkata 700 064, India}

\date{\today}
\begin{abstract}
We study the correlations of heavy quarks produced in relativistic heavy
ion collisions and find them to be quite sensitive to the effects of the medium and the 
production mechanisms. In order to put this on a quantitative footing,  as a first step,
we analyze the azimuthal, transverse momentum, and rapidity correlations of  heavy quark-
anti quark ($Q\overline{Q}$) pairs in $pp$ collisions at $\cal{O}$($\alpha_{s}^{3}$).
This sets the stage for the identification and
study of medium modification of similar
correlations in relativistic collision of heavy nuclei at the Large Hadron
Collider. Next we study the additional production of charm quarks in heavy ion
collisions due to multiple scatterings, {\it viz.}, jet-jet collisions, jet-thermal
collisions, and thermal interactions. We find that these
give rise to azimuthal correlations which are quite different from those 
arising from prompt initial production at leading order and at next to leading order.

\end{abstract}

\noindent{\em PACS}: 14.65.Dw, 25.75.-q, 25.75.Cj, 12.38.Mh, 14.40.Pq, 14.65.Fy, 12.38.-t\\

\noindent{\em Keywords}: heavy quarks, relativistic heavy ion collisions,
                         pp collisions, 
                         quark gluon plasma, NLO pQCD, correlations, D-mesons,
$J/\psi$.

\maketitle

\section{Introduction}

The study of relativistic heavy ion collisions and quark gluon plasma (QGP) is
approaching its zenith with the first experiments performed
at the Large Hadron Collider
at CERN Geneva (though not yet at the top energy) involving lead nuclei.
Together with the wealth of data already accumulated at the Relativistic Heavy
Ion Collider at Brookhaven National Laboratory, we now have an enormous
task to decipher, analyze, and quantitatively explain these observations and
extract information about the properties of the QGP. These
analyses are also paving the way for additional measurements, some of which
can already be performed using the present detector set-ups, while others
will become amenable to studies with the upgrades planned for all the
major experiments, ALICE, PHENIX, and STAR, etc.
 Taken in its entirety, this represents the most important and
fruitful international collaboration in high energy nuclear physics to date.

The focus has now progressed from models to theories and from qualitative
to quantitative determination of various properties of quark gluon plasma.
Enormous strides made towards exploring the shear viscosity ~\cite{bass} 
of the matter produced in these collisions is one such example.

In the present work we consider using heavy quarks to probe the QGP.
The heavy quarks (only charm and bottom quarks are considered here) offer several
unique advantages. The conservation of flavour in strong interaction
 dictates that they are produced in pairs $(Q\overline{Q})$. Their large
mass provides that $Q^2$ necessary for their production is large and thus
one may confidently use pQCD, for these studies. Their large masses ensure 
that the hadrons containing the heavy quarks stand
out in the swarm of pions.

\begin{figure*}[h]
\begin{center}
\includegraphics[width=4in,angle=270]{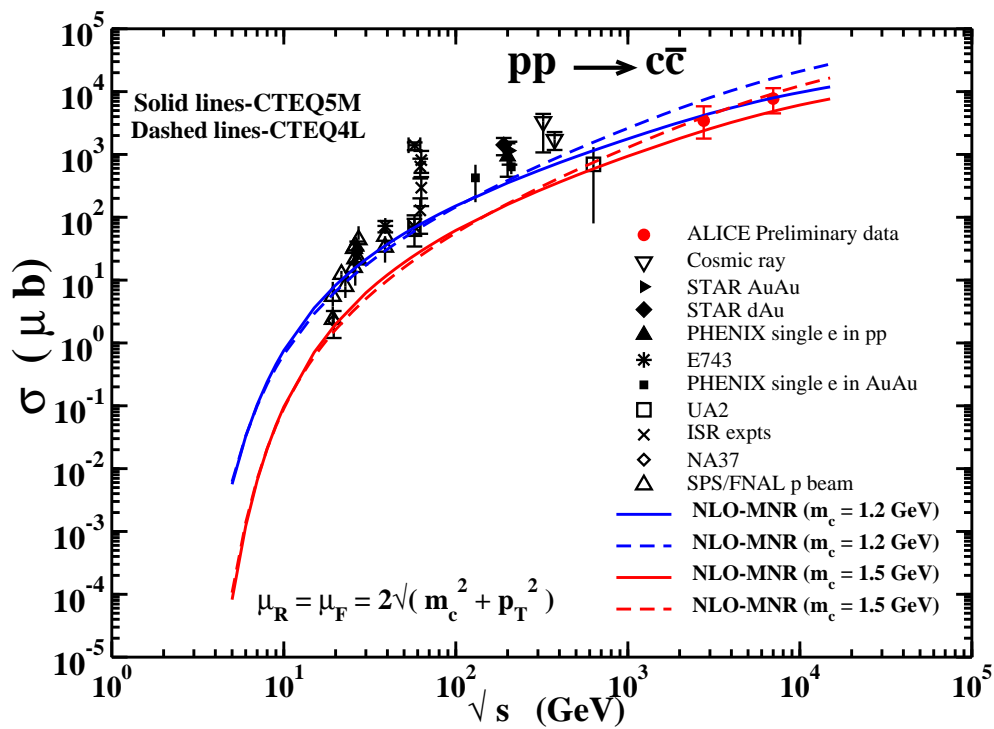}
\caption{Energy dependence of the charm quark production in $pp$ collisions.}
\label{sqrts}
\end{center}
\end{figure*}

\begin{figure*}[h]
\begin{center}
\includegraphics[width=4in,angle=270]{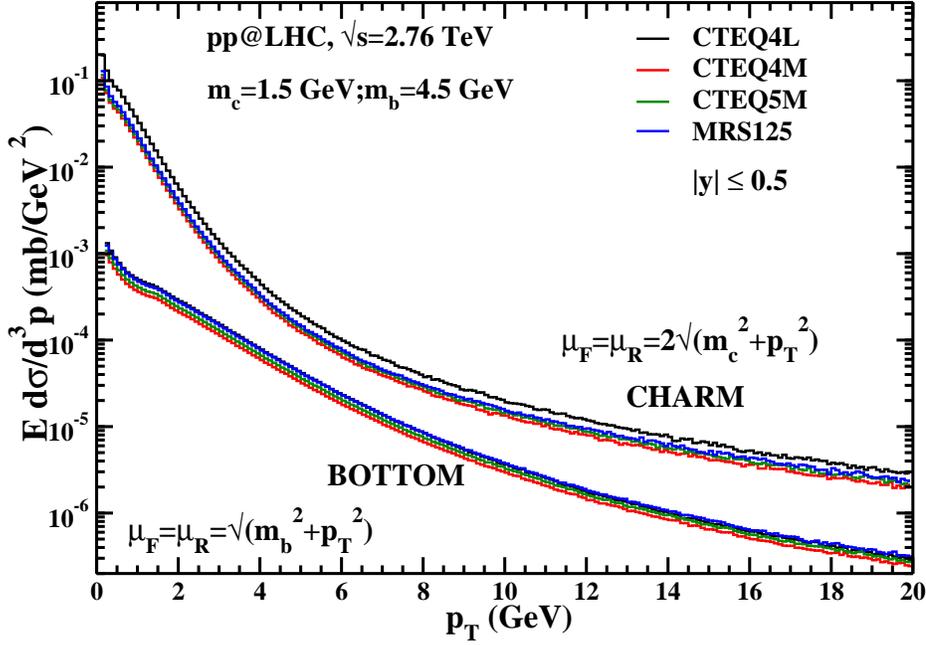}
\caption{Transverse momentum distribution of the heavy quarks in the central rapidity
region in $pp$ collisions at $\sqrt{s}$=2.76 TeV, for different structure functions 
using NLO pQCD.}
\label{ptdistr}
\end{center}
\end{figure*}
\begin{figure*}[h]
\begin{center}
\includegraphics[width=2.35in,angle=270]{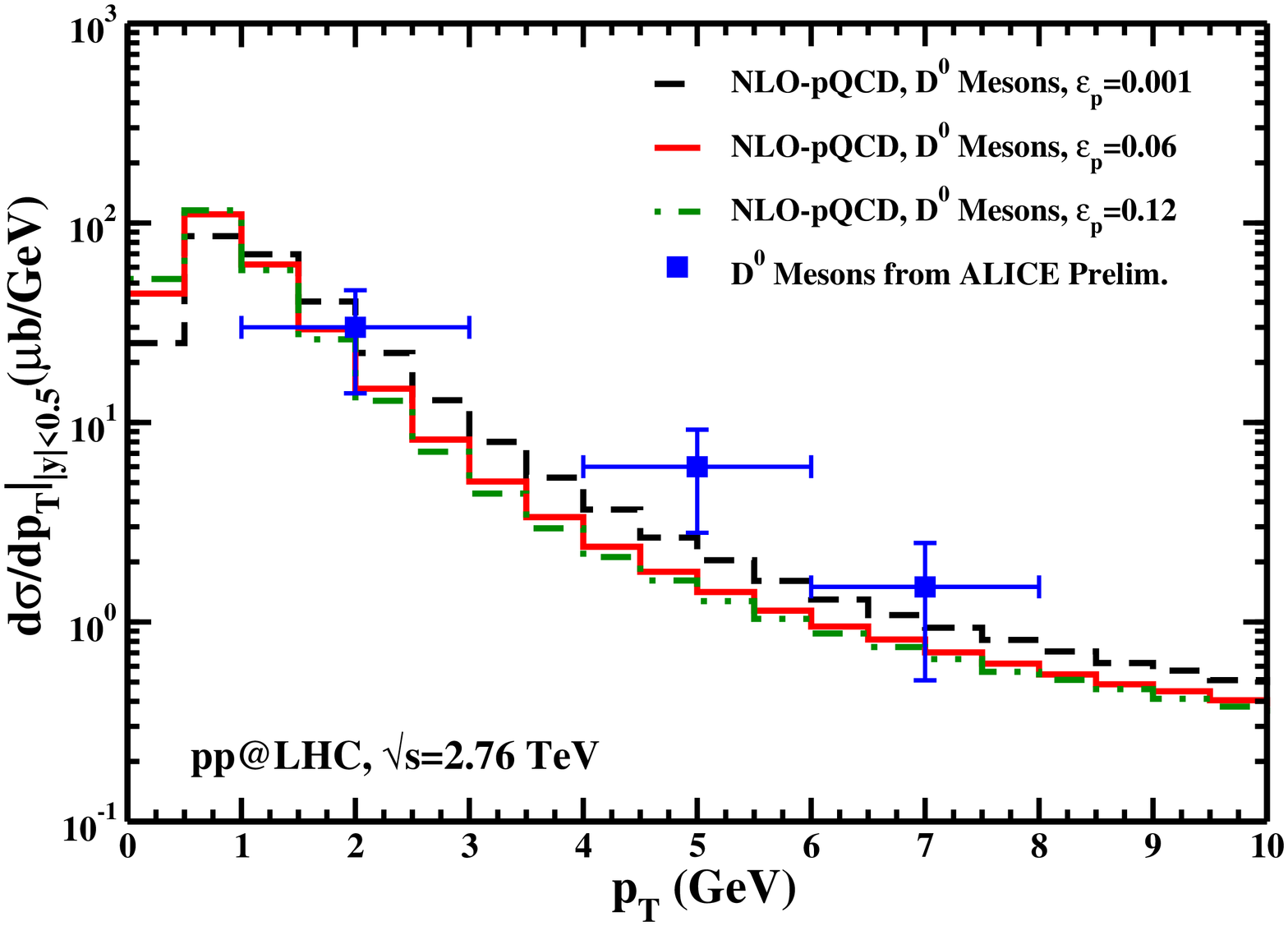}
\includegraphics[width=2.35in,angle=270]{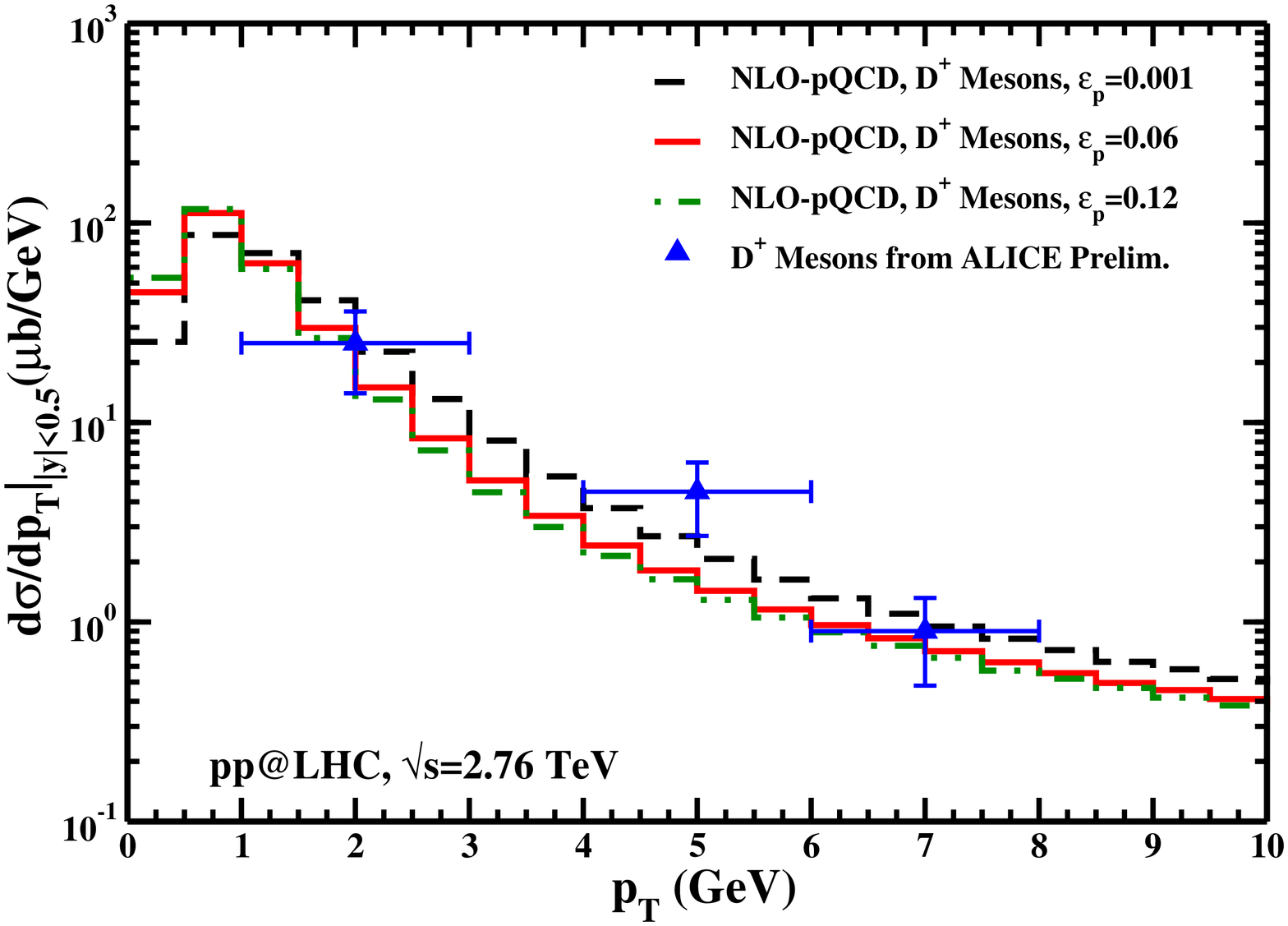}
\caption{(left) Transverse momentum distribution of D$^{0}$-mesons and (right) 
 of D$^{+}$-mesons, in pp collisions for $\sqrt{s}$= 2.76 TeV.} 
\label{dmesons}
\end{center}
\end{figure*}

\begin{figure*}[h]
\begin{center}
\includegraphics[width=2.35in,angle=270]{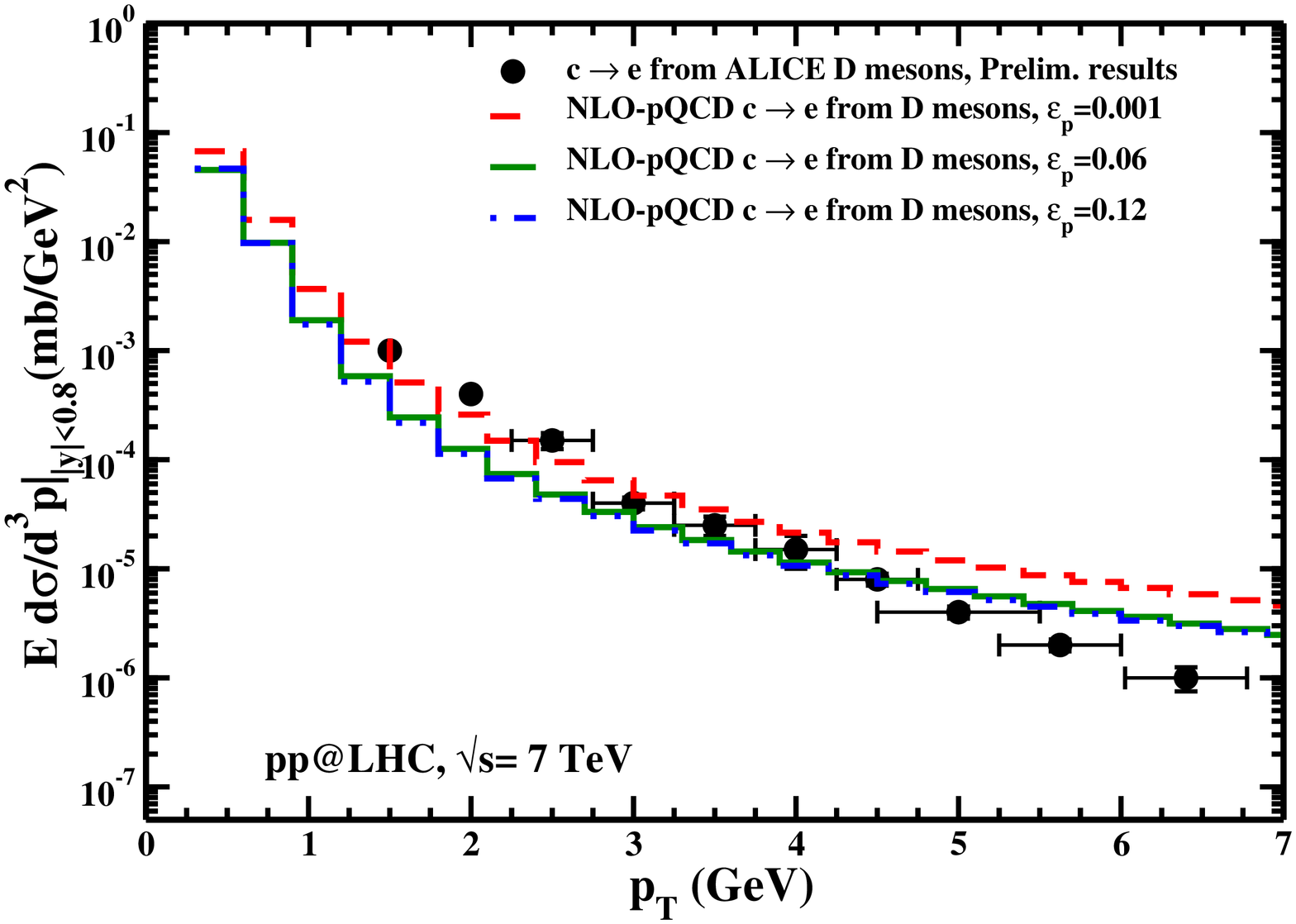}
\caption{Transverse momentum distribution of single electrons 
from $pp$ collisions at $\sqrt{s}$ = 7 TeV.} 
\label{electrons}
\end{center}
\end{figure*}

 Their large mass also provides that, even though
buffeted by light quarks and gluons during their passage through the
quark gluon plasma, the direction of their motion may not change 
substantially. This should make them 
a valuable probe for the properties of the plasma which depend 
on the reaction plane. Our understanding of the 
effect of the dead cone~\cite{deadcone1, deadcone2, deadcone3} on the suppression
of radiation has undergone quite some evolution since it was proposed
earlier.  It is also not yet clearly established that heavy
quarks will completely thermalize in the plasma formed at RHIC 
and LHC energies (see Ref.~\cite{hvqtherm1}).
However it must be safe to assume that the drag~\cite{drag} suffered by heavy quarks
will mostly slow it down and the so-called diffusion~\cite{diffusion}  processes
 will not alter its direction
considerably. Thus, the azimuthal correlation of heavy quarks integrated
over $p_T$ may be reasonably immune to the energy loss suffered by them. 

\begin{figure*}[h]
\begin{center}
\includegraphics[width=2.35in,angle=270]{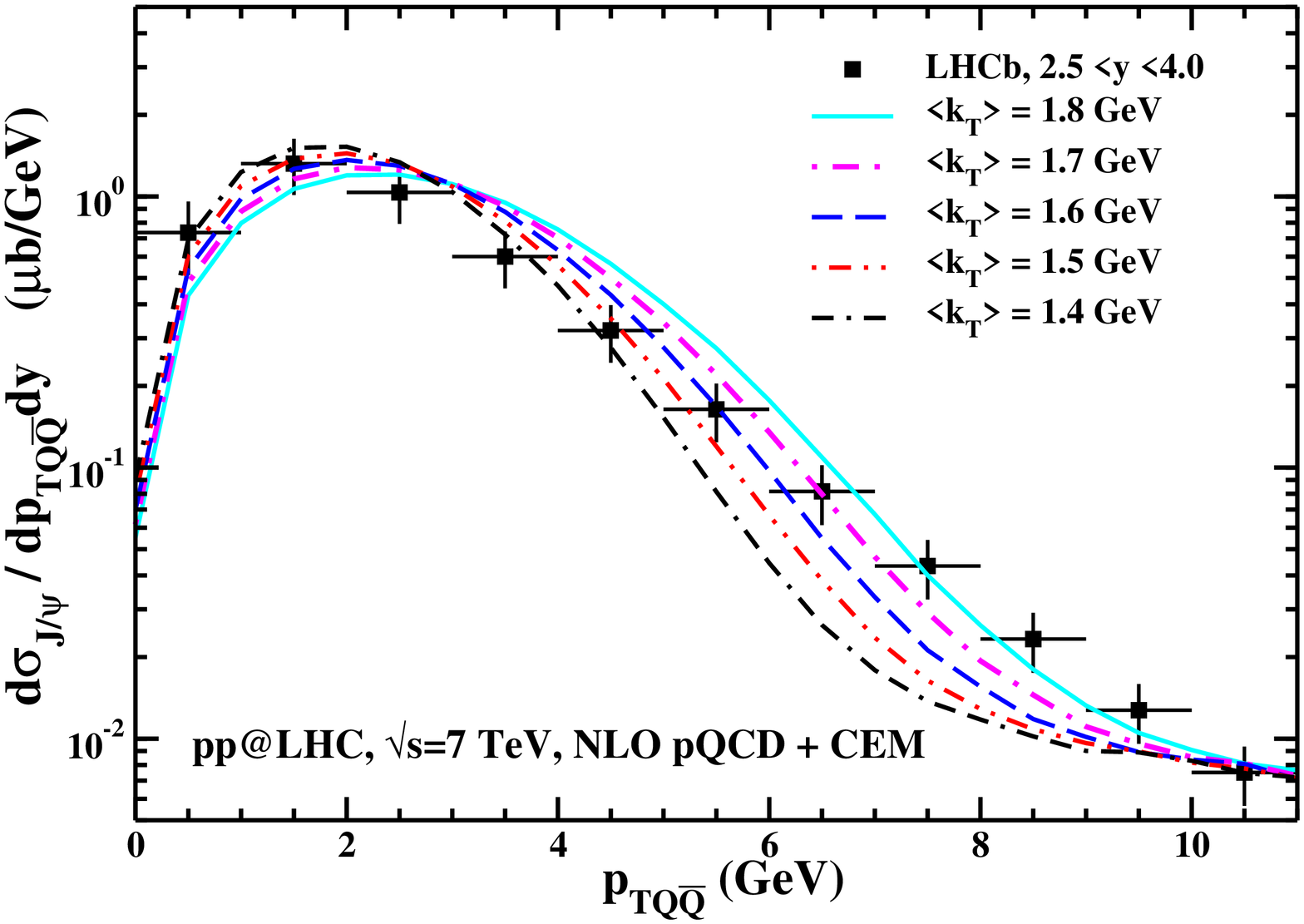}
\includegraphics[width=2.35in,angle=270]{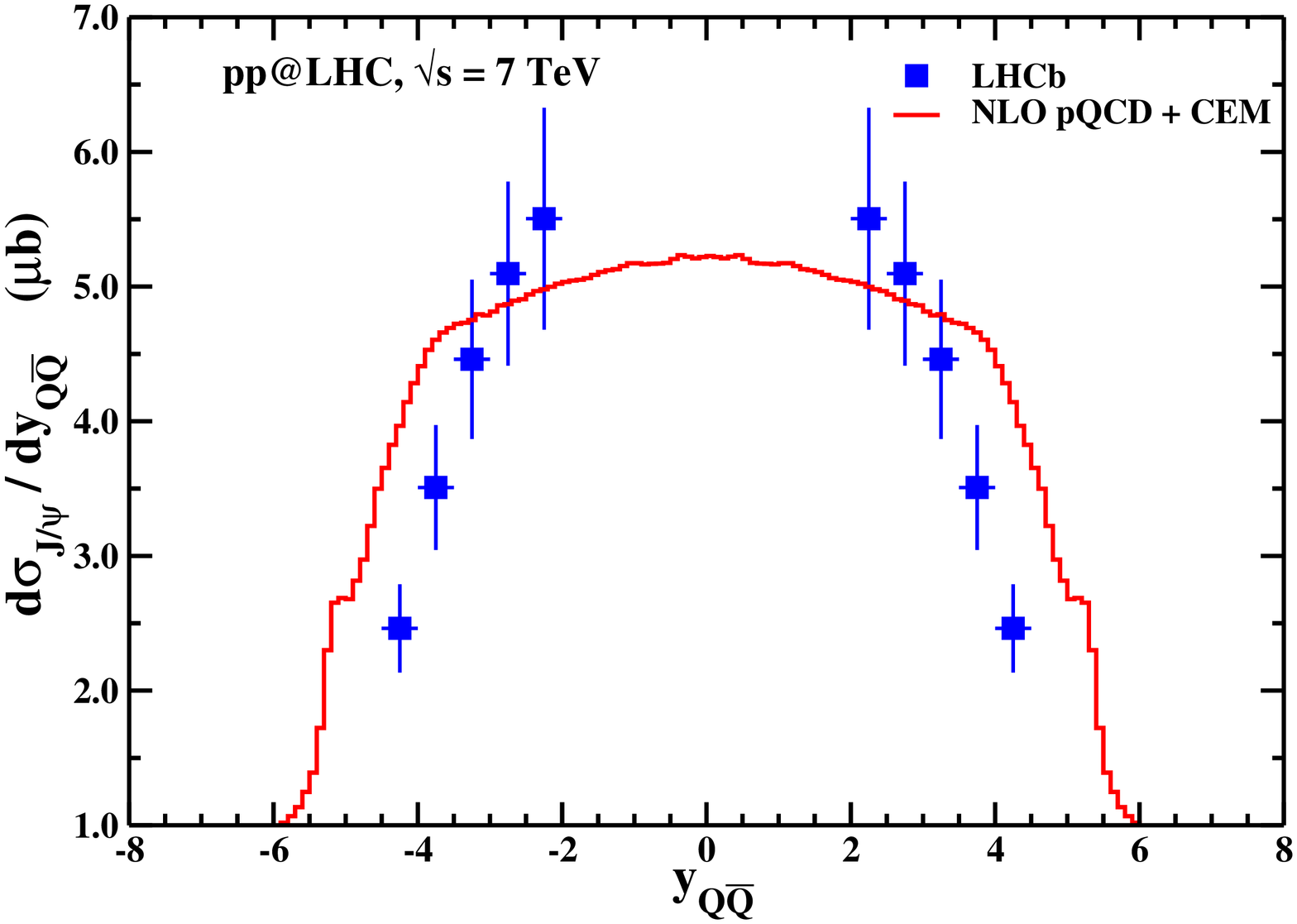}
\caption{Transverse momentum (left panel) and rapidity distribution 
(right panel) of $J/\psi$ from $pp$ collision at $\sqrt{s}$= 7 TeV, using color
evaporation model.} 
\label{jpsi}
\end{center}
\end{figure*}

The heavy quarks could be influenced by the flow~\cite{nuxu1} generated
in such collisions. If this is true, then a very interesting situation
may arise for heavy quarks which is not possible for light quarks or gluons.
Consider a $Q\overline{Q}$ pair produced in a central collision having $y=0$.
At leading order, their transverse momenta would be equal in magnitude
and point towards opposite directions. 
Consider a heavy quark Q moving
away from the centre  with momentum $\bf{p_T}$. Then its
partner $\overline{Q}$ would move with momentum $-\bf{p_T}$ towards the
centre. Their velocities would be $\bf{v_Q}=\pm \bf{p_T}$ $/M_T$,
where $M_T=\sqrt{p_T^2+M_Q^2}$,
and $M_Q$ is the mass of the heavy quark.  Let the radial flow velocity
be $\bf{v_f}$. Now if $|\bf{v_f}| \geq |\bf{v_Q}|$, the $\overline{Q}$ will turn
back and start moving away from the centre! Thus the $Q\overline{Q}$ pair,
which should have appeared back-to-back would appear as moving in the
same direction. This would drastically  alter the azimuthal correlation
 of the pair.  A similar change of direction of motion is not possible for
light quarks and gluons as they move with the speed of light. 
Taking, for example, $|\bf{v_f}|\approx$ 0.6, 
(see Ref.~\cite{nuxu1}) one can see that the azimuthal
correlation of charm quarks for $p_T \leq $ 1.2 GeV and for bottom quarks having
$p_T \leq $ 3.5 GeV could be considerably modified from their primordial
 value. Recalling that non-back-to-back heavy quarks are produced from NLO
processes (see later also), this would introduce an interesting richness
in these studies.
\begin{figure*}[h]
\begin{center}
\includegraphics[width=2.35in,angle=270]{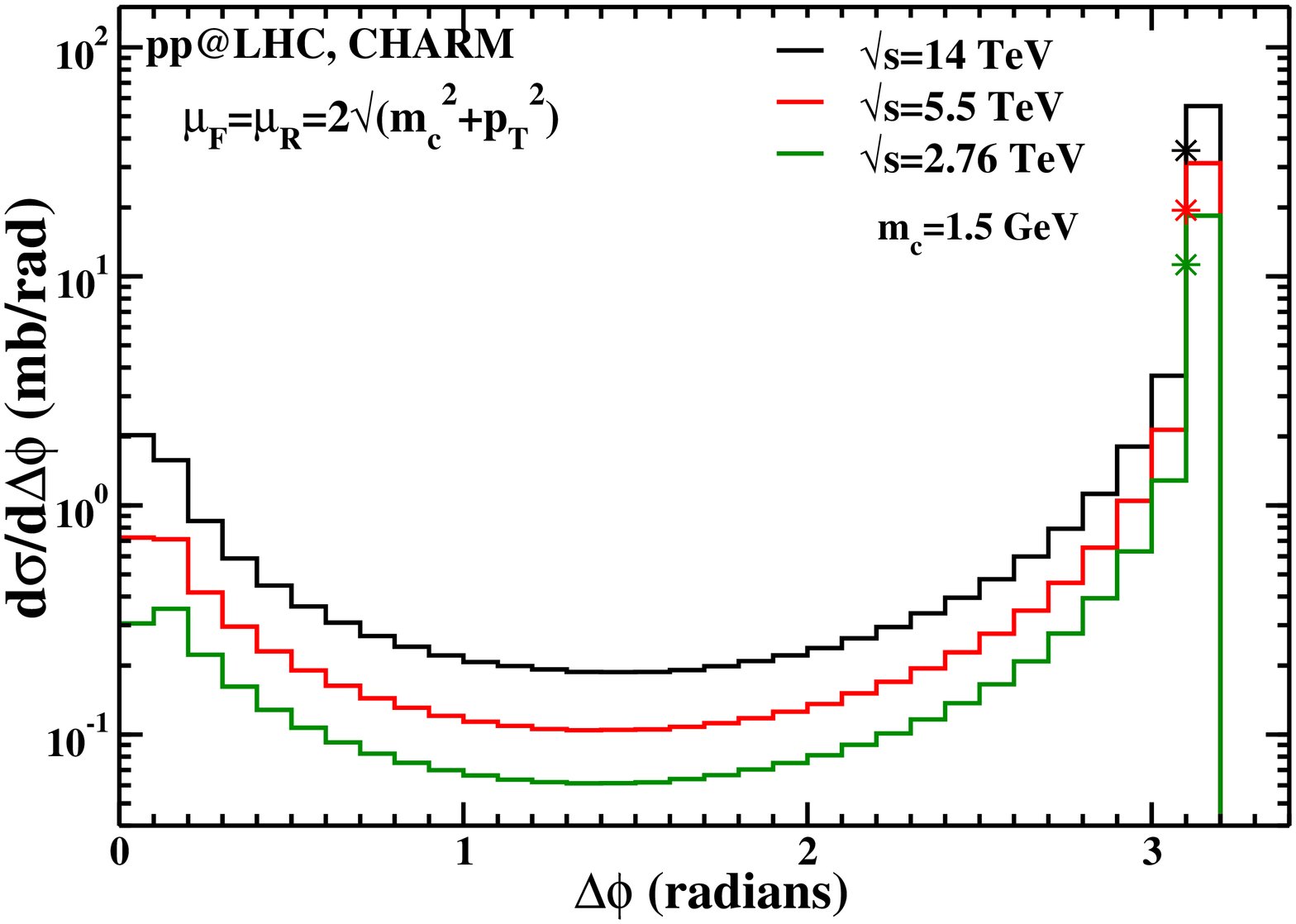}
\includegraphics[width=2.35in,angle=270]{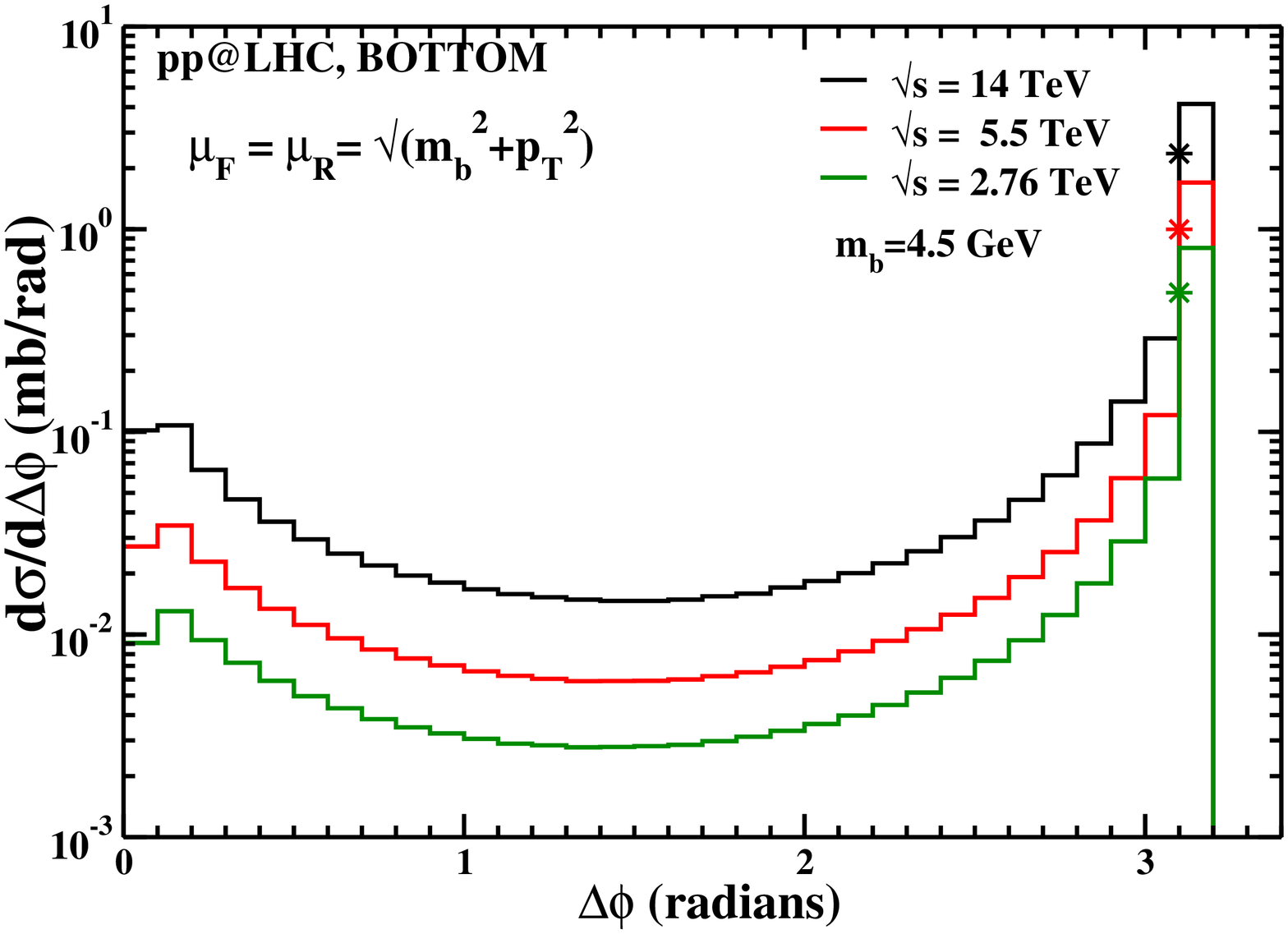}
\caption{Azimuthal correlation of charm (left panel) and
bottom (right panel) quarks at 2.76, 5.5 and 7 TeV for
$pp$ collisions. The symbols give the LO values for
 $d\sigma/d\phi= \sigma_{\textrm{LO}}/\delta(\Delta \phi)$
where $\delta(\Delta \phi)$ is the size of $\phi$ bin.}
\label{phi}
\end{center}
\end{figure*}

Now consider charm quarks (say) produced from the primary processes
 $gg \rightarrow Q\overline{Q}$ at leading order and 
$gg \rightarrow gQ\overline{Q}$ at next-to-leading order.
In the absence of any intrinsic $k_T$ for partons, the quarks from the first process
will be produced back-to-back,
while those from the second process will be mostly collinear and
 will additionally
be accompanied with a recoiling parton. A comparison of the energy loss
suffered by the recoiling parton and the heavy-quarks will allow
us to obtain flavour dependence of the energy loss. 
A considerable richness to this picture 
is added by the realization
that the splitting $g \rightarrow Q\overline{Q}$, would produce collinear heavy quarks, 
while the process $gg \rightarrow Q\overline{Q}g$, where
a gluon is radiated by one of the heavy quarks will essentially give rise to a 
flat azimuthal correlation.

So far we have discussed only the azimuthal correlation of the heavy quarks.
A study of the transverse momentum of the pair and the rapidity-difference of the pair
can help us disentangle the LO and the NLO processes. Recall that the transverse 
momentum of the $Q\overline{Q}$ pair would be
identically zero at LO and equal to that of the recoiling parton at NLO. Deviations from the
results for $pp$ collisions at the corresponding centre of mass energy in nuclear collisions 
will provide a measure of medium modifications as usual.

The entire discussion so far assumes that there may be no additional production
of heavy quarks after the initial prompt production. Often this production for
$pp$ collisions is taken  as a baseline for the study of nuclear modification
factor $R_{AA}$. It is obvious that any additional production of heavy quarks,
for example due to multiple scattering of high momentum quarks and gluons
produced similarly, see Refs.~\cite{lingyu,mdy1,lmw,liufries1} or due to passage of 
a high energy quarks or gluons through
the QGP ~\cite{mdy1,liufries1}, or due to scattering among the thermalized partons, 
if the temperature
is sufficiently large ~\cite{mdy1,lmw,shor}, will necessitate revision of our estimates
for the energy loss suffered by heavy quarks as they traverse the QGP, obtained
by analyzing the nuclear modification function $R_{AA}$, (see Ref.~\cite{energyloss}).

By now there is also a growing realization that $R_{AA}$ is not able to seriously 
discriminate between different mechanisms of energy loss and evolution of the
system~\cite{renk} and the correlation of the leading hadrons are slowly emerging as 
more discerning probes~\cite{wang}. Consider a simple example. We need to know the transverse
momentum of heavy quarks in $pp$ collisions in order to have a base-line to estimate
the nuclear modifications. The NLO pQCD results for these are easily approximated
by a $K$ factor multiplying the results for LO pQCD (see eg. Ref.~\cite{jamiln}).
Now consider the azimuthal correlations of heavy quarks  produced in similar collisions.  
As we discussed above, the LO pQCD results for the correlation is a delta function around
$\Delta\phi=\pi$. However, we shall see that the correlation function estimated at NLO, though
still peaking at $\Delta\phi=\pi$ fills up the phase-space from zero to  $\pi$ with an interesting
catenary like structure. Considering that one uses deviations from $pp$ collisions to obtain results
for nuclear modifications, these will have to be quantitatively understood for $pp$ collisions
before we can accurately decipher the later.

The present work aims at investigating azimuthal, momentum, and rapidity
correlations for heavy quark-anti quark pairs for $pp$ collisions and setting
the stage for the study of the deviations in these due to medium modifications
in heavy ion collisions at the corresponding energies. We also discuss the
complexities arising from the additional production of heavy quarks
due to multiple scatterings.

The paper is organized as follows. In the next section we discuss various
correlations for $pp$ collisions using NLO pQCD.
In Sect. 3 we discuss the 
azimuthal correlations in Pb+Pb collisions due to initial production 
and various multiple collisions. Our results for $pp$ and Pb+Pb collisions 
are discussed in Sect. 4 followed by conclusion in Sect. 5.


\section{Proton Proton Collisions}

The results for particle and photon productions in $pp$ collisions serve as a baseline in search for 
quark-gluon-plasma and other
medium effects at the corresponding centre of mass energy/nucleon for collision of heavy nuclei. 
This paradigm may have to be modified
if the recent suggestions for formation of QGP (perhaps only in high multiplicity events), Ref.~\cite{werner}
 in $pp$ collisions turn out to be valid. An alternative
criterion of comparing results for peripheral collisions to those for central collisions has also 
been used with considerable success,
 with the understanding that the peripheral collisions may be considered as 
a superposition of $pp$ collisions.

The correlation of heavy quarks produced in $pp$ collisions is defined in general as:
\begin{equation}
E_1 E_2\frac{d\sigma}{d^{3}p_1 d^{3}p_2}=
\frac{d\sigma}{dy_1 dy_2 d^{2}p_{T_1} d^{2}p_{T_2}}
=C~,
\label{ini1}
\end{equation}
where $y_1$ and $y_2$ are the rapidities of heavy quark and anti-quark and 
$\bf{p_T}_i$ are their transverse momenta.

At the leading order, the differential cross-section for the charm correlation from proton-proton 
collision can be written as:
\begin{equation}
C_{LO}=\frac{d\sigma}{d^{2}p_{T}dy_1 dy_2}
\delta{(\bf{p_T}_1+\bf{p_T}_2)}~.
\end{equation}

In the above $\bf{p_T}_1=\bf{p_T}_2=\bf{p_T}$ and
\begin{eqnarray}
\frac{d\sigma}{dy_1 dy_2 dp_{T}} &=& 2 x_{a}x_{b}p_{T}\sum_{ij}
\left[f^{(a)}_{i}(x_{a},Q^{2})f_{j}^{(b)}(x_{b},Q^{2})
\frac{d\hat{\sigma}_{ij}(\hat{s},\hat{t},\hat{u})}{d\hat{t}}
\right.\nonumber\\
&+& \left.f_{j}^{(a)}(x_{a},Q^{2})f_{i}^{(b)}(x_{b},Q^{2})
\frac{d\hat{\sigma}_{ij}(\hat{s},\hat{u},\hat{t})}{d\hat{t}}\right]
/(1+\delta_{ij})~,
\label{ini}
\end{eqnarray}
where $x_{a} $ and $x_{b} $ are the fractions of the momenta carried by the partons 
from their interacting parent hadrons.
These are given by
\begin{equation}
x_{a}=\frac{M_{T}}{\sqrt{s}}(e^{y_1}+e^{y_2});~ x_{b}=\frac{M_{T}}{\sqrt{s}}(e^{-y_1}+e^{-y_2})~.
\end{equation}
where $M_{T} $ is the transverse mass, $\sqrt{m_{Q}^{2}+p_{T}^{2}}$, of the produced heavy quark.
The subscripts $i$ and $j$ denote the interacting partons, and $f_{i}$ and $f_{j}$ 
are the partonic distribution functions for the nucleons. 
We shall use CTEQ5M structure function, though we have checked that similar results are obtained 
for other modern structure functions (see later).
 The differential cross-section for partonic interactions, $d\hat{\sigma}_{ij}/d\hat{t}$ 
is given by
\begin{equation}
\frac{d\hat{\sigma}_{ij}}{d\hat{t}} = \frac{\left|M\right|^{2}}{16\pi\hat{s}^{2}}~,
\label{dsdt}
\end{equation}
where $\left|M\right|^{2}$ is the invariant amplitude for different 
sub-processes as obtained from Ref.~\cite{combridge}.
The physical sub-processes included for the leading order, 
$\cal{O}$ $(\alpha_{s}^{2}) $ production of heavy quarks are:
\begin{eqnarray}
g+g \rightarrow Q+\overline{Q}\nonumber\\
q+\bar{q} \rightarrow Q+\overline{Q}~.
\end{eqnarray}

At next-to-leading order, $\cal{O}$ $(\alpha_{s}^{3})$ subprocesses 
included are as follows
\begin{eqnarray}
g+g \rightarrow Q+\overline{Q}+g\nonumber\\
q+\bar{q} \rightarrow Q+\overline{Q}+g\nonumber\\
g+q(\bar{q}) \rightarrow Q+\overline{Q}+q(\bar{q})~.
\end{eqnarray}

We show our results for azimuthal correlation $C(\Delta\phi)$, where 
$\Delta\phi$=$|\phi_1-\phi_2|$
as well as rapidity correlations, $C(\Delta y)$, where 
$\Delta y$=$y_1-y_2$, of produced heavy quarks. 
We also present $(\Delta\eta, \Delta \phi)$  correlations
in the jet radius parameter, $R$,  
where $R=\sqrt{(\Delta\eta)^2+(\Delta\phi)^2}$ along
with the transverse momentum, invariant mass, and rapidity of the pair.

We verify the accuracy of our results
by evaluating the production of $J/\psi$ and charm 
measured recently.

\section{\bf Lead Lead Collisions}

Let us now move towards Pb+Pb collisions under study at the
Large Hadron Collider (LHC).  
We have discussed that most of the heavy-quarks and so also
quarks and gluons having large transverse momenta are
produced in initial hard collisions. At the energies 
reached at the LHC, the sheer number of quarks and gluons 
produced in these collisions leads to vehement multiple
collisions and gluon multiplication. This then leads to
a quark-gluon plasma at a very large initial temperature.

As discussed earlier, we would like to know if 
these initial temperatures  are large enough to produce heavy quarks as well (see eg. Ref.~\cite{shor}). 
The multiple collisions among the very high momentum quarks and gluons
(the so called jet-jet collisions) have been seen earlier to produce 
substantial number of heavy quarks. These jets, produced at very early times $\tau \approx 1/p_T$
will have to necessarily pass through the QGP which will be formed only after 
$\tau \approx $ 0.1 fm/$c$. Do these lead to a substantial production of heavy quarks?
Some of these questions have been addressed earlier~\cite{lingyu,mdy1,lmw,liufries1}.

Since those early studies, several new developments have taken place.
We now know the particle rapidly density (see eg. Ref.~\cite{Aamodt}), important to 
calculate the initial conditions, for which only values estimated by
several authors were known earlier. There has, now, been a growing
realization that jet-quenching measured in terms of the nuclear modification
function $R_\textrm{AA}$ is not able to seriously discriminate between various
theories of evolution of the plasma and the mechanism of energy loss. Thus 
correlations are being studied more closely to help us in this enterprise.

Thus we extend our earlier study~\cite{mdy1} to explore the correlations of
heavy quarks in collision of heavy nuclei due to initial production and various
multiple collisions, e.g., jet-jet interactions, jet-plasma interactions,
and the scattering of thermal partons, to see if these processes make large
contribution to the correlation. This is important as, at least the jet-jet
collision was found to make a large contribution to the production of
heavy quarks~\cite{mdy1}.

\subsection{Prompt Interactions}

The basic formulation which gives the correlation of 
produced heavy quarks from initial 
fusion of gluons and quark-anti quark
annihilation in proton-proton collision is given by Eq.~\ref{ini1}. 
Thus the azimuthal distribution of heavy quark for Pb+Pb collision 
at $b=0$ is given by
\begin{equation}
 E_1 E_2\frac{dN}{d^3p_1\,d^3p_2}=T_{AA} E_1 E_2 \frac{d\sigma_{pp}}{d^3p_1\,d^3p_2}~.
\end{equation}

For central collisions of lead nuclei, the nuclear thickness function is taken as 
$T_{AA}$= 292 fm$^{-2}$. In the above     
$\bf{p_1}$ and $\bf{p_2}$ are the momenta of the 
heavy quarks produced.

\subsection{Jet-Jet Interaction}
The initial hard scattering will produce massless gluons 
and light quarks in large numbers. 
These partons have large transverse
momenta. These quarks and gluons may ultimately 
thermalize because of frequent interactions among themselves 
and if sufficient energy is available, 
their interactions may lead to the production of heavy quarks 
as well. Here we give the formulation for  
azimuthal distribution of produced heavy quarks pair from jet-jet 
interaction.  
Since the jet-jet contribution to the heavy 
quark production is comparable to that of primary 
production~\cite{lingyu,mdy1}, it should be interesting to
see if their azimuthal distributions differ.

As a first step we obtain the distribution of
light partons, having $p_T$ $>$ 2 GeV, from a LO pQCD calculation using CTEQ5M structure 
function, for $pp$ collisions at 2.76 TeV and 5.5 TeV.
We parametrize them as:
\begin{eqnarray}
\frac{dN}{dyd^{2}p_{T}} &=& \left.T_{\rm{AA}}
\frac{d\sigma^{\rm {jet}}_{pp}}{d^{2}p_{T}dy}\right|_{y=0}\nonumber\\
                    &=& K \,\frac{C}{(1+p_{T}/B)^{\beta}}\nonumber\\
&=&h_{\mathrm{jet}}(p_{T})~,
\label{jetdistr}
\end{eqnarray}
where the $K$ factor is taken as 2.5 to account 
for higher order effects 
and the parameters $C,\ B,$ and $\beta$ are given in Table~1.
The factorization and renormalization scales are chosen as 
$Q = p_T$


\begin{figure*}[h]
\begin{center}
\includegraphics[height=3in,width=2.5in,,angle=270]{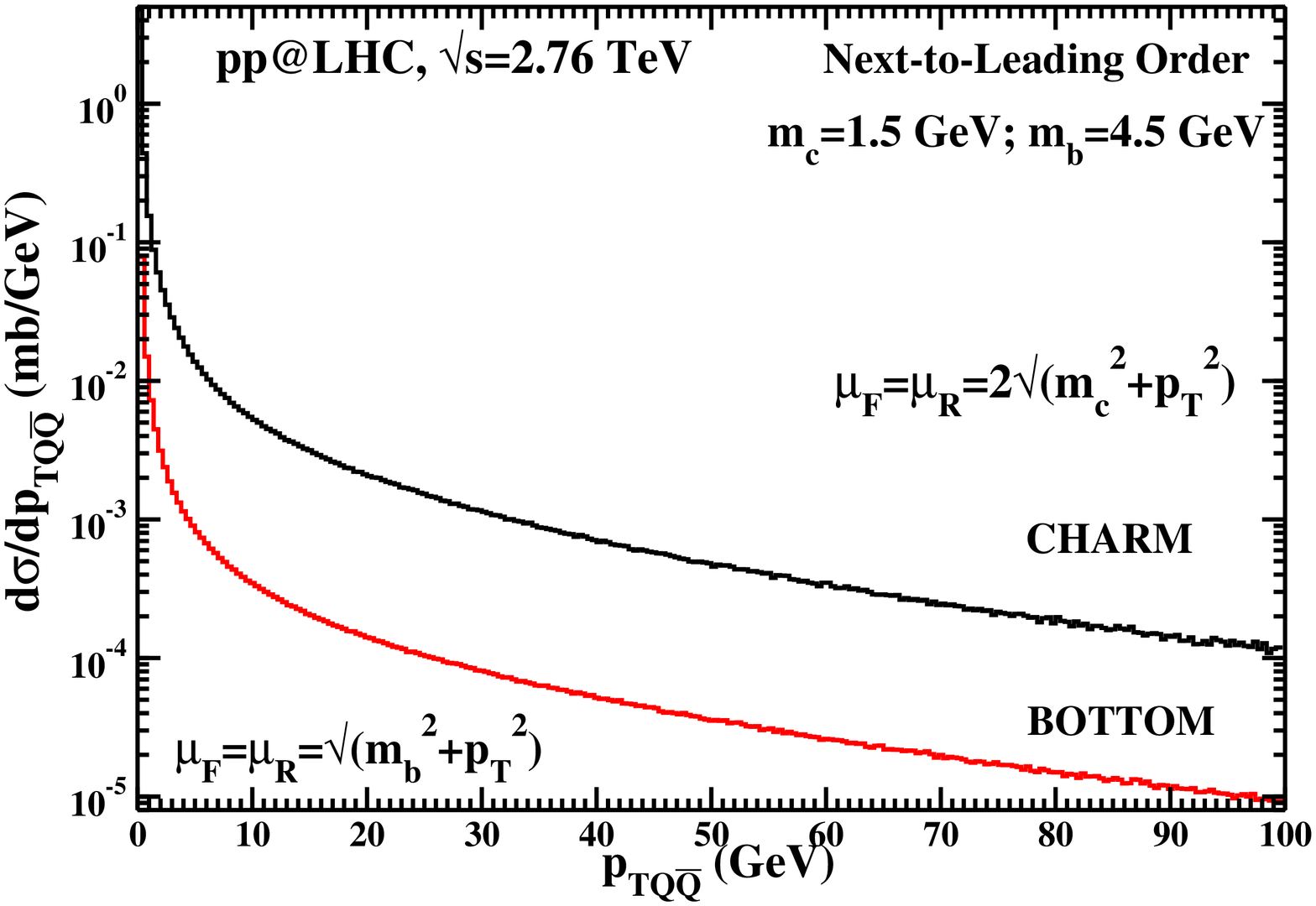}
\includegraphics[height=3in,width=2.5in,angle=270]{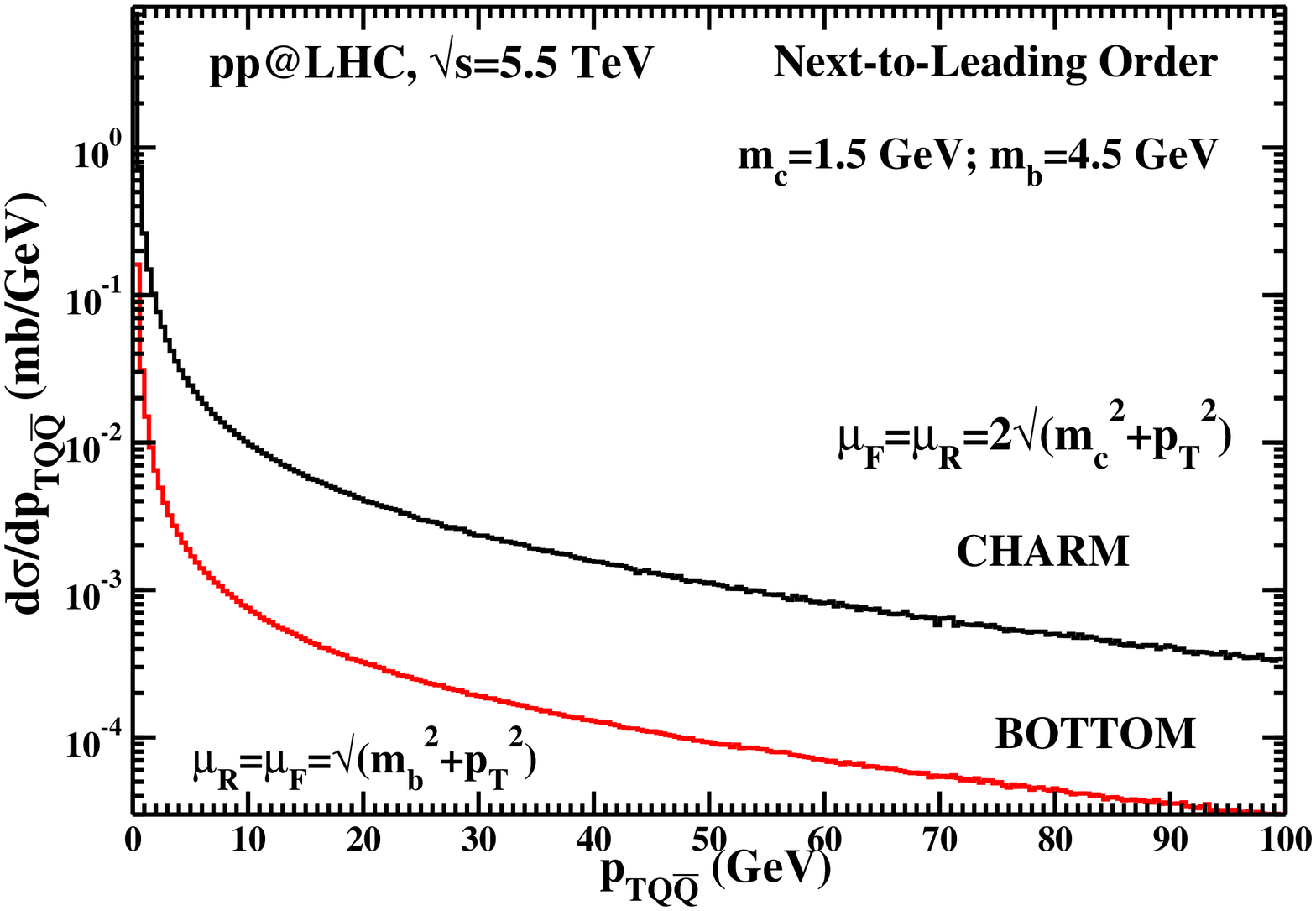}
\includegraphics[height=3in,width=2.5in,angle=270]{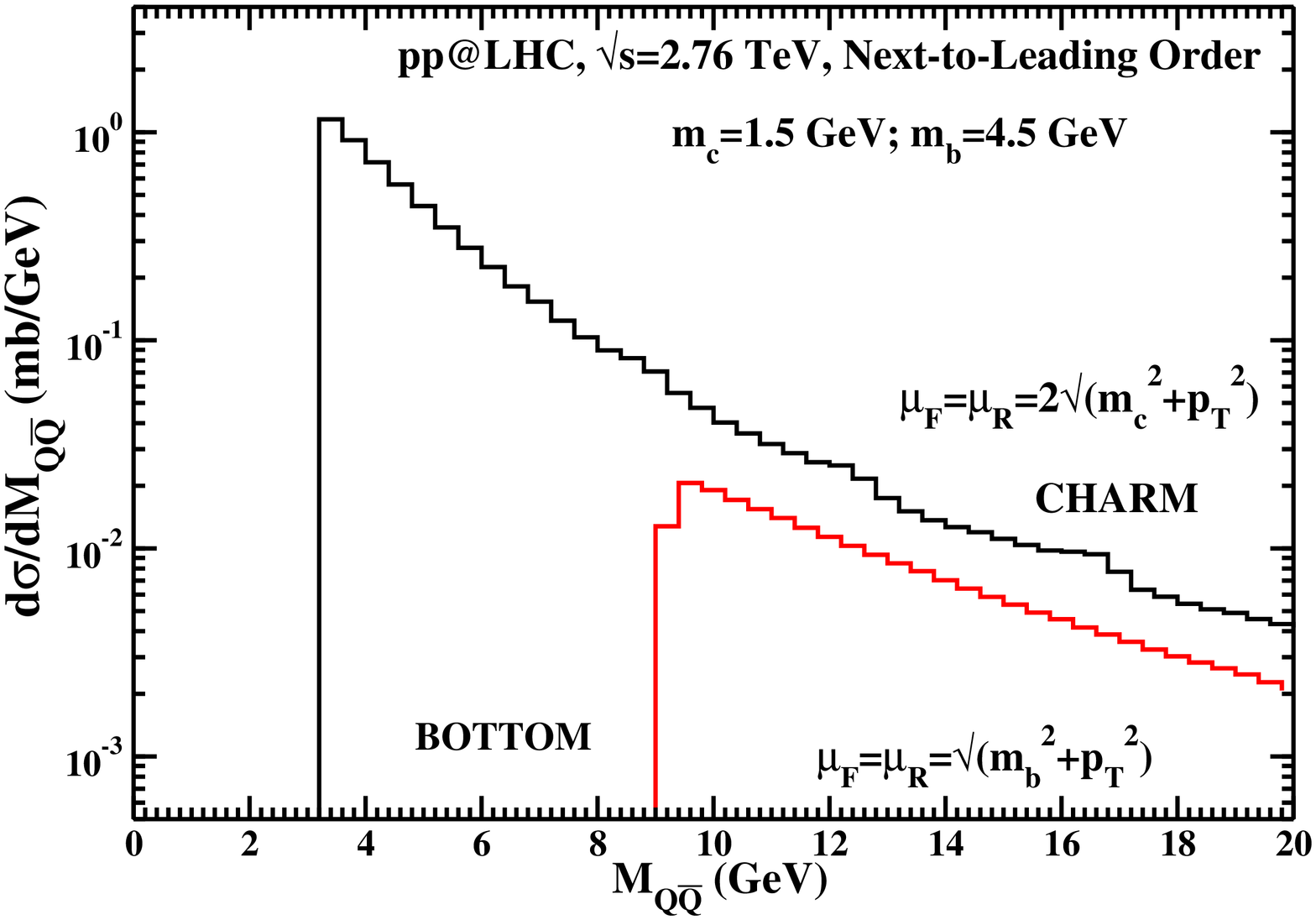}
\includegraphics[height=3in,width=2.5in,angle=270]{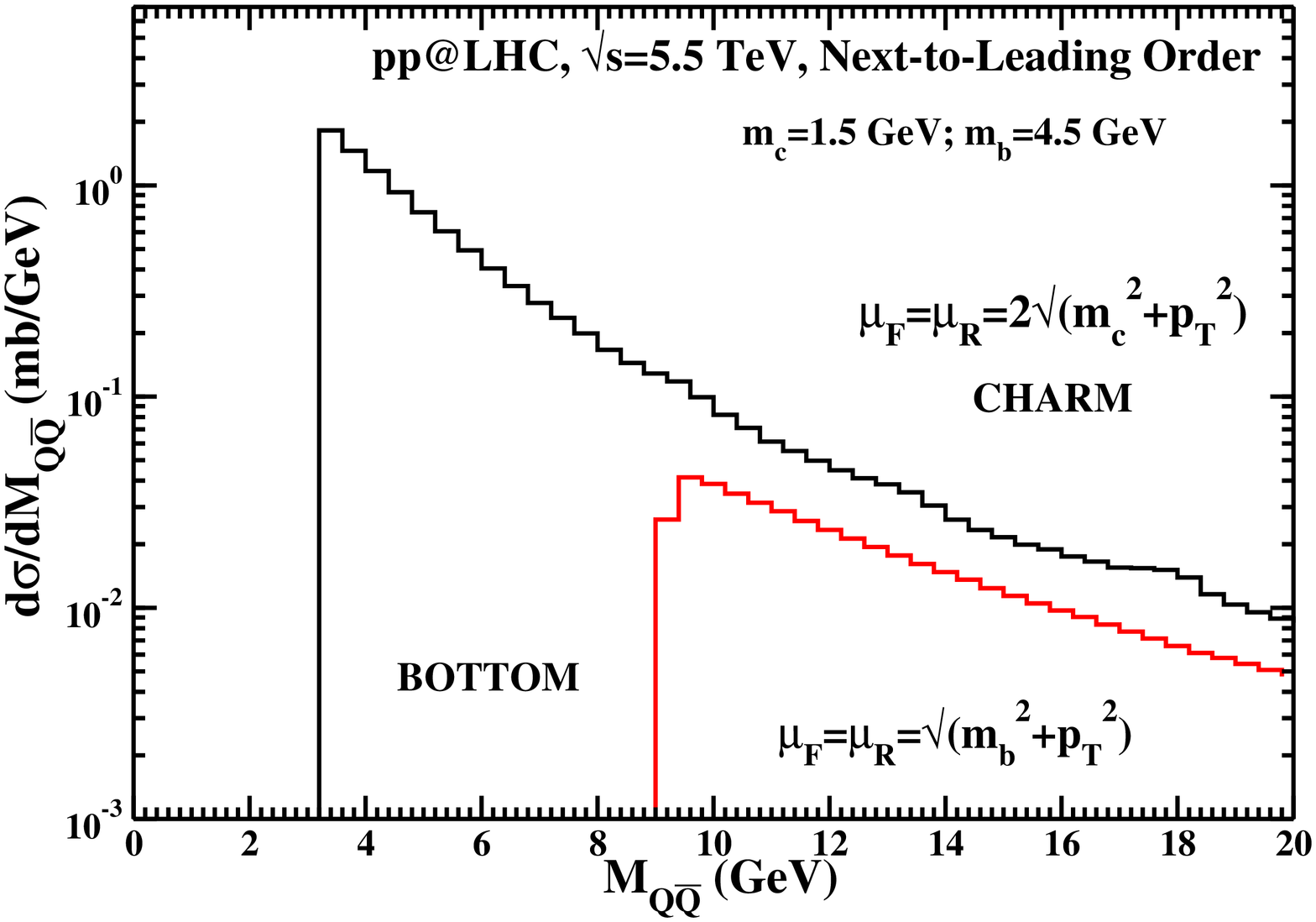}
\includegraphics[height=3in,width=2.5in,angle=270]{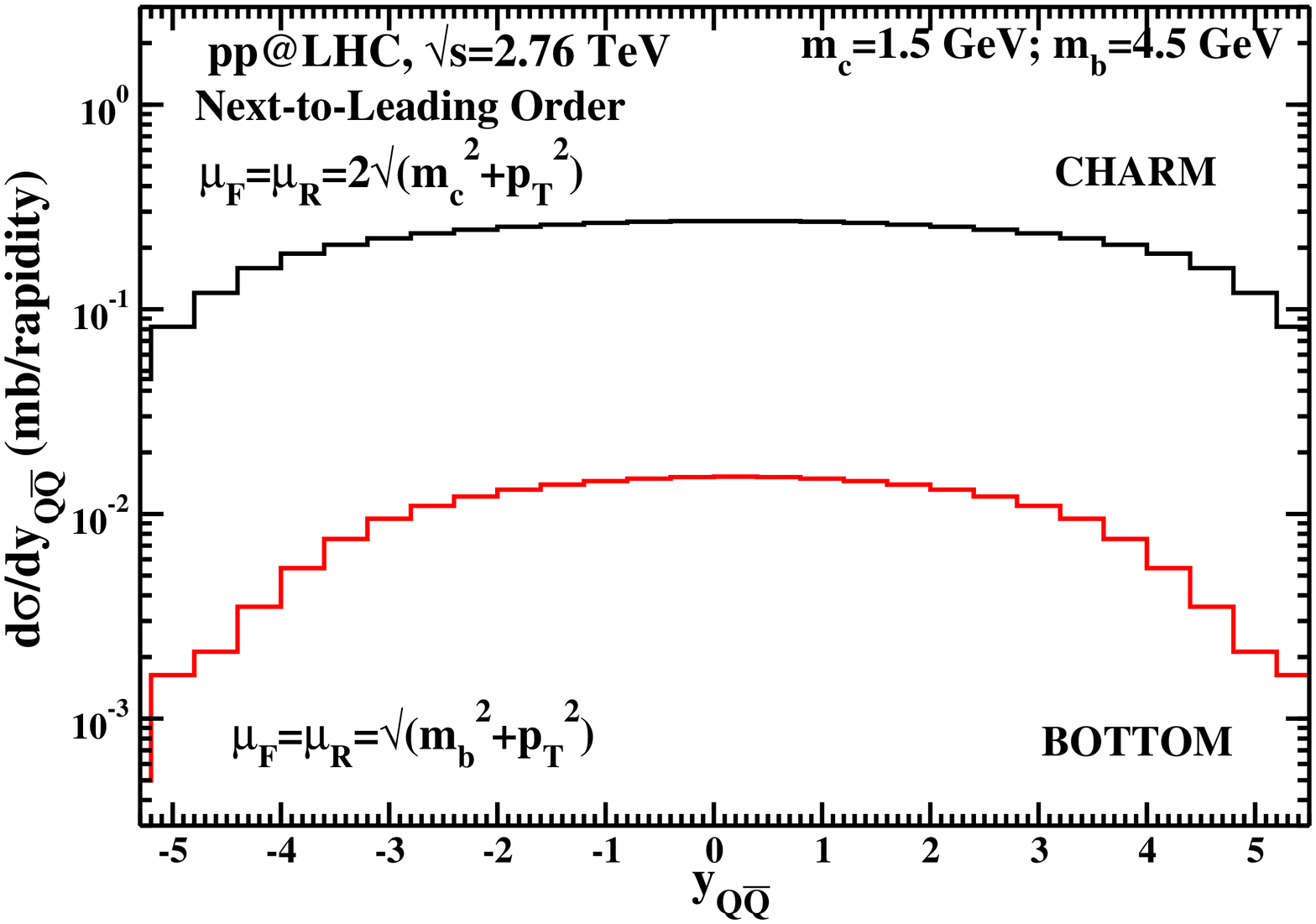}
\includegraphics[height=3in,width=2.5in,angle=270]{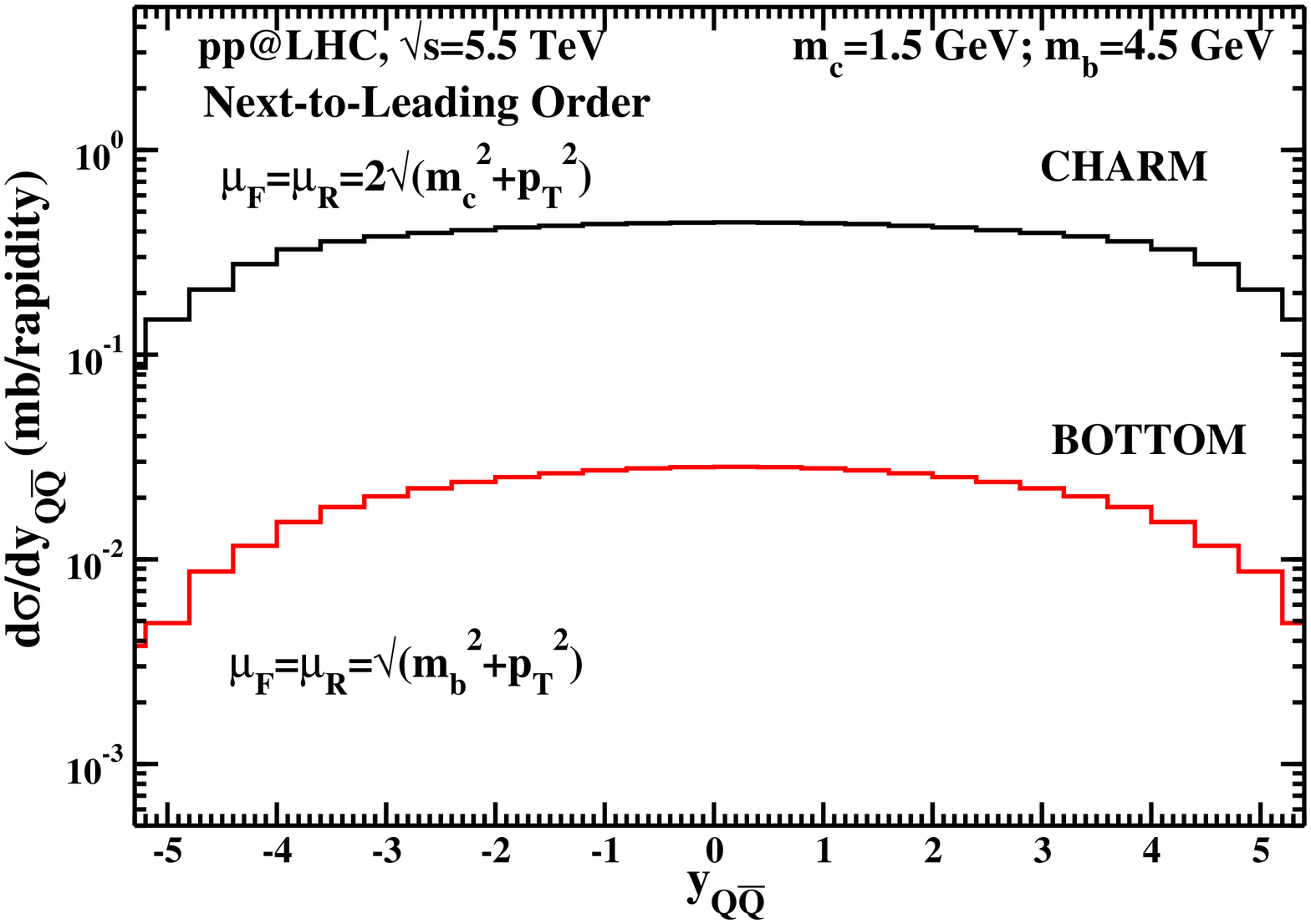}
\caption{Transverse momentum, invariant mass and rapidity distribution of 
charm and bottom quark pairs at LHC.}
\label{mqq}
\end{center}
\end{figure*}

\begin{figure*}[h]
\begin{center}
\includegraphics[height=3in,width=2.54in,angle=270]{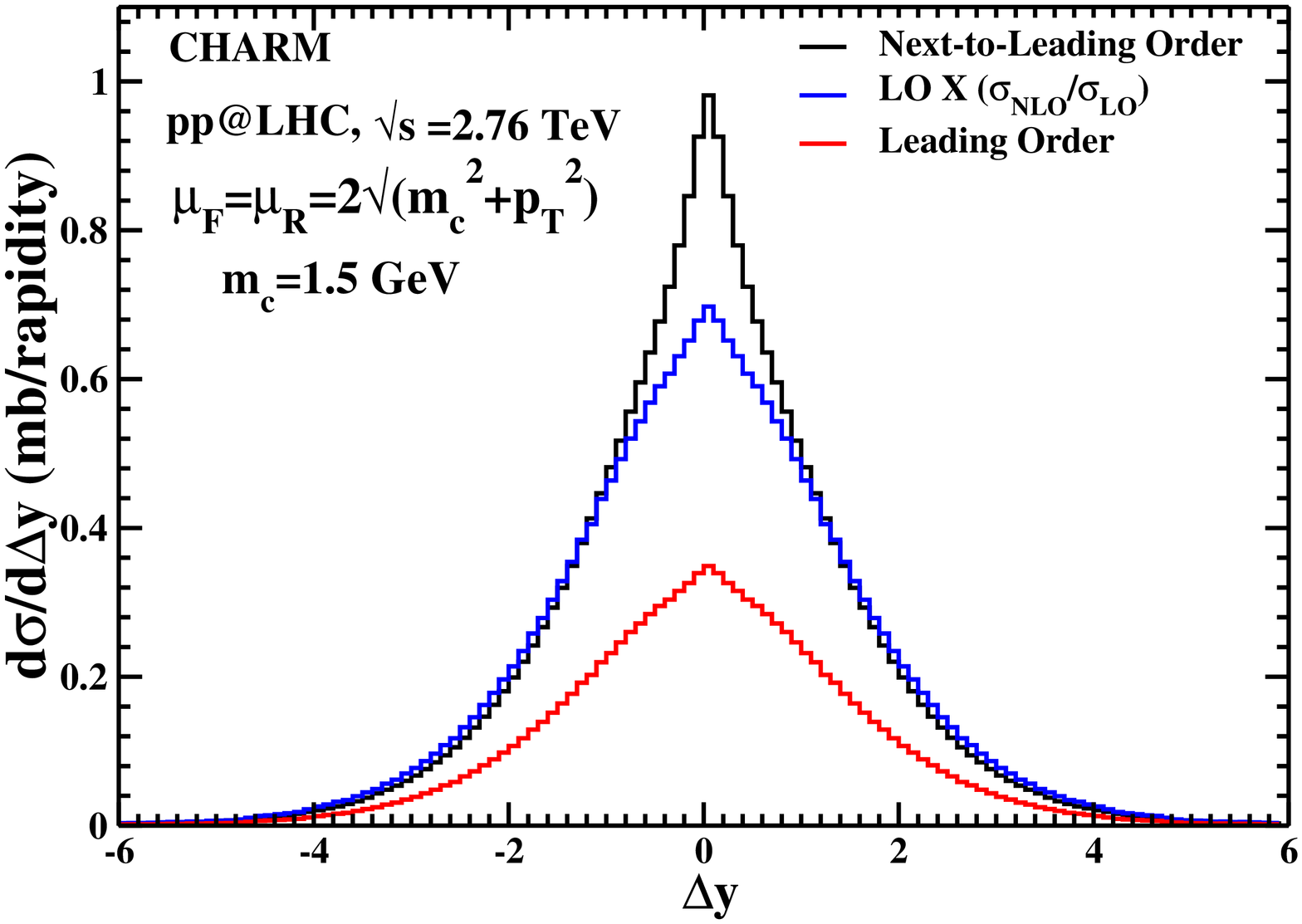}
\includegraphics[height=3in,width=2.54in,angle=270]{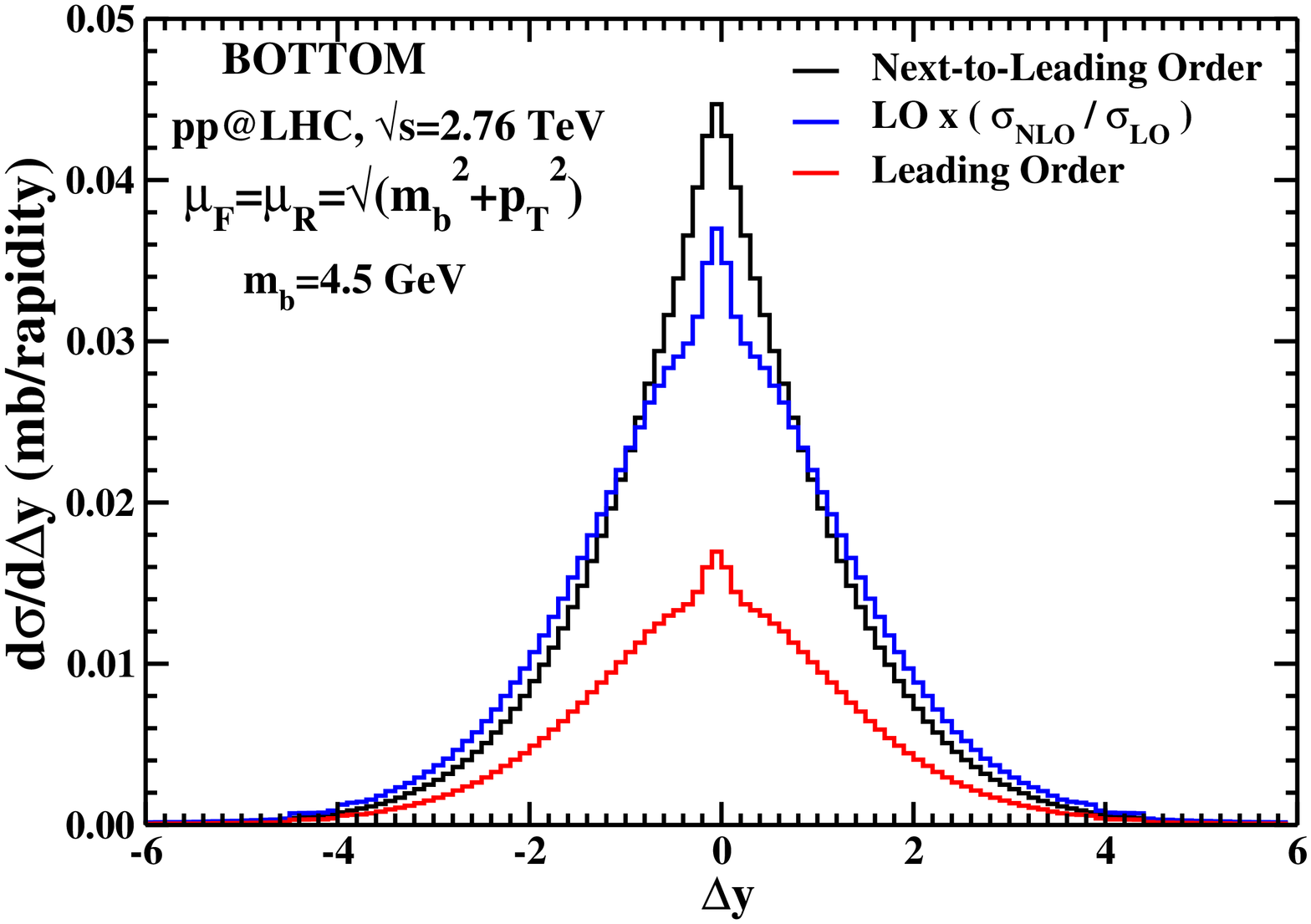}
\includegraphics[height=3in,width=2.54in,angle=270]{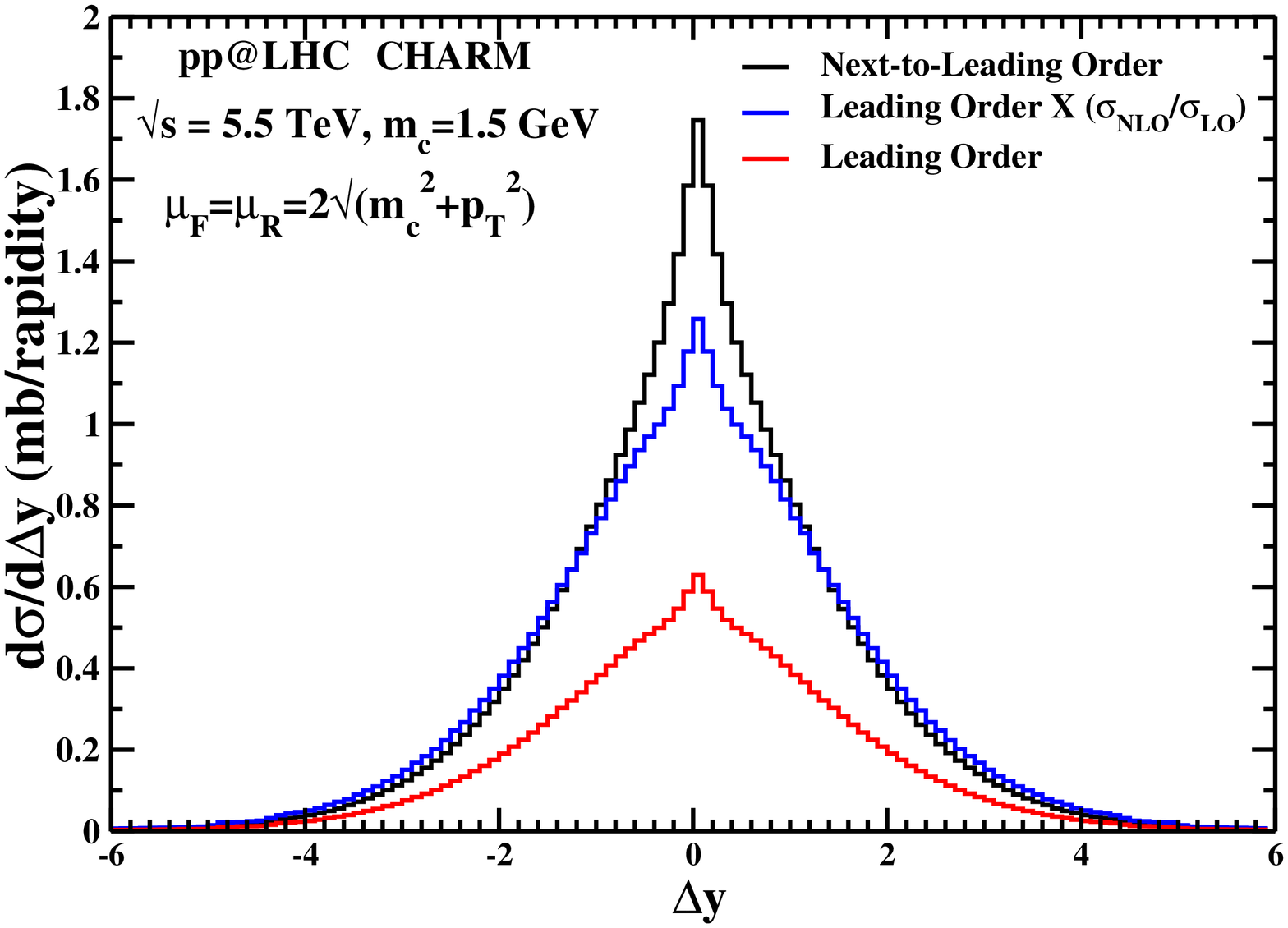}
\includegraphics[height=3in,width=2.54in,angle=270]{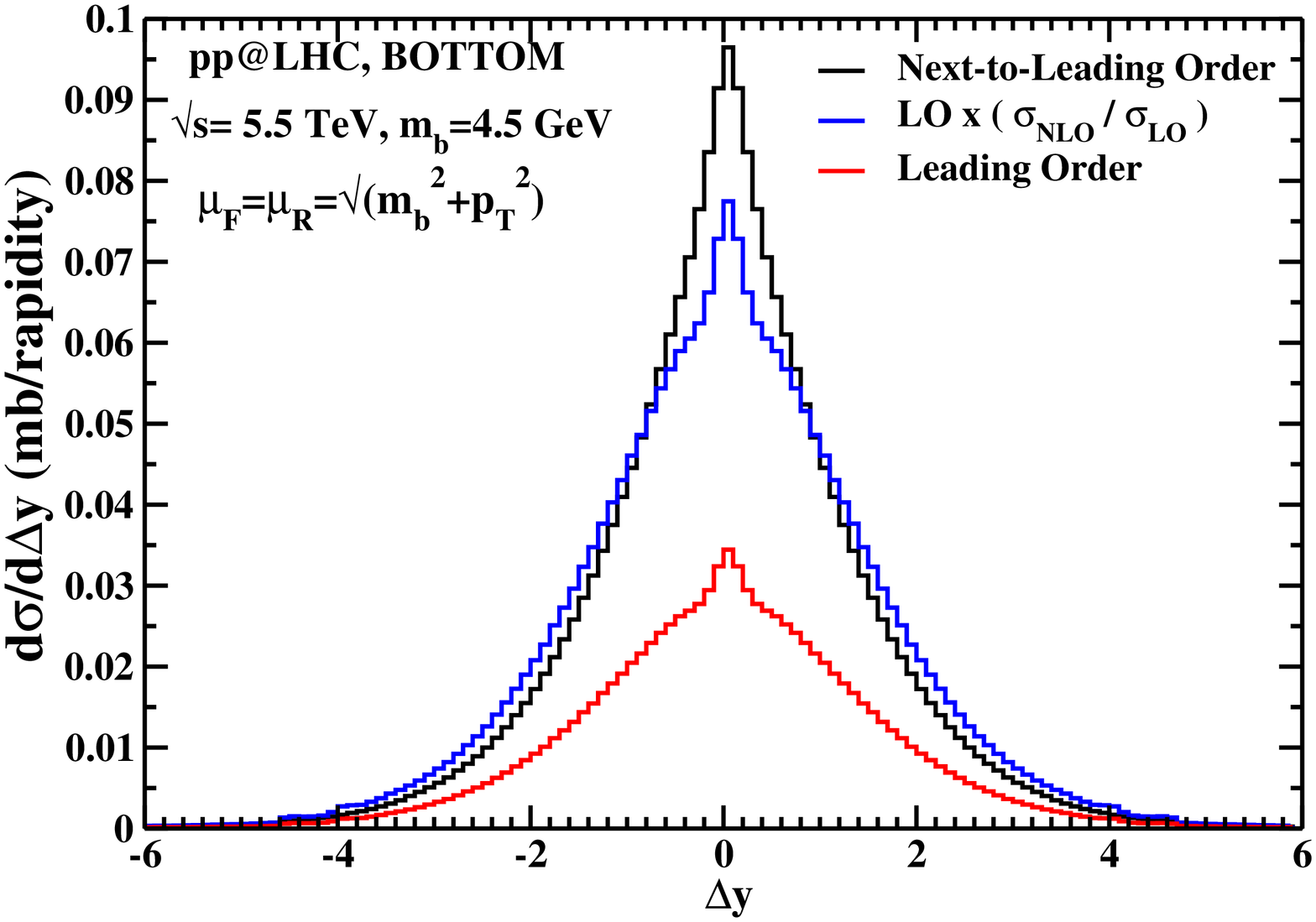}
\caption{Rapidity difference of $\Delta y=y_{Q}-y_{\bar{Q}}$, of charm and bottom quarks 
at LO and NLO in pp collisions.}
\label{Dy}
\end{center}
\end{figure*}

\begin{figure*}[h]
\begin{center}
\includegraphics[height=3in,width=2.54in,angle=270]{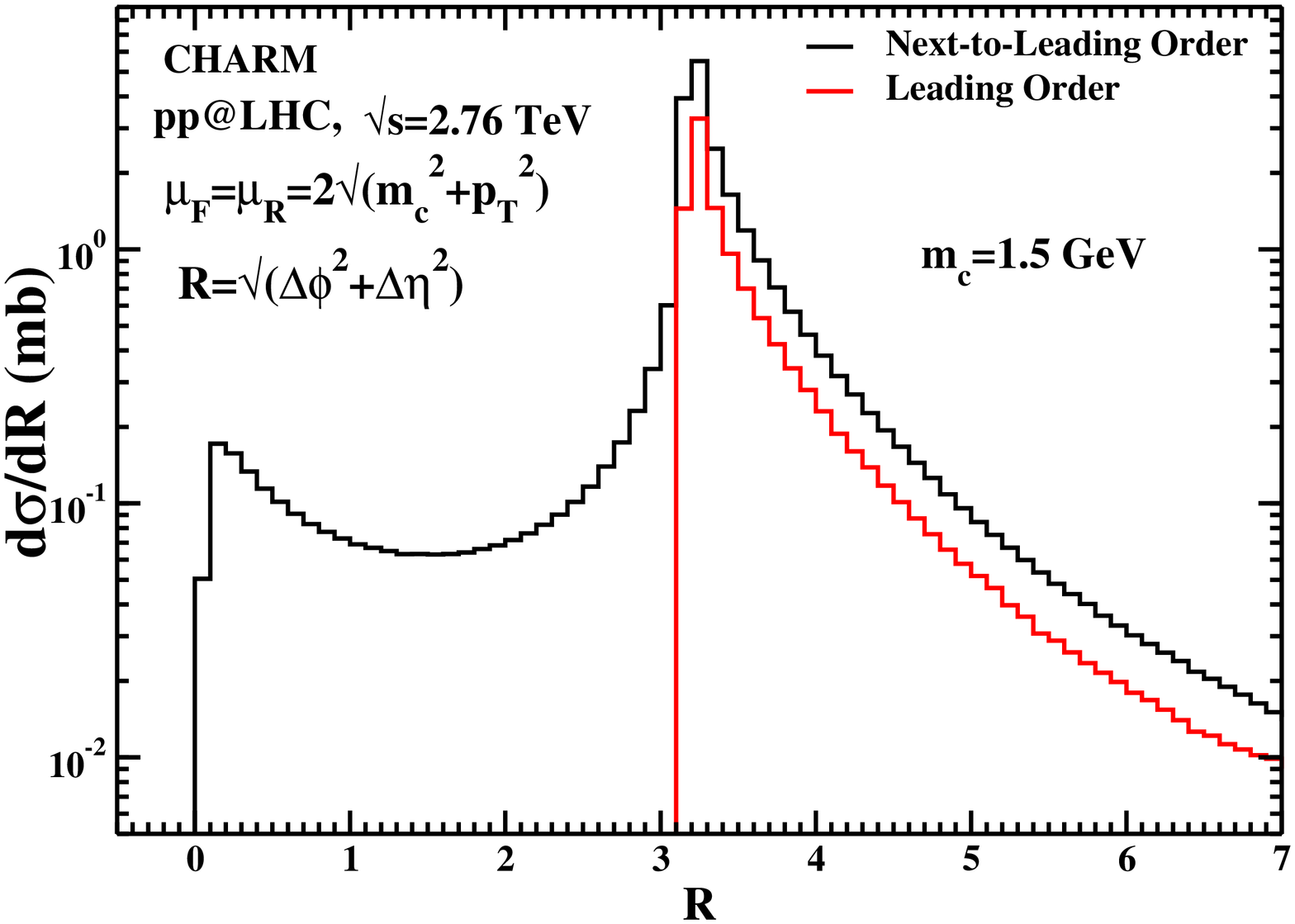}
\includegraphics[height=3in,width=2.54in,angle=270]{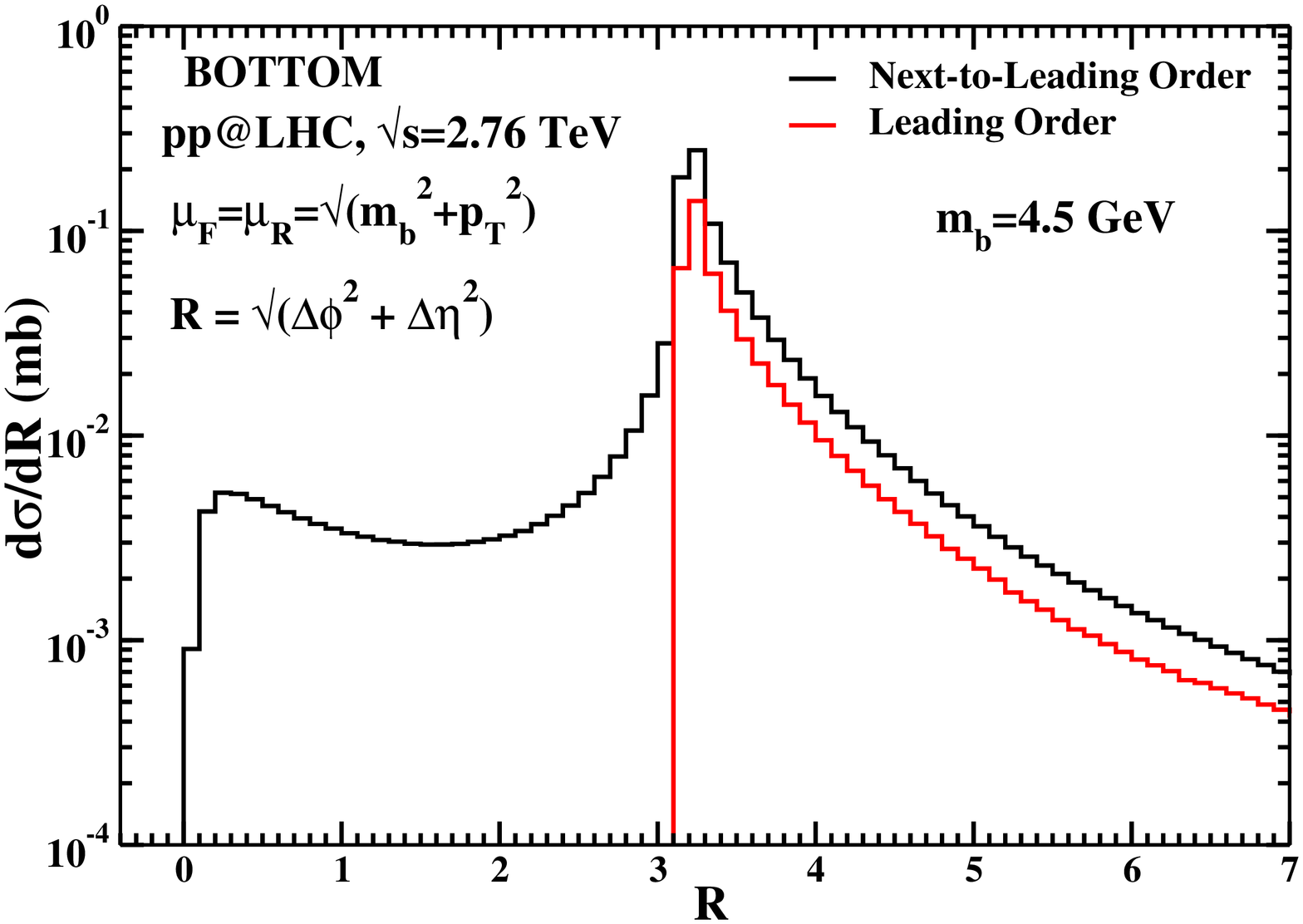}
\includegraphics[height=3in,width=2.54in,angle=270]{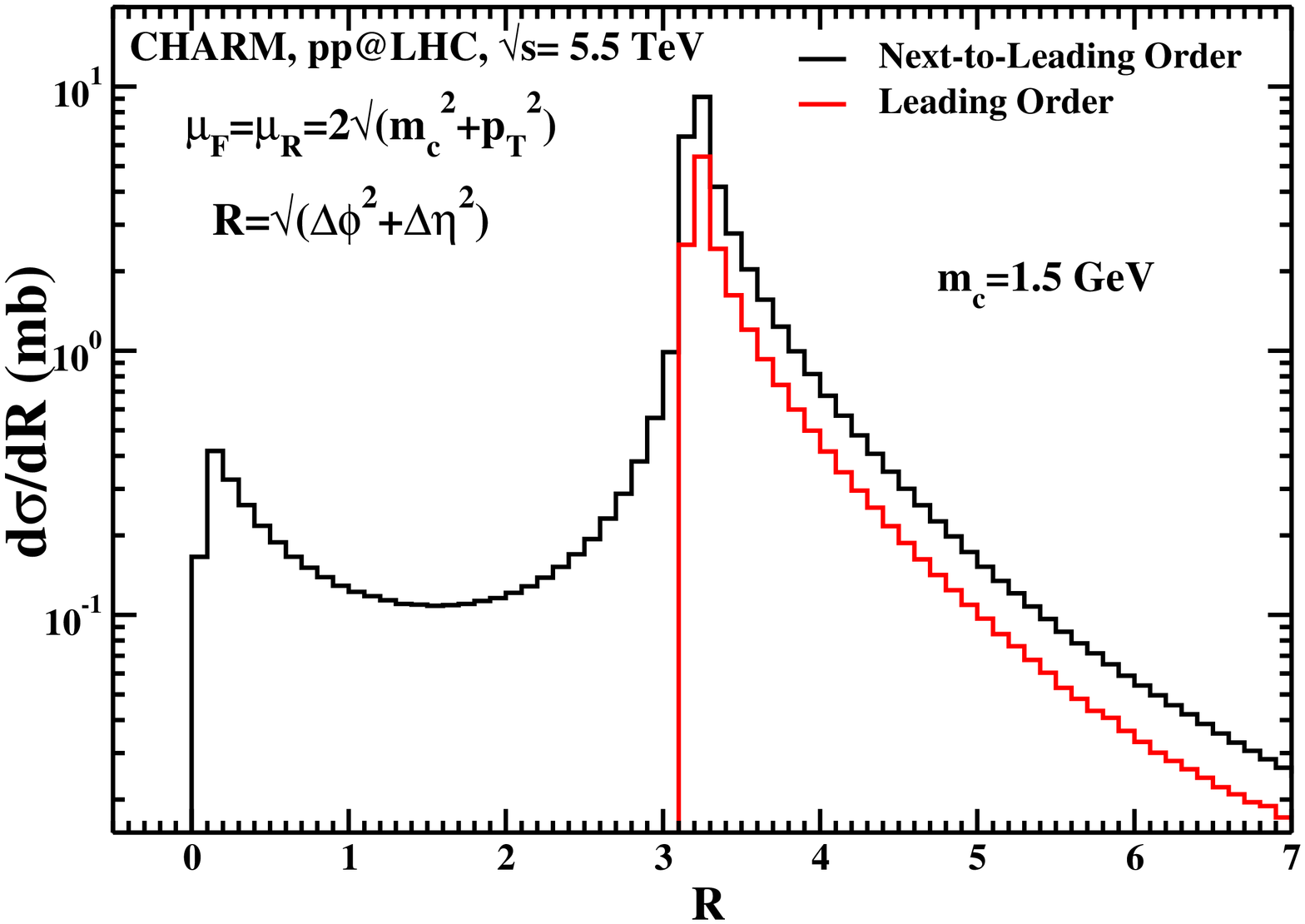}
\includegraphics[height=3in,width=2.54in,angle=270]{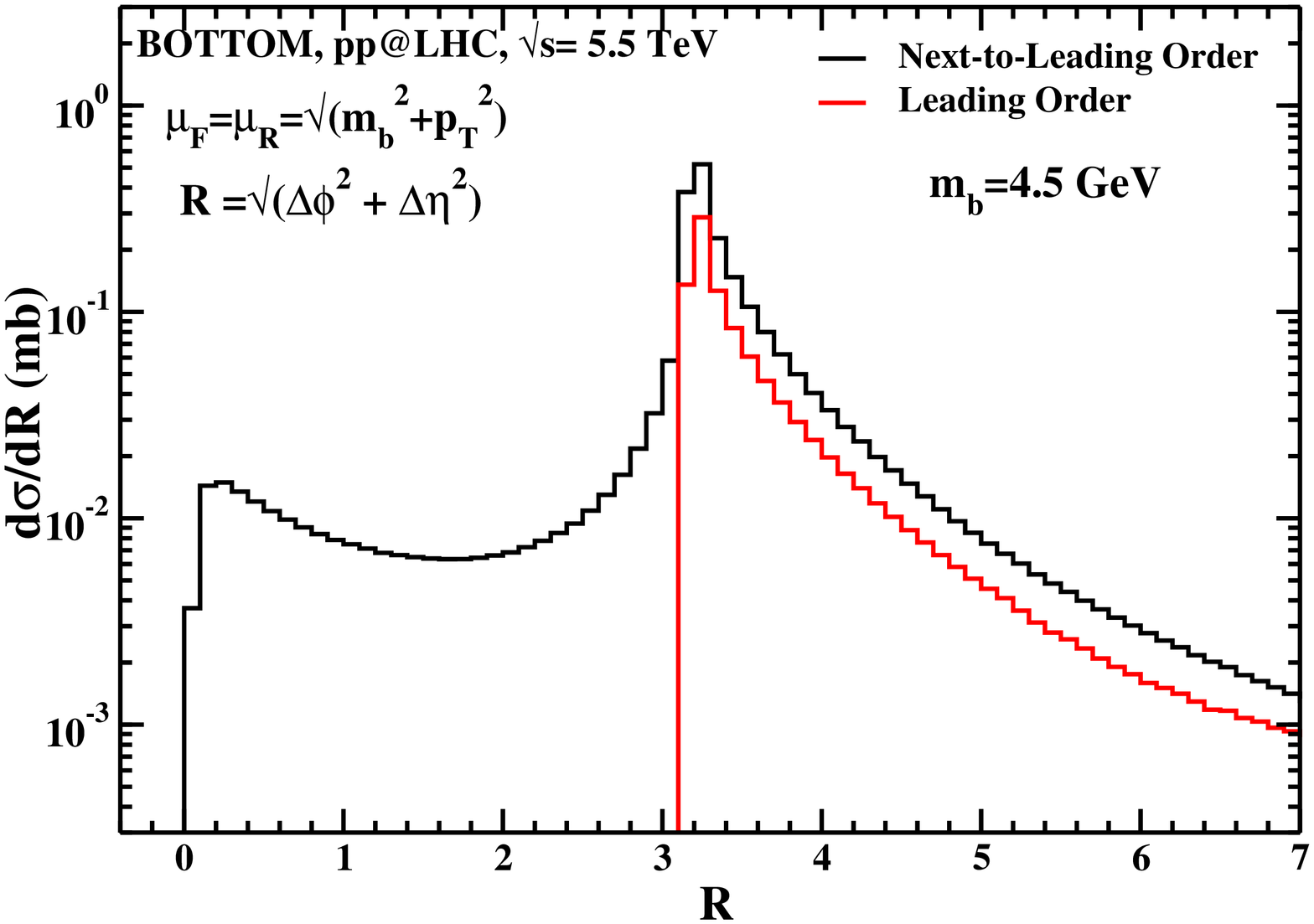}
\caption{($\Delta\eta$, $\Delta\phi$) correlations 
of heavy quarks produced in $pp$ collisions at $\sqrt{s}$=2.76 TeV  
and 5.5 TeV at LO and NLO.}
\label{R}
\end{center}
\end{figure*}


\begin{figure*}[h]
\begin{center}
\includegraphics[height=3in,width=2.5in,angle=270]{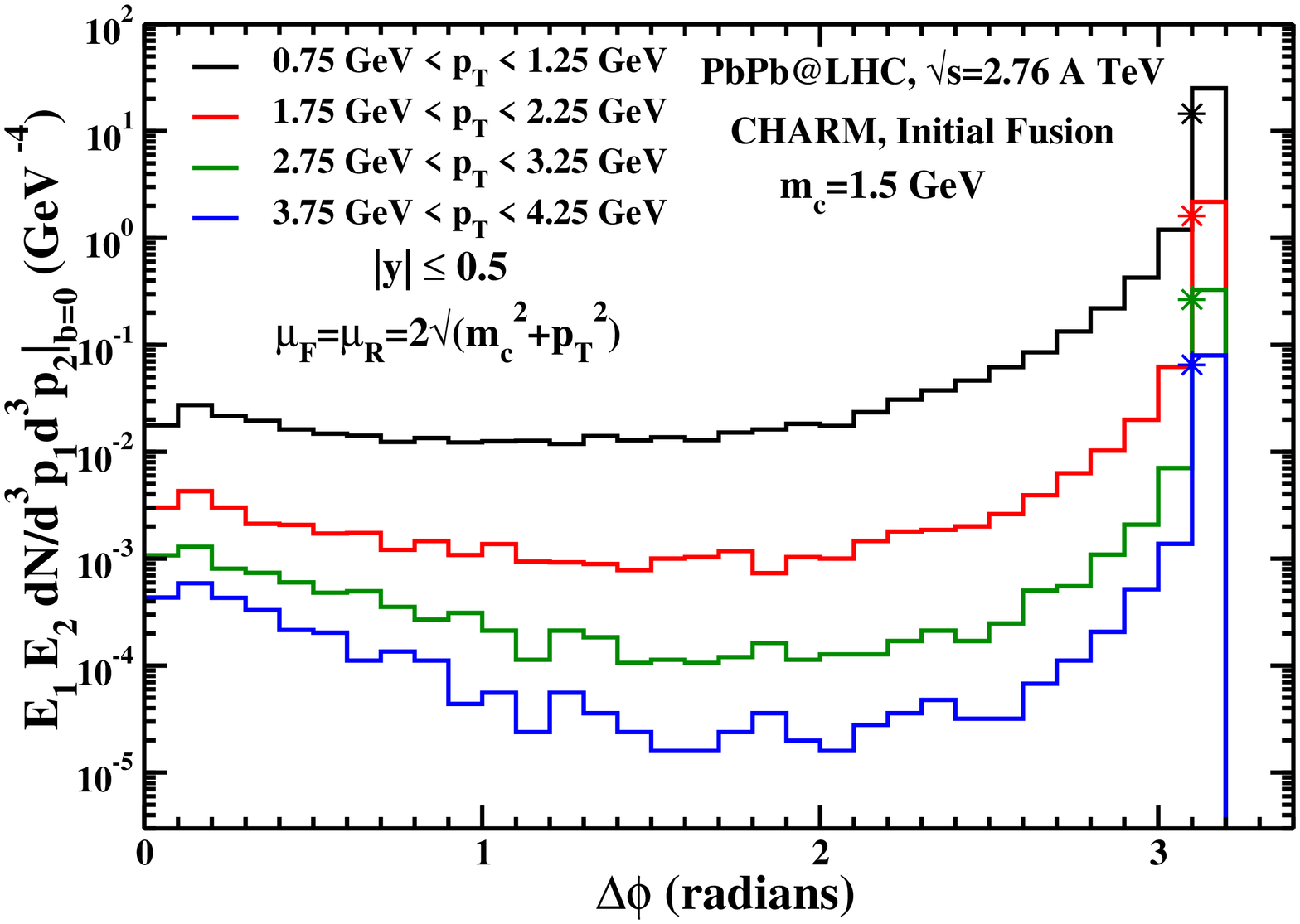}
\includegraphics[height=3in,width=2.5in,angle=270]{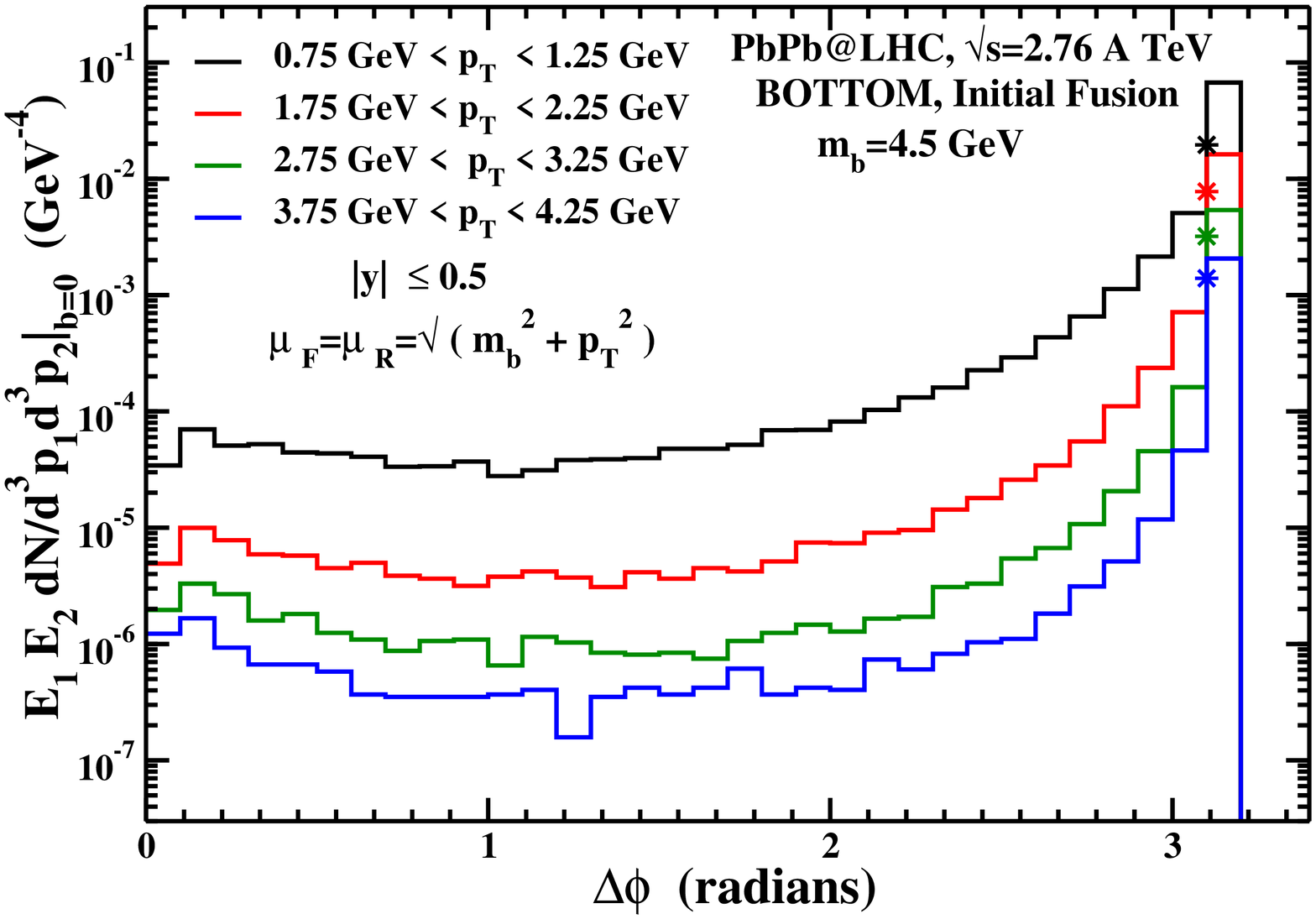}
\includegraphics[height=3in,width=2.5in,angle=270]{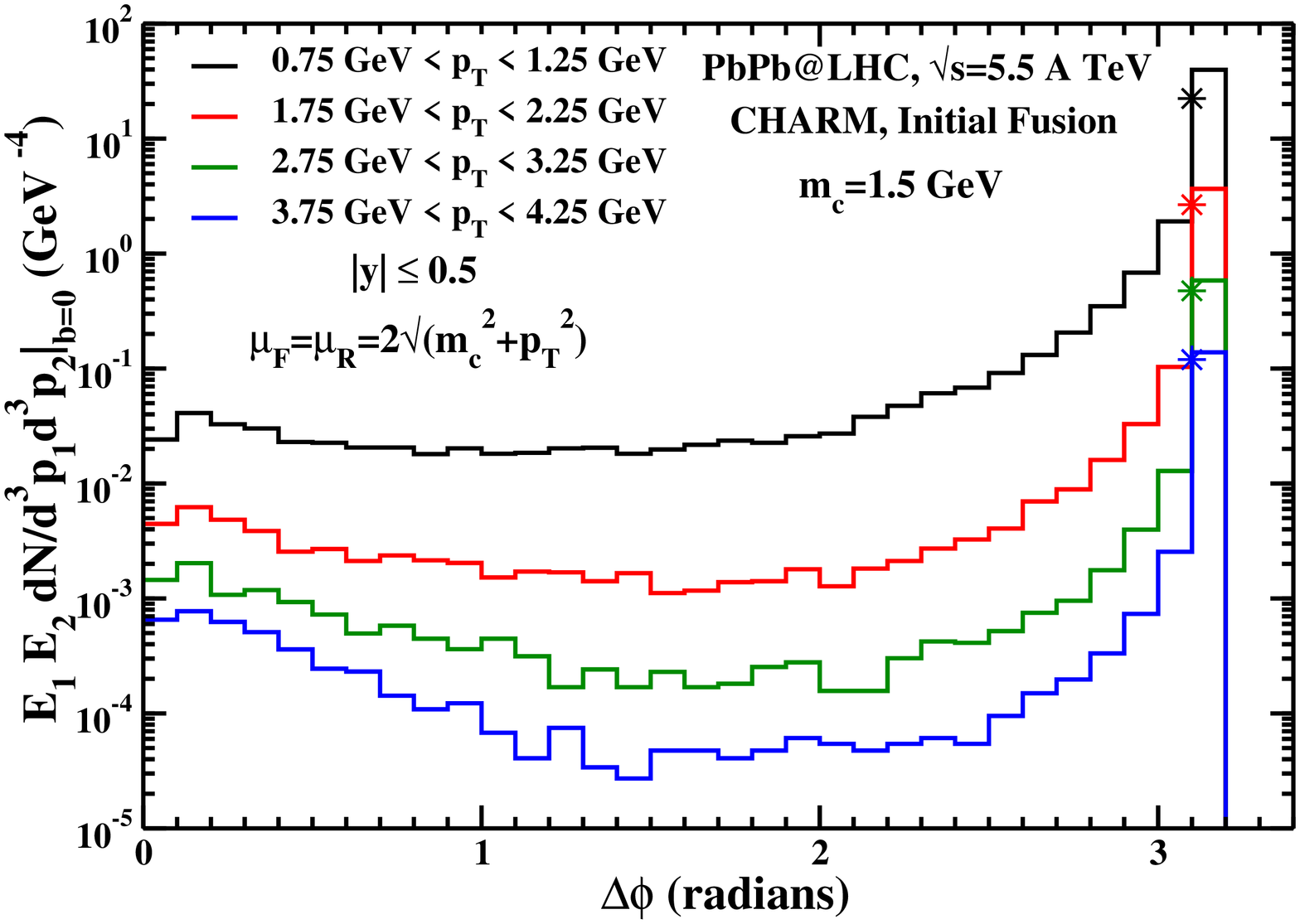}
\includegraphics[height=3in,width=2.5in,angle=270]{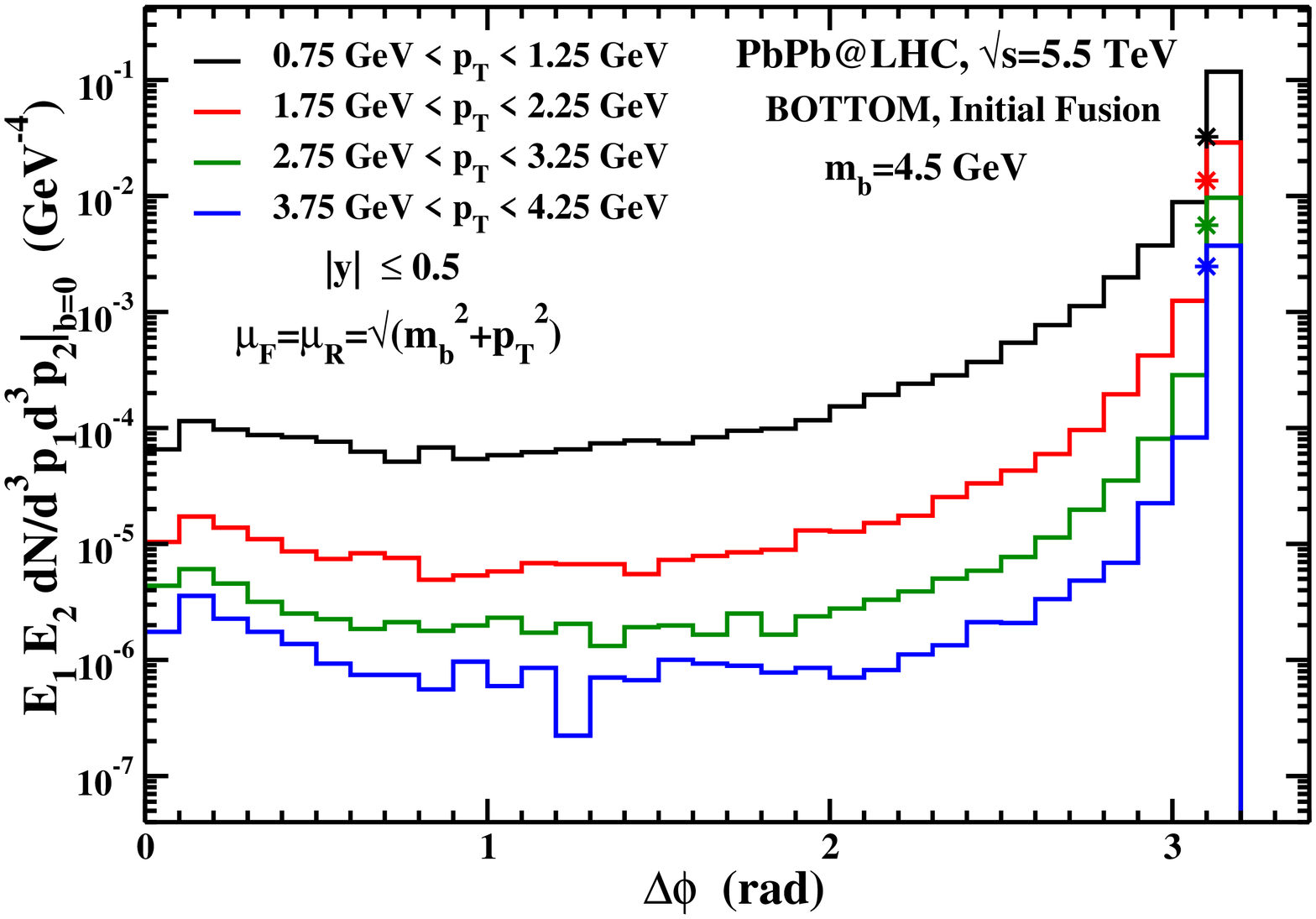}
\caption{(Colour on-line) Azimuthal correlation of heavy quarks from prompt interaction for 
lead on lead collisions 
at LHC, having different transverse momenta and rapidities close to zero. 
The symbols give the corresponding LO values, with the same bin-size for
$\Delta \phi$. The upper panels are for 2.76 ATeV while the lower
panels give results for 5.5 ATeV. The left panels give results for
charm quarks while the right panels give the results for bottom quarks.}
\label{azipbini}
\end{center}
\end{figure*}

\begin{figure*}[h]
\begin{center}
\includegraphics[height=3in,width=2.5in,angle=270]{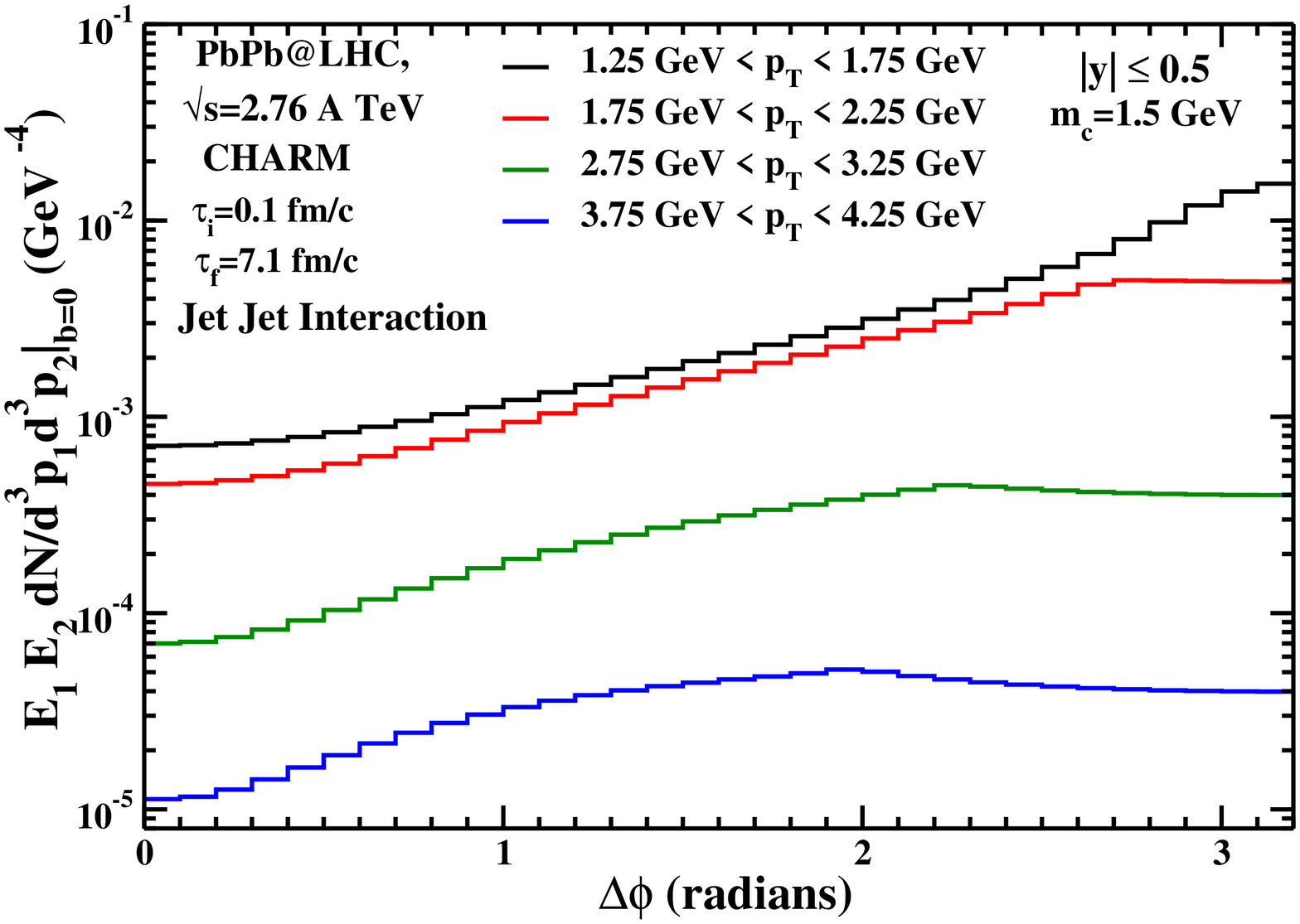}
\includegraphics[height=3in,width=2.5in,angle=270]{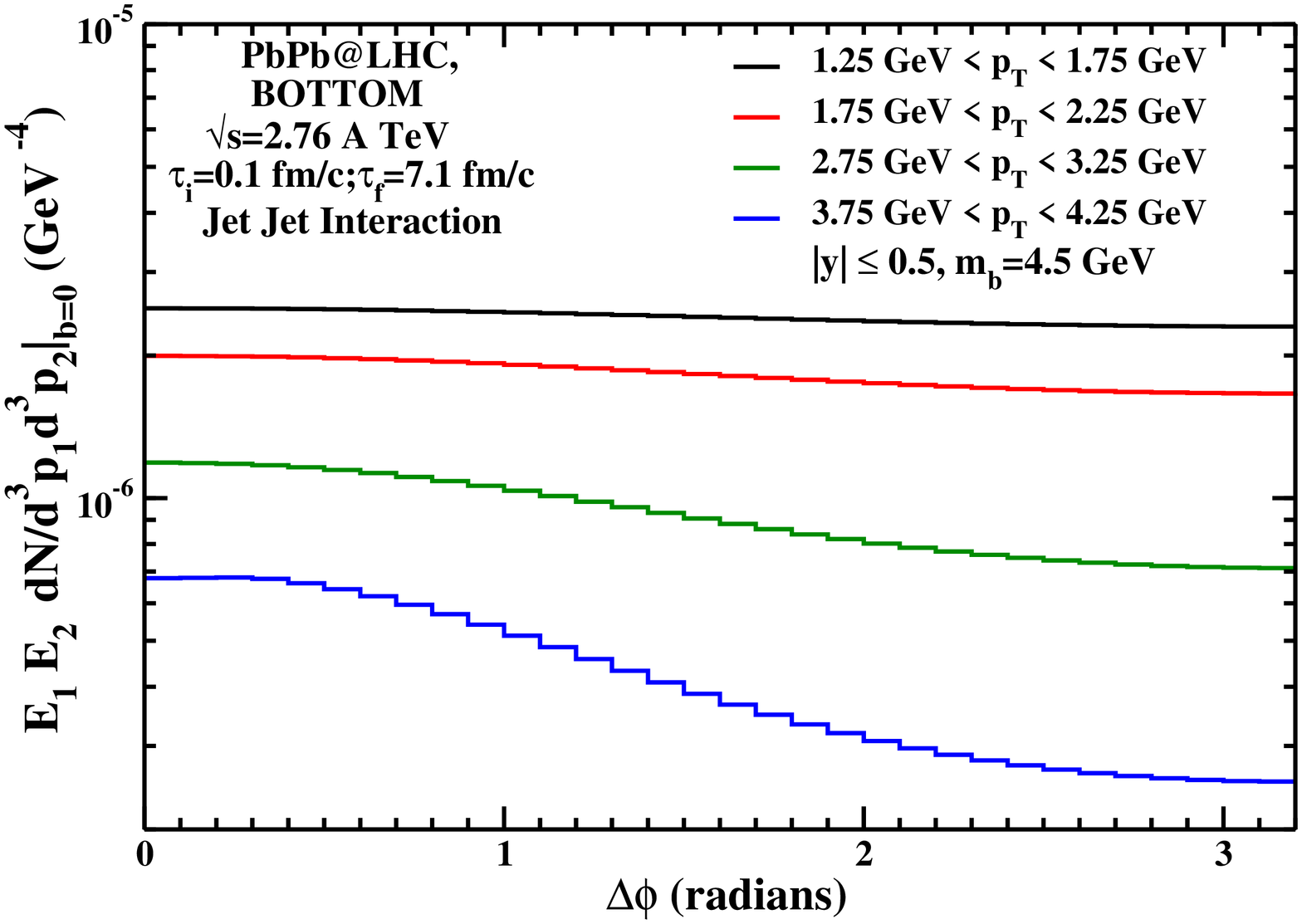}
\includegraphics[height=3in,width=2.5in,angle=270]{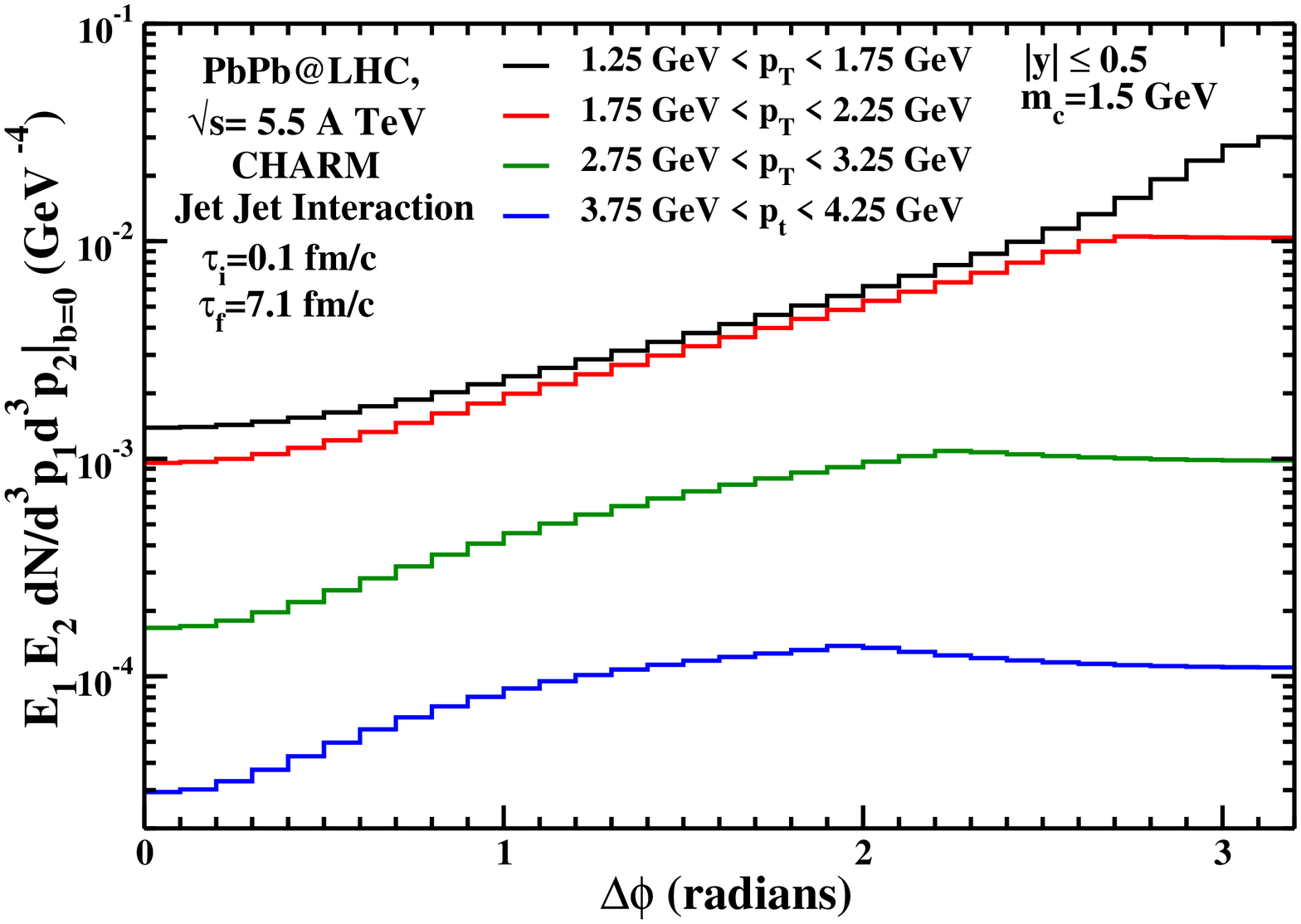}
\includegraphics[height=3in,width=2.5in,angle=270]{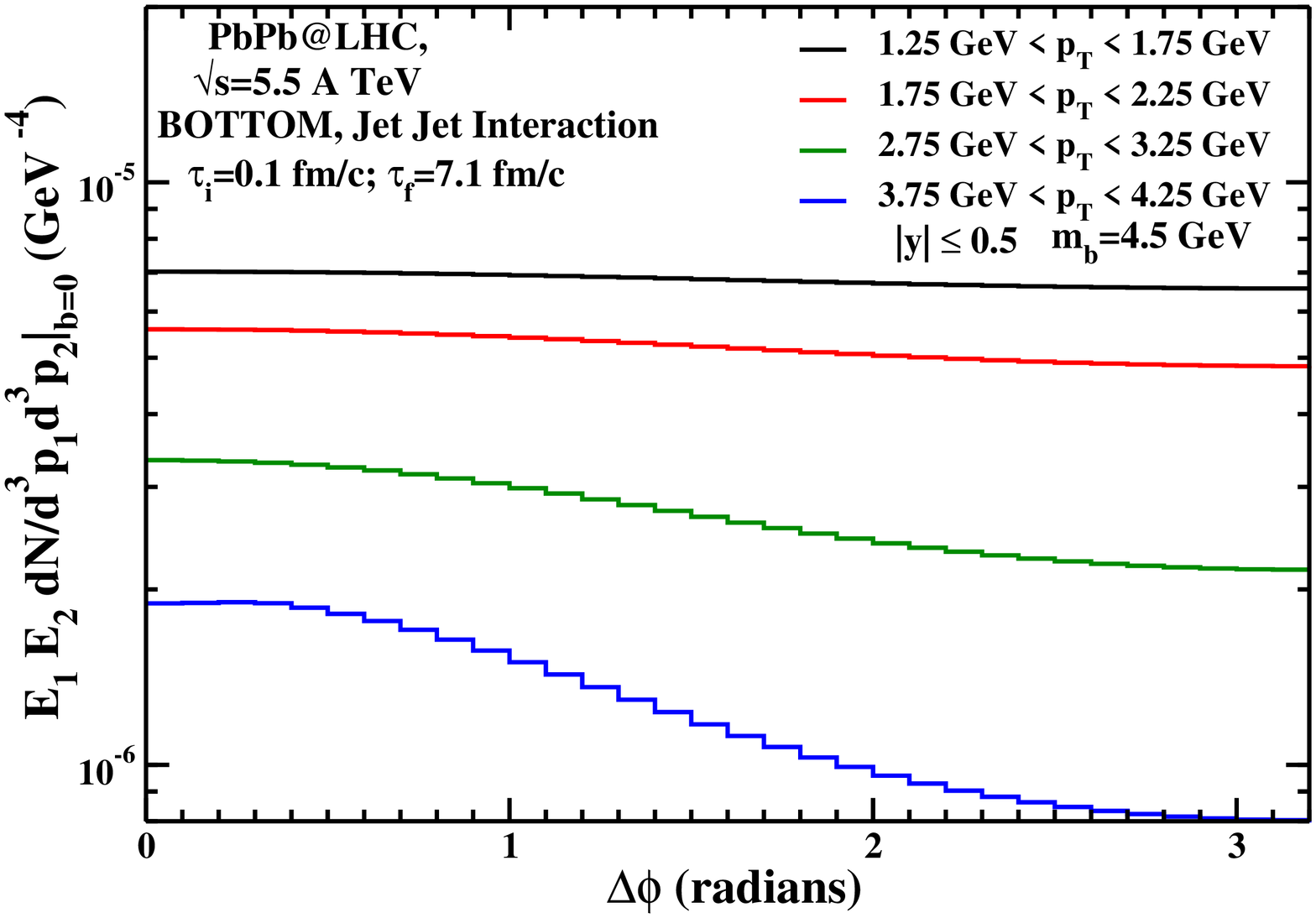}
\caption{Azimuthal correlation of heavy quarks from jet-jet interaction for
 lead on lead collisions 
at LHC, for different transverse momenta.}
\label{azipbjet}
\end{center}
\end{figure*}

\begin{figure*}[h]
\begin{center}
\includegraphics[height=3in,width=2.5in,angle=270]{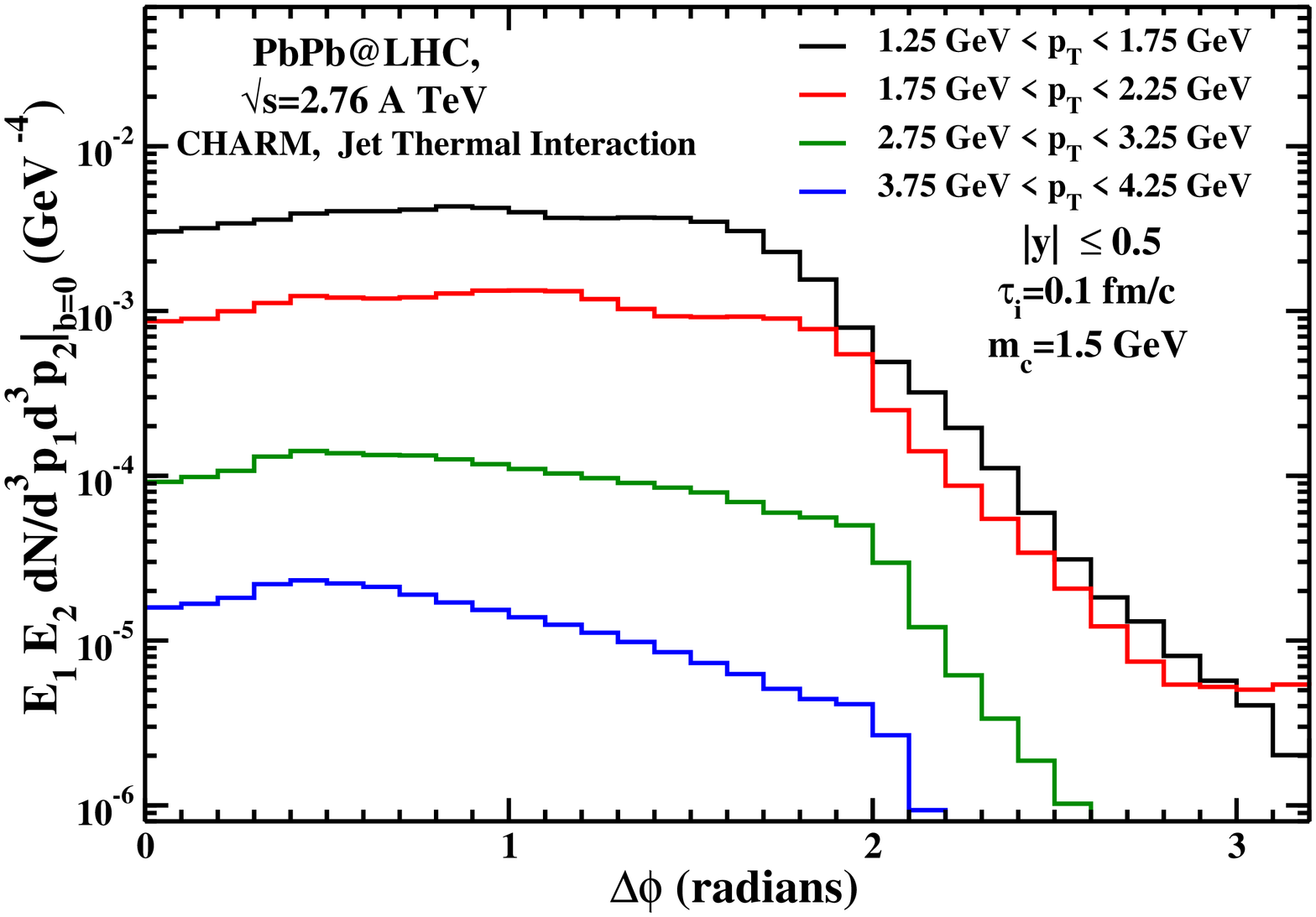}
\includegraphics[height=3in,width=2.5in,angle=270]{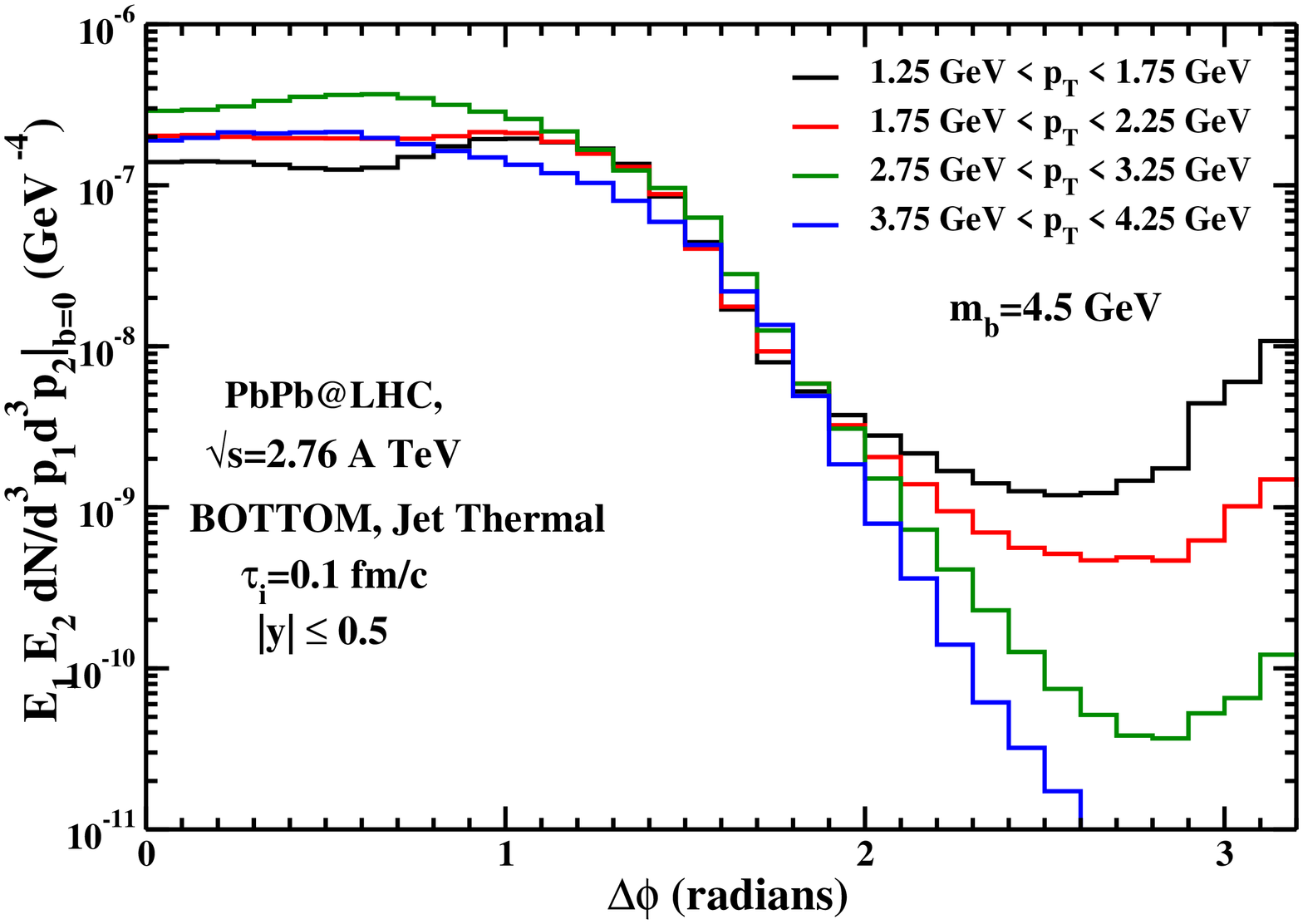}
\includegraphics[height=3in,width=2.5in,angle=270]{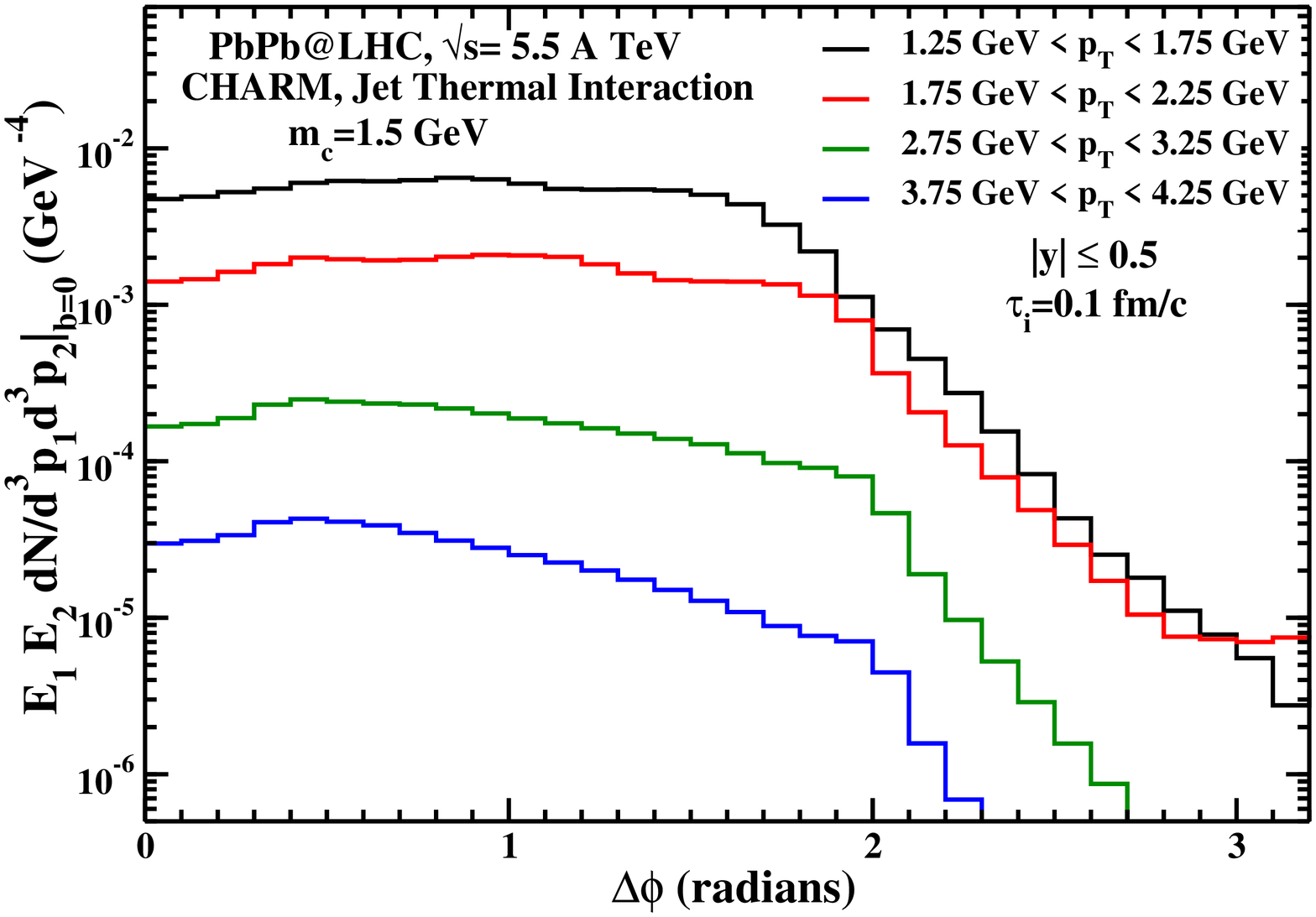}
\includegraphics[height=3in,width=2.5in,angle=270]{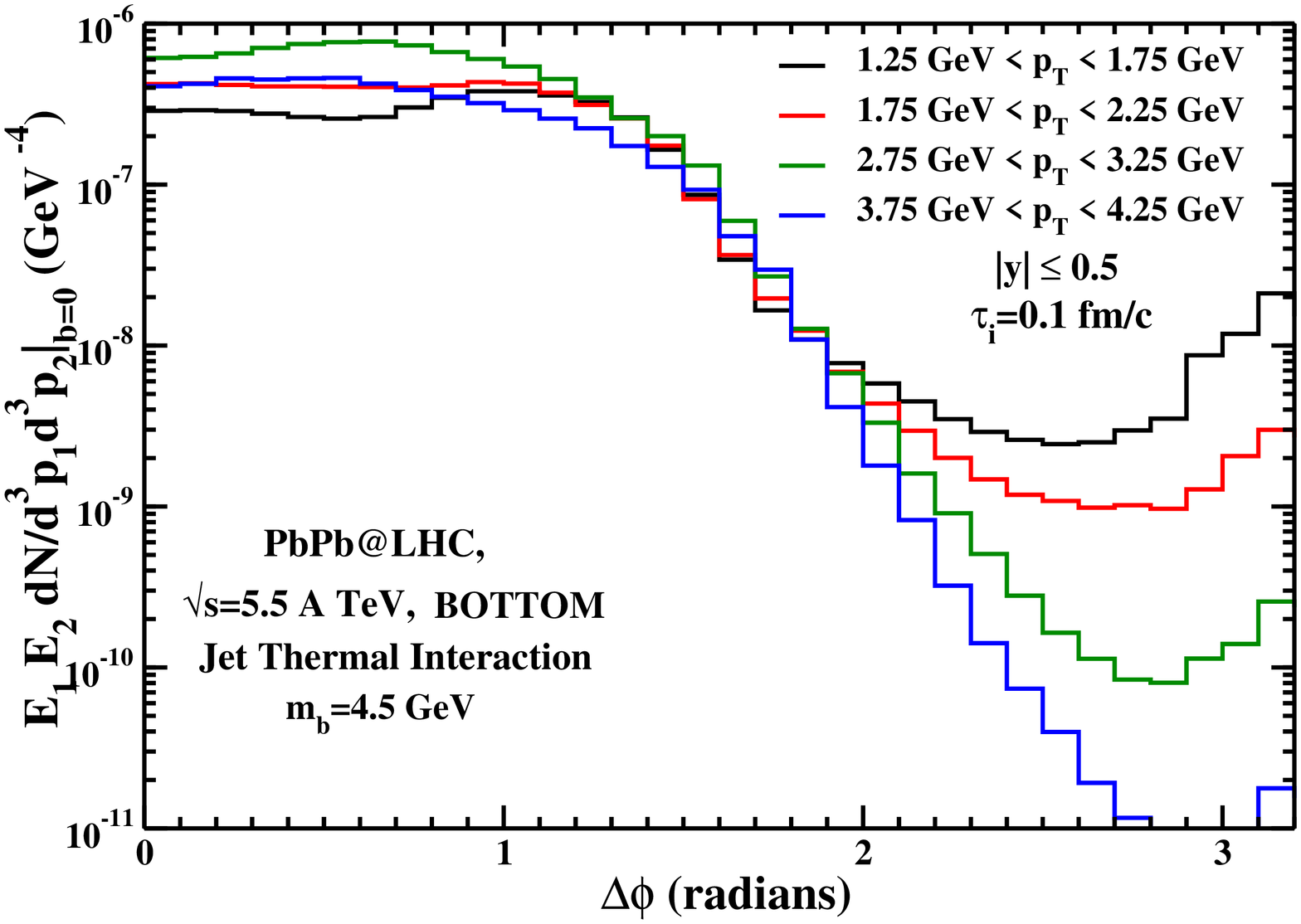}
\caption{Azimuthal correlation of heavy quarks from jet-thermal interaction for lead on lead collisions 
at LHC, for different transverse momenta.}
\label{azipbjetth}
\end{center}
\end{figure*}

\begin{figure*}[h]
\begin{center}
\includegraphics[height=3in,width=2.5in,angle=270]{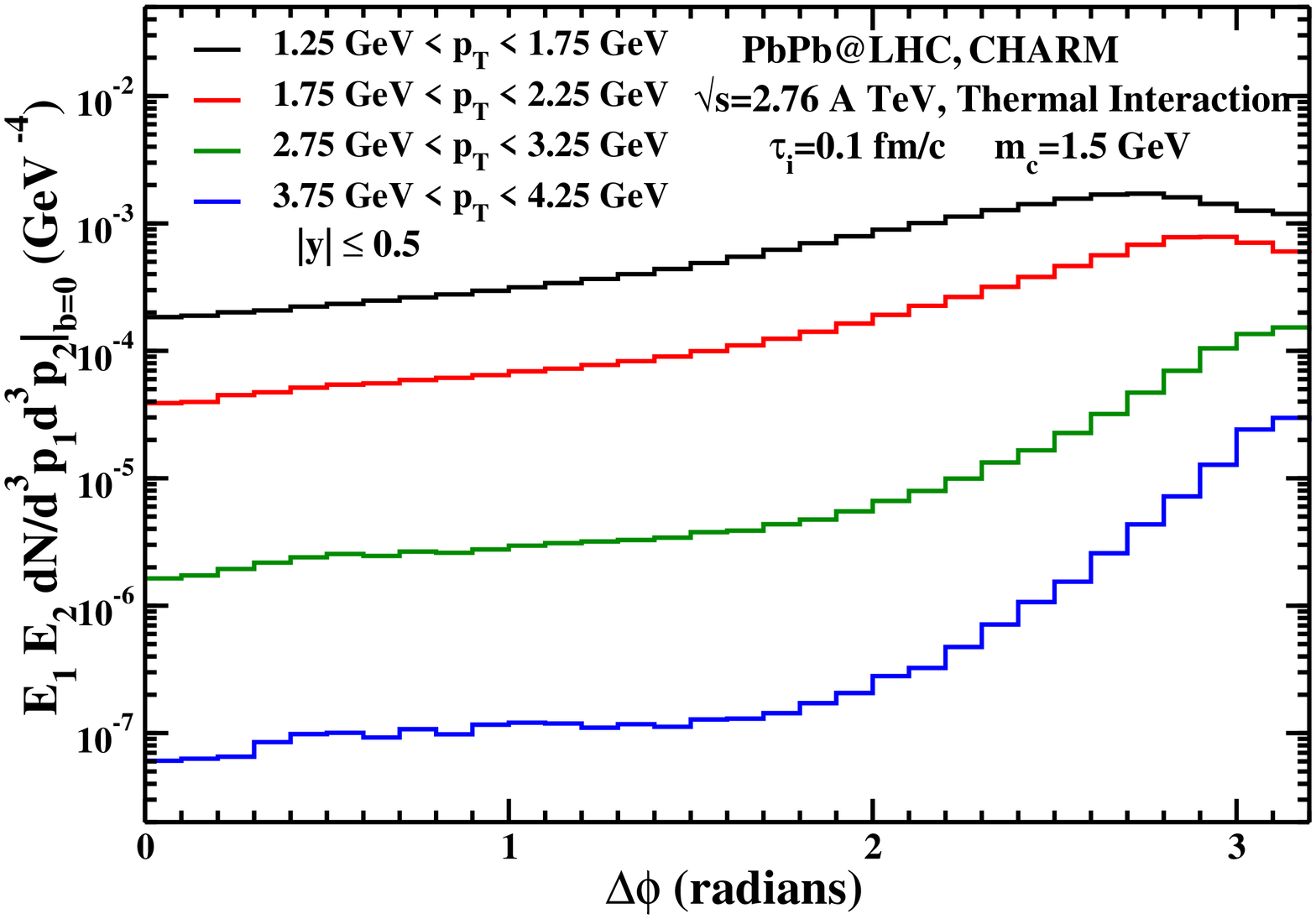}
\includegraphics[height=3in,width=2.5in,angle=270]{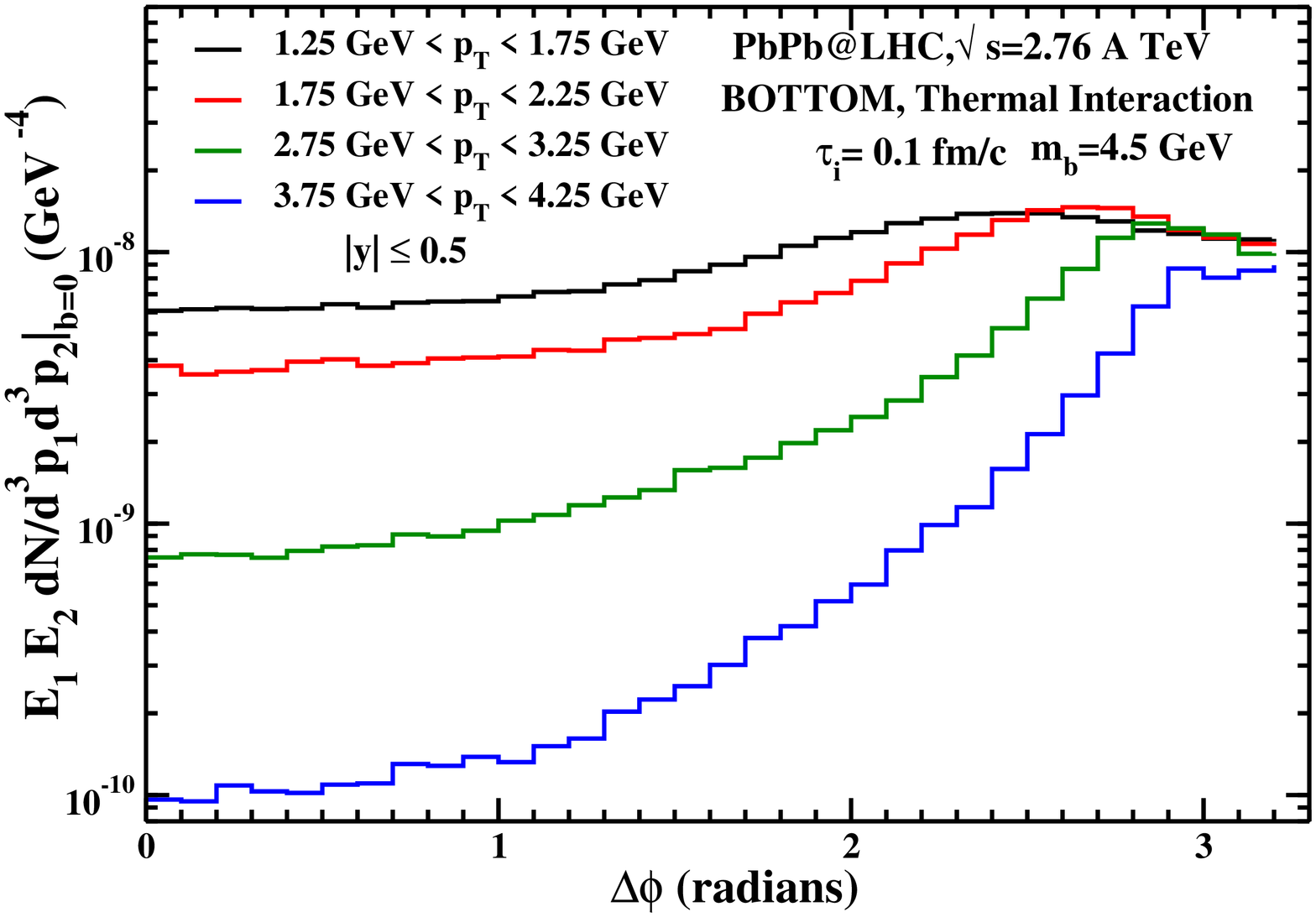}
\includegraphics[height=3in,width=2.5in,angle=270]{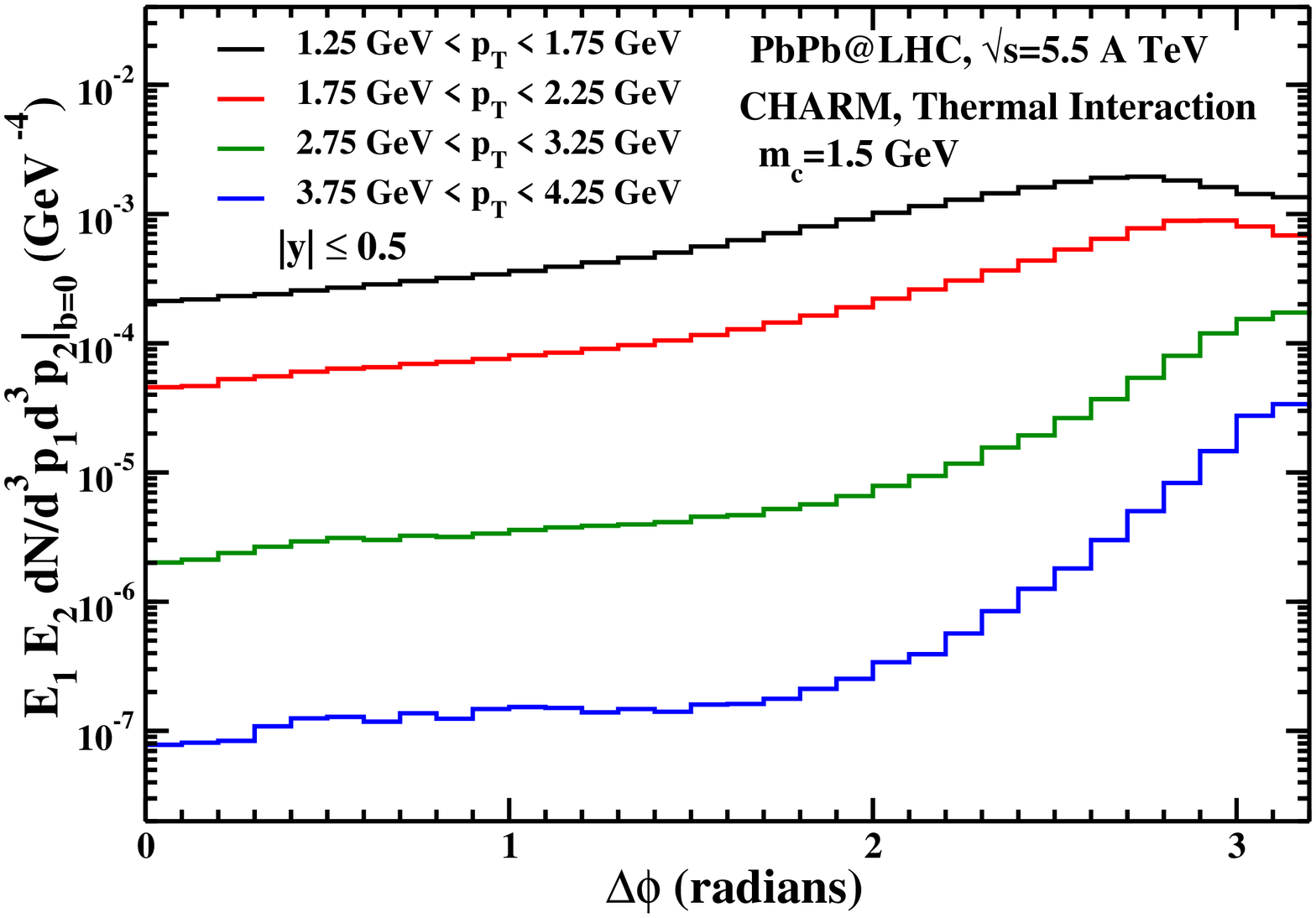}
\includegraphics[height=3in,width=2.5in,angle=270]{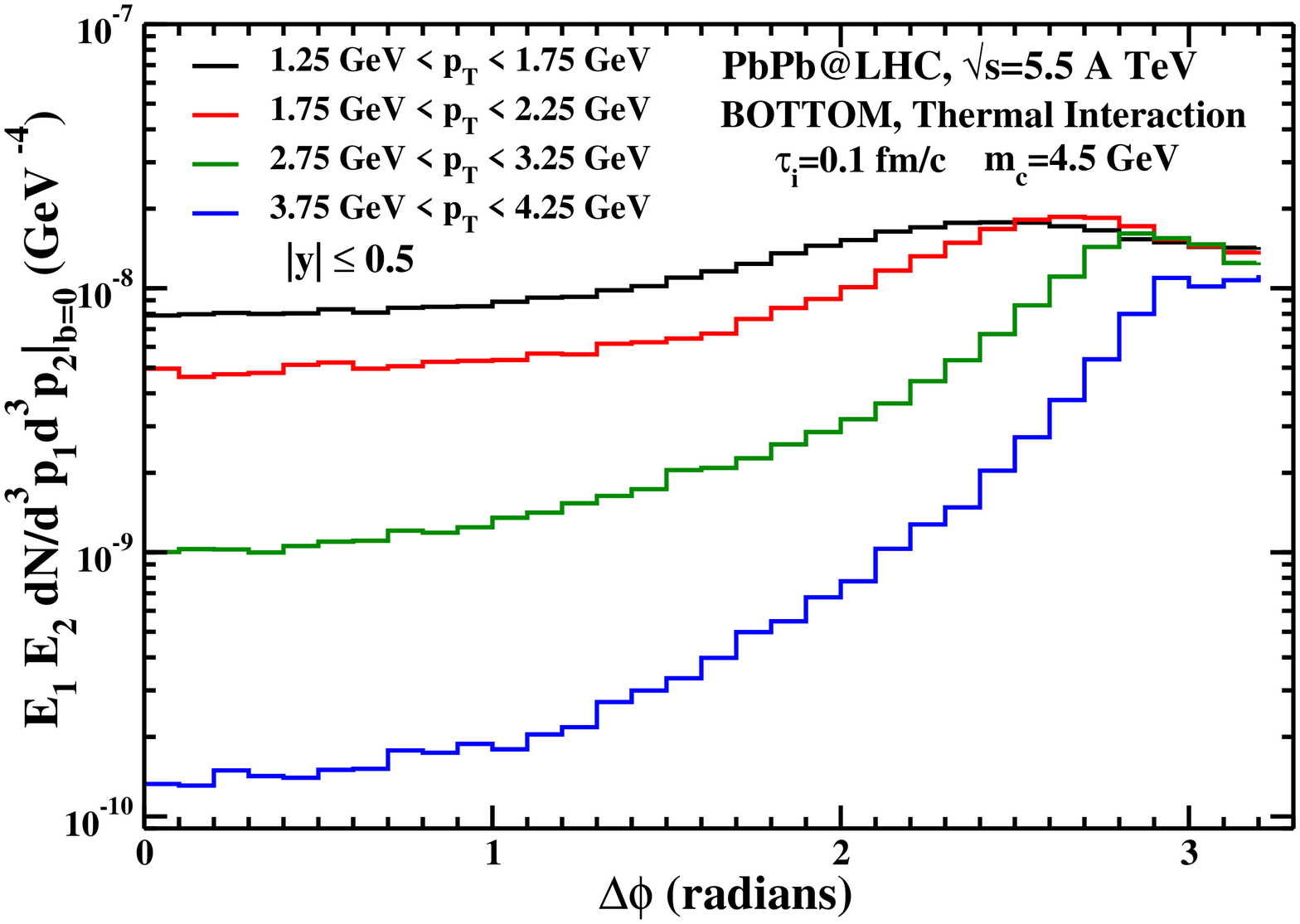}
\caption{Azimuthal correlation of heavy quarks from thermal interaction for lead on lead collisions 
at LHC, for different transverse  momenta.}
\label{azipbth}
\end{center}
\end{figure*}

\begin{figure*}[h]
\begin{center}
\includegraphics[width=2.33in,angle=270]{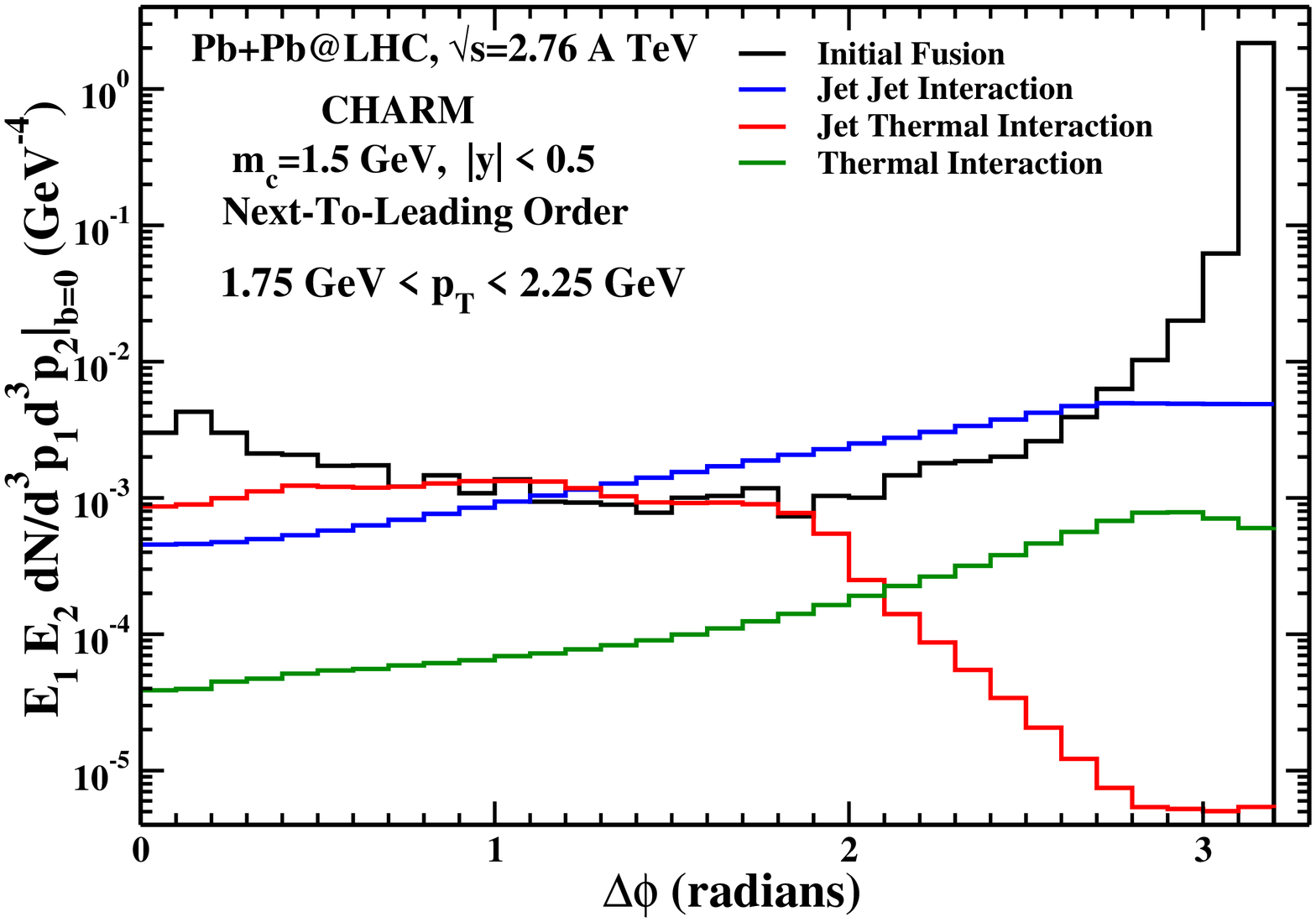}
\includegraphics[width=2.33in,angle=270]{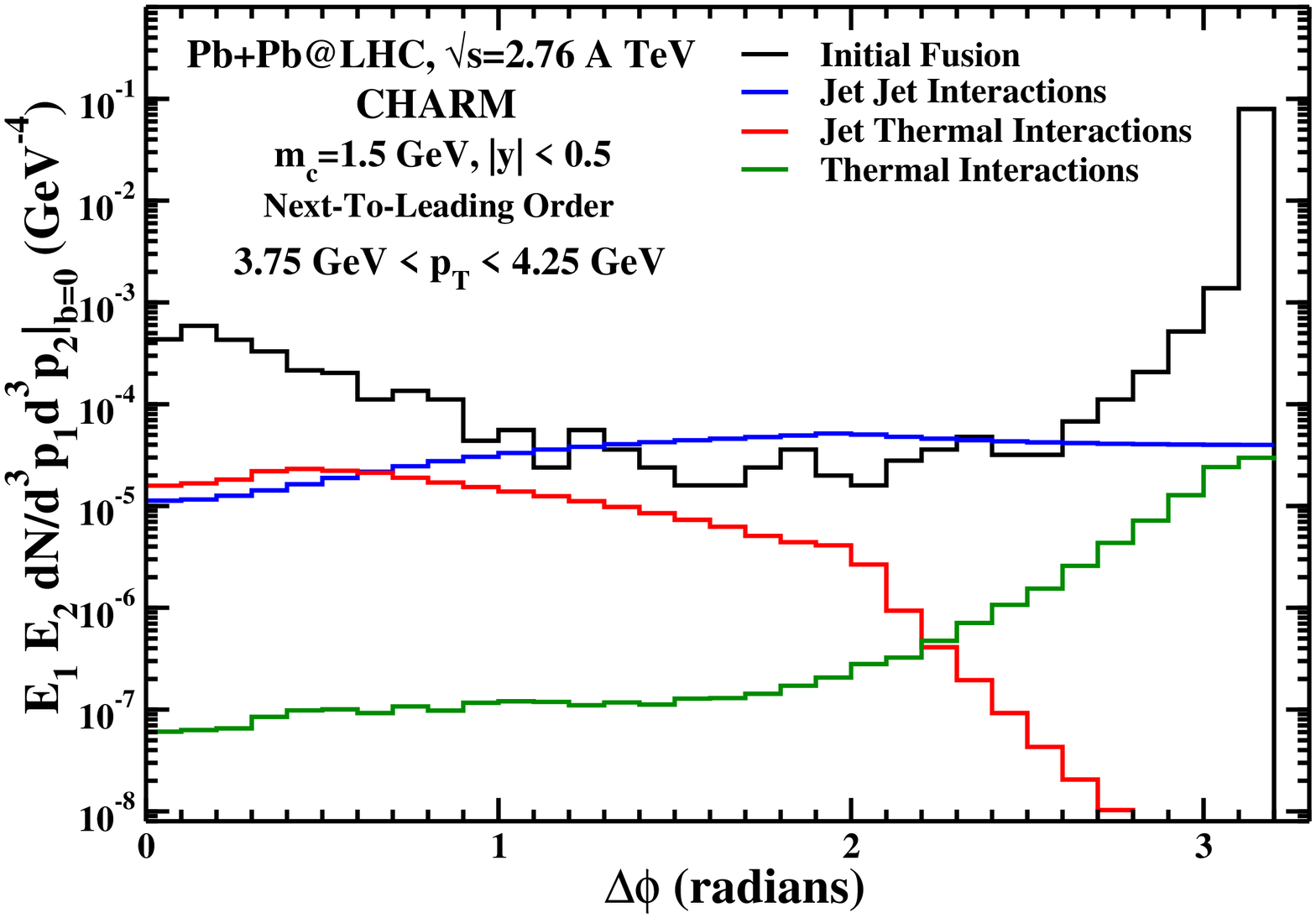}
\caption{Relative contributions of prompt, jet-jet, jet-thermal, and thermal productions
of charm quarks having transverse momenta of about 2 GeV/$c$ (left panel) and
4 GeV/$c$ (right panel) to their azimuthal correlation.}
\label{compare}
\end{center}
\end{figure*}

\begin{figure*}[b]
\begin{center}
\includegraphics[height=3in,width=2.5in,angle=270]{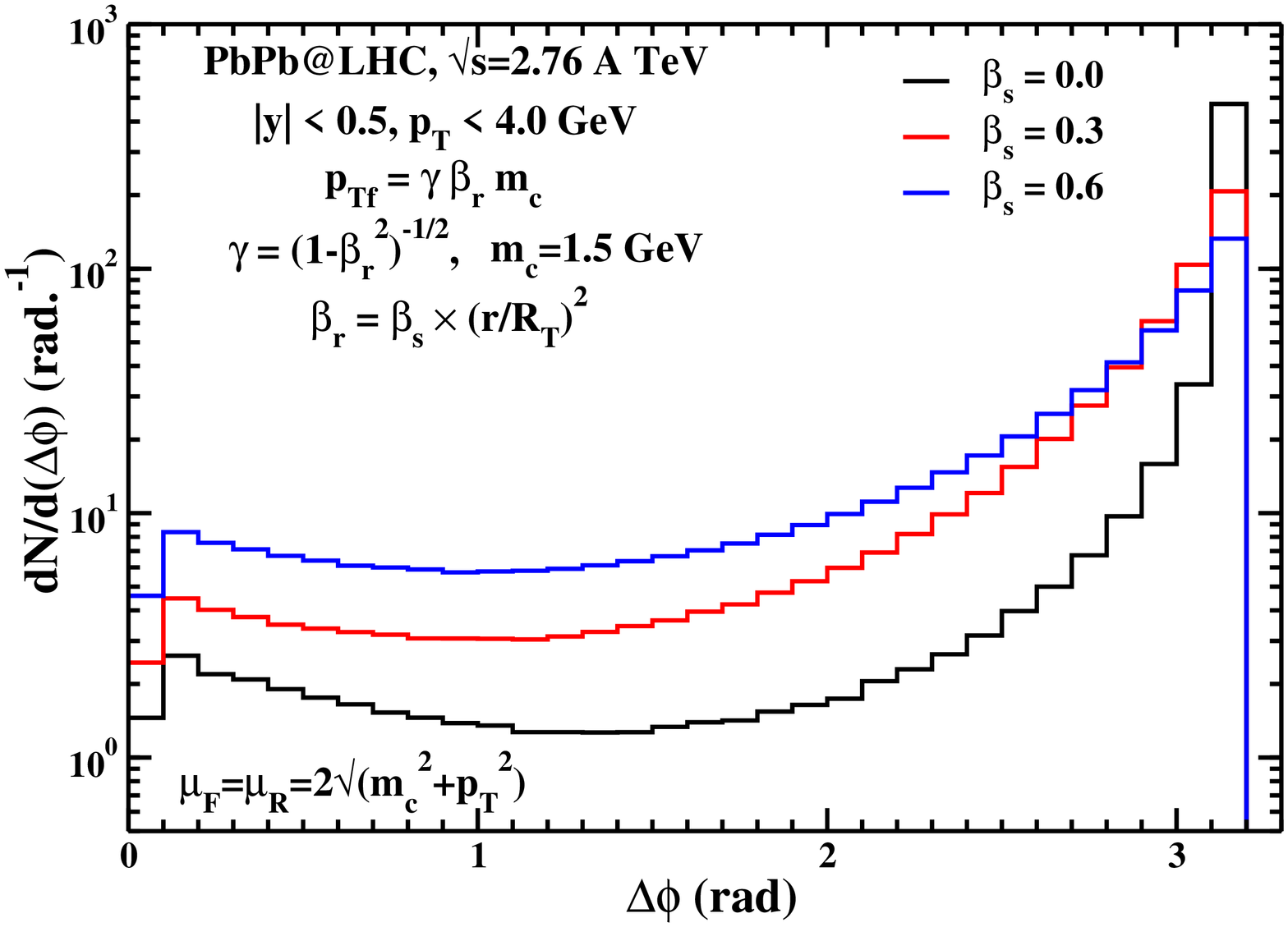}
\includegraphics[height=3in,width=2.5in,angle=270]{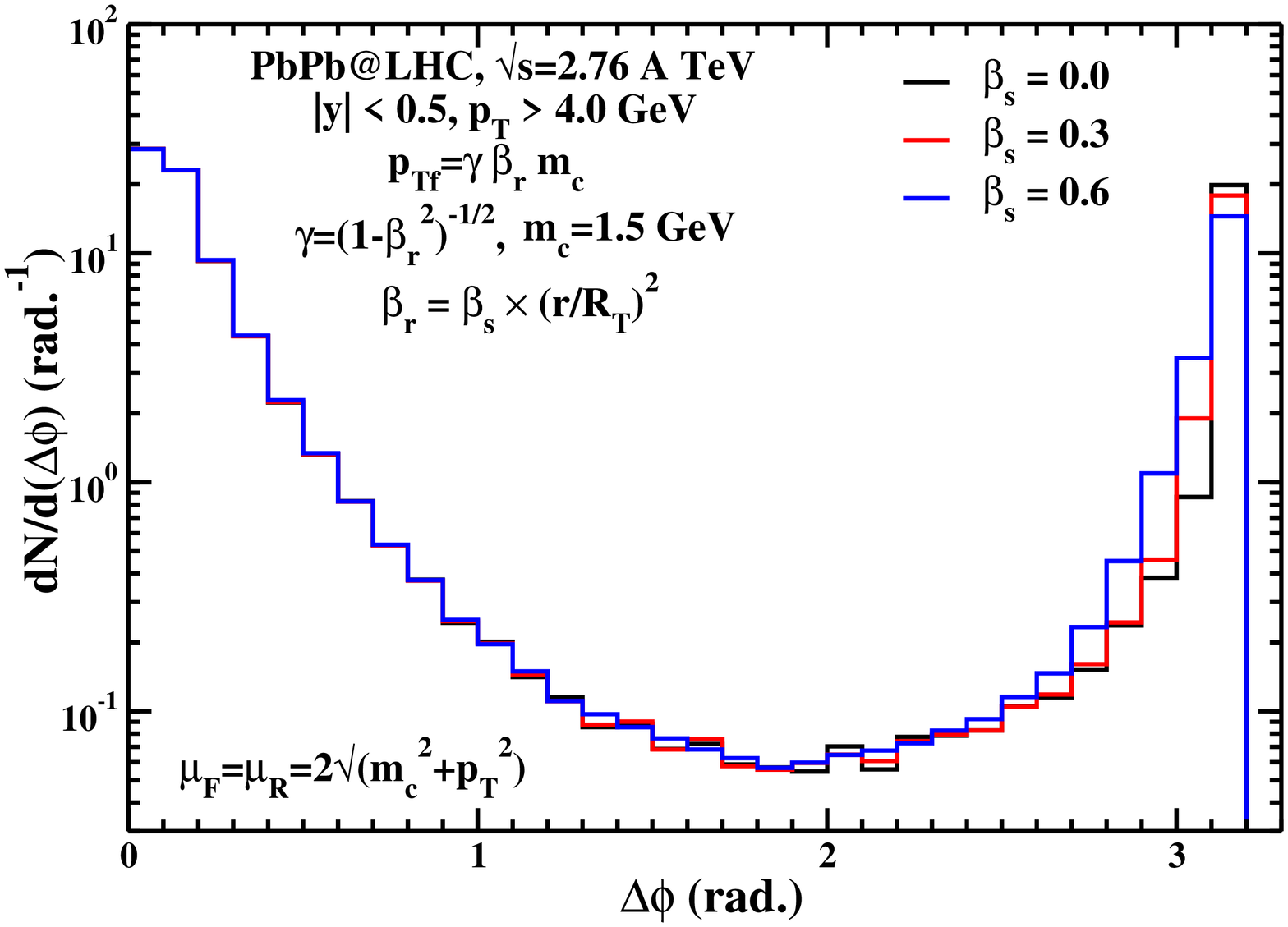}
\caption{Azimuthal correlation of heavy quarks from prompt interaction for lead on lead collisions 
at LHC, with flow}
\label{flow1}
\end{center}
\end{figure*}


\begin{table}[h]
\caption{Parametrization of the quark and gluon distributions from initial scattering 
of partons 
at 2.76 and 5.5 TeV in $pp$ collisions, using LO pQCD and CTEQ5M structure functons,
for $p_T >$ 2 GeV.}
\begin{center}
\vskip 0.2in
\begin{tabular}{c  c  c  c  c  c}
  \hline
$\sqrt{s}$[TeV]   &  & C [1/GeV$^{2}$] & B [GeV]  & $\beta$ & \\
  \hline
  & u & 1.078$\times$10$^{3}$ & 1.127 & 5.615 &  \\
  & d & 1.279$\times$10$^{3}$ & 1.099 & 5.579 &  \\
2.76~ & s & 1.395$\times$10$^{2}$ & 1.899 & 6.432 &  \\
  & $\overline{u}$ & 3.371$\times$10$^{2}$ & 1.434 & 5.999 &  \\
  & $\overline{d}$ & 3.734$\times$10$^{2}$ & 1.401 & 5.953 &  \\
  & g & 2.947$\times$10$^{3}$ & 1.892 & 6.523 &  \\
  \hline
   & u & 7.961$\times$10$^{2}$ & 1.293 & 5.580 &  \\
   & d &~~~~~~~ 9.478$\times$10$^{2}$~~~~~~~~~&~~~~~~1.254~~~~~~~& 5.539&  \\
5.5~   & s & 1.228$\times$10$^{2}$ & 2.174 & 6.418 &  \\
   & $\overline{u}$ & 2.659$\times$10$^{2}$ & 1.663 & 5.966 &  \\
   & $\overline{d}$ & 2.908$\times$10$^{2}$ & 1.624 & 5.924 &  \\
   & g & 2.449$\times$10$^{3}$ & 2.192 & 6.519 &  \\
  \hline
\end{tabular}
\end{center}

\end{table}

Now the azimuthal distribution of heavy quarks for collisions having an
impact parameter, $b=0$, due to  
jet-jet interaction can be written as:
\begin{eqnarray}
 E_1\ E_2\frac{dN}{d^{3}p_1d^{3}p_2}&=&\frac{1}{16(2\pi)^{8}}\int d^{4}x\int \frac{d^{3}p_{a}d^{3}p_{b}}{\omega_{a}\omega_{b}}
\delta^{4}(\Sigma p^{\mu})\nonumber\\
&\times&\left[\frac{1}{2}g_{g}^{2}f_{\textrm{jet}}^{g}(p_{Ta})f_{\textrm{jet}}^{g}(p_{Tb})\left|M_{gg\rightarrow Q\overline{Q}}\right|^{2}\right.\nonumber\\
&+& \left. g_{q}^{2}\sum_i \left \{f_{\textrm{jet}}^{q_i}(p_{Ta})f_{\textrm{jet}}^{\bar{q_i}}(p_{Tb})\left|M_{q\bar{q}\rightarrow Q\overline{Q}}\right|^{2}
+(q_i\leftrightarrow\bar{q_i})\right \}\right]~,\nonumber\\
\label{jetn}
\end{eqnarray}
where $p_{a}$, $p_{b}$ are the four momenta of the incoming partons 
and $p_1$ and $p_2$ are the same for the outgoing heavy quarks, and 
$q_i$ stands for the flavour of the light quarks. 
The jet distribution function $f_{\textrm{jet}}(p_{T})$ is given by
\begin{equation}
f^{i}_{\rm {jet}}(p_{T}) = \frac{(2\pi)^{3}}{g_{i}
\tau\pi R_{T}^{2}p_{T}} \frac{dN_{i}}{d^{2}p_{T}\,dy}
\, \delta(y-\eta)\,\Theta(\tau_{f}-\tau)
\, \Theta(\tau-\tau_{i})~.
\end{equation}
This follows the Bjorken space-time correlation used earlier in 
Refs.~\cite{lingyu,mdy1,rainer}.  
In the above, $p_T$ is the transverse momentum, $y$ is 
rapidity, $\eta$ is the space-time rapidity, and $g_i$ is
the spin-colour degeneracy of the partons, which is $2\times8$ 
for gluons and $2\times3$ for quarks. As indicated
earlier, $\tau_i$ and $\tau_f$ are the the formation time of 
the jet and the life time of the plasma. $R_\textrm{T}$ is
the transverse size of the system. 
Now the Eq.~\ref{jetn} reduces to:
\begin{eqnarray}
 E_1\ E_2\frac{dN}{d^{3}p_1d^{3}p_2}&=&\frac{1}{16(2\pi)^{8}}\int d^{4}x\int d^{2}p_{Tb}\, dy_{b}\frac{\delta(\Sigma{E})}{\omega_{a}}\nonumber\\
&\times&\left[\frac{1}{2}g_{g}^{2}f_{\textrm{jet}}^{g}(p_{Ta})f_{\textrm{jet}}^{g}(p_{Tb})\left|M_{gg\rightarrow Q\overline{Q}}\right|^{2}+ g_{q}^2 \times \right.\nonumber\\
&&\left. \sum_i 
\left \{f_{\textrm{jet}}^{q_i}(p_{Ta})
f_{\textrm{jet}}^{\bar{q_i}}(p_{Tb})\left|M_{q\bar{q}\rightarrow Q\overline{Q}}\right|^{2}+(q_i\leftrightarrow\bar{q_i})\right \}\right],
\label{jeteq}
\end{eqnarray}
where $d^{4}x$=$\tau d\tau\,rdr\,d\eta\,d\phi_r$ and 
$d^{3}p$=$p_{T}dp_{T}\,d\phi_p\,Edy$. Further 
\begin{eqnarray}
\frac{\delta(\Sigma{E})}{\omega_a}=\nonumber\\\frac{\delta(p_{Tb}-p_{Tb0})}
{[p_{T1}\cos(\phi_1-\phi_b)+p_{T2}\cos(\phi_2-\phi_b)-M_{T1}\cosh(y_1-\eta)-M_{T2}\cosh(y_2-\eta)]}~.\nonumber\\
\end{eqnarray}

Thus the final integration obtained by putting the above expression in Eq.~\ref{jeteq} 
reduces to:

\begin{eqnarray}
E_1\ E_2\frac{dN}{d^{3}p_1d^{3}p_2}=\frac{\ln(\tau_{f}/\tau_{i})}{16(2\pi)^{2}\pi R_{T}^{2}}\int d\eta\,d\phi_{b}\nonumber\\\times
\frac{1}{p_{Ta}[p_{T1}\cos(\phi_1-\phi_b)+p_{T2}\cos(\phi_2-\phi_b)-M_{T1}\cosh(y_1-\eta)-M_{T2}\cosh(y_2-\eta)]}\nonumber\\
\times \left[\frac{1}{2}\,h_{\textrm{jet}}^{g}(p_{Ta})\,h_{\textrm{jet}}^{g}(p_{Tb0})\left|M_{gg\rightarrow Q\overline{Q}}\right|^{2}\right.\nonumber\\
+\left.\sum_i\left \{h_{\textrm{jet}}^{q_i}(p_{Ta})\,h^{\bar{q_i}}_{\textrm{jet}}(p_{Tb0})
\left|M_{q\bar{q}\rightarrow Q\overline{Q}}\right|^{2}+(q_i\leftrightarrow\bar{q_i})\right \}\right]~.
\label{jet}
\end{eqnarray}
with $p_{Tb0}$ given by,
\begin{eqnarray}
 p_{Tb0}&=&\nonumber\\
&&\frac{p_{T1}p_{T2}\cos(\phi_1-\phi_2)-M_{T1}M_{T2}\cosh(y_1-y_2)-m_{Q}^{2}}{p_{T1}\cos(\phi_1-\phi_b)+p_{T2}\cos(\phi_2-\phi_b)
-M_{T1}\cosh(y_1-\eta)-M_{T2}\cosh(y_2-\eta)}\nonumber\\
\end{eqnarray}
These are similar to the expressions obtained earlier by 
Levai et al~\cite{lmw}.
 
The formation time for the jets (light $p_T$ partons) is taken as
$\tau_{i}$=0.1 fm/$c$, as we count those having $p_T >$ 2 GeV, as jets,
We take $\tau_{f}\approx R_{\textrm{T}}$, of the system and perform rest
 of the integration numerically. Note, however, that $\tau_i$ and $\tau_f$ appear in
the above expression only through the term $\ln(\tau_f/\tau_i)$ and thus reasonable 
variations in their values will lead to only a modest variation in the results. 

\subsection{Jet-Thermal Interaction}

Now, we consider passage of high energy energy jets through 
quark gluon plasma and estimate azimuthal dependence of 
the produced heavy quarks. 
  
Suppose, a light parton is produced at position $\bf{r}$, 
and moves at an angle $\alpha$ 
where $\cos{\alpha}$=$\hat{r}.\hat{d}$, 
then the distance $d$ travelled by the jet along the direction 
$\hat{d}$, before it reaches the surface is 
given by:
\begin{equation}
 d=-r\cos{\alpha}+\sqrt{R^{2}+r^{2}\sin^{2}{\alpha}}~,
\end{equation}
Hence the time available for heavy quark production is the minimum of [$\tau_d,\,\tau_f$], 
where 
$\tau_d$ is time taken by the parton to cover the distance
 $d$ and $\tau_f$ is the time when 
quark gluon plasma hadronizes.

The azimuthal distribution of the produced heavy quark from jet-thermal interaction is given by
\begin{eqnarray}
 E_1\ E_2\frac{dN}{d^3p_1d^3p_2}=\frac{1}{16(2\pi)^{8}}\int d^{4}x\int d\phi_b\,dy_b\nonumber\\
\times \frac{p_{Tb0}}{[p_{T1}\cos(\phi_1-\phi_b)+p_{T2}\cos(\phi_2-\phi_b)-M_{T1}\cosh(y_1-y_b)-M_{T2}\cosh(y_2-y_b)]}\nonumber\\
\times \left[\frac{1}{2}g_{g}^{2}f_{\textrm{jet}}^{g}(p_{Ta})f_{\textrm{th}}^{g}(p_{Tb0})\left|M_{gg \rightarrow Q\overline{Q}}\right|^{2}\right.\nonumber\\
+\left.g_{q}^{2}\sum_{i} \left
\{ f_{\textrm{jet}}^{q_i}(p_{Ta})
f_{\textrm{th}}^{\bar{q_i}}(p_{Tb0})\left|M_{q\bar{q}\rightarrow Q\overline{Q}}\right|^{2}+(q_i\leftrightarrow\bar{q_i})\right \}\right]~,
\end{eqnarray}
The $p_{T}$ distribution for jet partons is given in Eq.~\ref{jetdistr}. The 
distribution of the thermal partons and the cooling of the plasma is given in 
the Sect.~3.4.

After some simplifications, the final result is given by
\begin{eqnarray}
E_1\ E_2\frac{dN}{d^3p_1d^3p_2}=\frac{1}{16(2\pi)^{4}\pi R_{T}^{2}}\int rdr\,d\tau\,d\eta\,d\phi_b\,dy_b\nonumber\\
\times \frac{p_{Tb0}}{p_{Ta}[p_{T1}\cos(\phi_1-\phi_b)+p_{T2}\cos(\phi_2-\phi_b)-M_{T1}\cosh(y_1-y_b)-M_{T2}\cosh(y_2-y_b)]}\nonumber\\
\times \left[\frac{1}{2}g_{g}\,h_{\textrm{jet}}^{g}(p_{Ta})f_{\textrm{th}}^{g}(p_{Tb0})
\left|M_{gg \rightarrow Q\overline{Q}}\right|^{2}\nonumber\right.\\
\left.+g_{q}\sum_i \left \{h_{\textrm{jet}}^{q_i}(p_{Ta})f_{\textrm{th}}^{\bar{q_i}}(p_{Tb0})
\left|M_{q\bar{q} \rightarrow Q\overline{Q}}\right|^{2}+(q_i\leftrightarrow\bar{q_i})\right \}\right]~.
\label{thermjet}
\end{eqnarray}
which is then evaluated numerically.

\subsection{Thermal Interaction}

We have discussed earlier that the multiple scatterings among the
quarks and gluons leads to the formation of quark gluon plasma at a
large initial temperature. An interaction among the thermalized partons 
may also lead to charm production provided the initial 
temperature of quark 
gluon plasma is high. Using the recent 
results from ALICE at $\sqrt{s}$=2.76 A TeV for central collisions of lead-lead nuclei,
we take particle multiplicity density to be $dN/dy$=2850 at $\sqrt{s}$=2.76 TeV/nucleon,~\cite{Aamodt}
and extrapolate it to 3000 
for $\sqrt {s}=$ 5.5 TeV/nucleon. Now using the relation~\cite{bjorken}
\begin{equation}
 \frac{2\pi^{4}}{45\zeta(3)\pi R_{T}^{2}}\,\frac{dN}{dy}=4aT_{0}^{3}\tau_{0}
\end{equation}
and initial formation time for QGP, $\tau_{i}$=0.1 fm/c, 
we estimate $T_{0}$ to be 653 MeV at 2.76 TeV/nucleon
and 664 MeV at 5.5 TeV/nucleon respectively.

Recall also that at RHIC energies, $\tau_i$ up to 0.6 fm/$c$ have been used, specially to
for the part of the evolution which could be described using hydrodynamics. One
may imagine $\tau_i$ getting smaller at LHC energies, due to increased
activity of minijets, etc. Thus for example, the parton saturation models~\cite{kari} suggest that
$p_{\textrm{sat}}$ at LHC energies is close to 2 GeV, which suggests that the initial
time $\tau_i$ for the plasma would be $\approx 1/p_\textrm{sat}$ or about 0.1 fm/$c$. We shall
discuss the consequences of taking large formation times (see later).  

Thus the azimuthal distribution of heavy quarks produced from interactions 
of thermalized partons is given by
\begin{eqnarray}
 E_1\ E_2\frac{dN}{d^3p_1d^3p_2}=\frac{1}{16(2\pi)^{8}}\int d^{4}x\int d\phi_b\,dy_b\nonumber\\
\times \frac{p_{Tb0}}{[p_{T1}\cos(\phi_1-\phi_b)+p_{T2}\cos(\phi_2-\phi_b)-M_{T1}\cosh(y_1-y_b)-M_{T2}\cosh(y_2-y_b)]}\nonumber\\
\times \left[\frac{1}{2}g_{g}^{2}f_{\textrm{th}}^{g}(p_{Ta})f_{\textrm{th}}^{g}(p_{Tb0})\left|M_{gg\rightarrow Q\overline{Q}}\right|^{2}\right.\nonumber\\
+\left.g_{q}^{2}\sum_{i} \left
\{f_{\textrm{th}}^{q_i}(p_{Ta})f_{\textrm{th}}^{\bar{q_i}}(p_{Tb0})
\left|M_{q\bar{q}\rightarrow Q\overline{Q}}\right|^{2}+(q_i\leftrightarrow\bar{q_i})\right \}\right]~,
\label{therm1}
\end{eqnarray}
where (the boosted) thermal distribution of partons is approximated as
\begin{equation}
 f_{\textrm{th}}(p_T,y,\eta)=\exp\left[-\frac{p_{T}}{T}\cosh(y-\eta)\right]~.
\label{therm2}
\end{equation}

The above integration is done numerically, with the temperature varying according to
Bjorken's cooling law, i.e. $T^3\tau =$ constant, till the temperature drops to
about 160 MeV.

\section{Results}
\subsection{Proton Proton Collisions}

In the results to be reported in the following, we shall use the CTEQ5M structure function,
though some results are also given for other structure functions.
The mass of the charm quarks is kept fixed at $m_c=$ 1.5 GeV, while that for bottom quarks
is $m_b= $ 4.5 GeV.
The factorization
and renormalization scales are taken as $C\,\sqrt{m_Q^2+p_T^2}$
with factor $C=$ 2 for charm quarks and 1 for bottom quarks. 
The NLO pQCD code (NLO-MNR) developed by Mangano et al.~\cite{mangano,mangano2}
has been used for the initial production of heavy quarks.

\subsubsection{Production of heavy quarks, charmed mesons, and $J/\psi$:}
The results for charm production along
with recent results obtained at LHC for $pp$ collisions are shown in Fig.~\ref{sqrts}.
For the sake of exploration we have also included results for $m_c=$ 1.2 GeV and the
structure function CTEQ5M. A very good description of the data Ref.~\cite{dainese}, 
without any adjustment of parameters is seen 
(see also Refs.~\cite{jamiln,vogt}).

We have given the results of our calculations using several structure
functions in Fig.~\ref{ptdistr} for the production of charm and bottom quarks at
central rapidities in $pp$ collisions at 2.76 TeV. We see that use of any of the more modern 
structure functions gives results which differ by just a few percent from each other.


One may also consider the production of D-mesons by writing schematically:
\begin{equation}
E\frac{d^3\sigma}{d^3p}=E_Q\frac{d^3\sigma(Q)}{d^3p_Q}\otimes D(Q\rightarrow H_Q)~,
\label{d-meson}
\end{equation}
where  the fragmentation of the heavy quark $Q$ into the heavy-meson $H_Q$ is
described by the function $D$. We have assumed that the shape of
$D(z)$, where $z=p_D/p_c$, is identical for all the $D$-mesons~\cite{pete},
\begin{equation}
D^{(c)}_D(z)=\frac{n_D}{z[1-1/z-\epsilon_p/(1-z)]^2}~,
\end{equation}
$\epsilon_p$ is the Peterson parameter and
\begin{equation}
\int_0^1 \, dz \, D(z)=1~.
\end{equation}
The production of a particular $D$-meson is then obtained by using the
fraction for it, determined experimentally~\cite{H1,ZEUS}. 

A comparison of our results for  $D^0$ and $D^+$ production
with the preliminary data obtained by ALICE 
experiment~\cite{alice-d}
is shown in Fig.~\ref{dmesons}.  We give results for $\epsilon_p=$ 0.001, 0.06, and
0.12 to show the sensitivity of our calculations to this variation. Considering
that no parameters have been adjusted, the results seem to be satisfactory. 
More detailed and accurate data will definitely put stringent constraints on all the
inputs.

Note that the semi-leptonic decay of $D$-mesons has been extensively used to study 
the production of charm and bottom quarks, as well as the energy loss 
suffered by them. The electrons coming from charm decay, for example,
are obtained by convoluting the distribution of $D$-mesons (Eq.~\ref{d-meson})
with the electron decay spectrum~\cite{alta} and accounting for the branching
to a particular D-meson~\cite{H1,ZEUS}. In case the
contributions of the $B$ and the $D$ mesons can not be distinguished, 
one should use the $B$ and $D$-meson mixtures,  with appropriate branchings,
$B \rightarrow e$, $D \rightarrow e$ and $B \rightarrow D \rightarrow e$.
The semileptonic decay of B-mesons becomes important at higher $p_T$ in spite of
their reduced production, though the contribution of the
 $B \rightarrow D \rightarrow e$ channel drops rapidly with
 increase in $pT$ (see e.g., Ref.~\cite{cnvprl95}).

The ALICE experiment has,  however, obtained the single electrons from the process
 $c \rightarrow D \rightarrow e$~\cite{electron}. The upgrades of STAR and PHENIX
experiments at RHIC will also be able to measure this. 

In Fig.~\ref{electrons}, we compare our results for the electrons measured 
by the ALICE experiment with the decay of charm and a reasonable 
agreement is seen. In a future publication, we shall report  on the
consequences of introducing an intrinsic $k_T$ for the partons and also
using different parametrizations of the decay spectrum of the electrons. 

The production of $J/\psi$ in $pp$ collisions is yet another important observable, which is
closely related to the production of charm quarks. For example, using the colour
evaporation model, one can write:
\begin{equation}
\frac{d\sigma_{J/\psi}}{dy}=F \int_{2m_c}^{2m_D} 
    \, dM \,\frac{d\sigma_{c\overline{c}}}{dM \,dy}~.
\end{equation}
where $M$ is the invariant mass of the pair, $y$ is its rapidity, $m_D$ is the
mass of $D$-mesons, and $F$ is the (constant) colour-evaporation factor which should be fixed by
evaluation at some energy. There is one small detail which should be mentioned here; 
the LO pQCD calculations 
for heavy quark production produce  $c\overline{c}$ pairs with pair-momentum identically 
equal to zero (though
the NLO processes do provide them with a net transverse momentum). 
This is corrected by imparting an intrinsic $k_T$ to the partons (see e.g.~\cite{braaten}).
 {\it {Only for these calculations}} we impart an intrinsic $k_T$ of 1.5 GeV/$c$ to the partons.

We show our results for the transverse momentum and the
rapidity distribution of $J/\psi$ in Fig.~\ref{jpsi} along with 
the experimental results for $pp$ collision at 7 TeV obtained for prompt $J/\psi$ 
by the LHCb experiment\cite{jpsi}. (Note that the ALICE collaboration has measured
the inclusive $J/\psi$ which includes the b-decays~\cite{jpsi_alice}. Even though
this contribution is of the order of 10\%, it is often accounted for by adding
the $b \rightarrow J/\psi$ contribution measured by the LHCb experiment.) 
We have explored the consequences of varying the intrinsic $k_T$
on the $p_T$ distribution of $J/\psi$ and as expected the slope of the $p_T$ 
distribution decreases with increase in $k_T$.  
A reasonable description of the
distribution of the transverse momentum and the 
rapidity distribution is seen. An accurate description of the
data will  involve  a more detailed exploration of the
parameters. For example, the colour evaporation coefficient is
kept fixed in these calculations, to magnify the effect of 
varying intrinsic $k_T$. Of-course the change of intrinsic $k_T$
will not affect the rapidity distribution.

It will be interest to continue with this study for the prompt production 
of higher resonances of $c\overline{c}$ as well as of $b\overline{b}$, 
when more accurate
and detailed data become available.

\subsubsection{Correlations:}

Having witnessed a good description of charm production as well as $J/\psi$ production,
we now move to the main topic of the present work. In the following we give our results
for azimuthal, rapidity-difference, transverse momentum, and jet-radius correlation for charm 
and bottom quarks at 2.76 and 5.5 TeV for $pp$ collisions. 
Deviations from these would signal medium modifications in case of nucleus-nucleus
collisions.

Fig.~\ref{phi} shows $p_T$ and rapidity integrated $\Delta\phi$ distribution for 
heavy quarks at $\sqrt{s}$ = 2.76 TeV, 5.5 TeV and 14.0 TeV for both
leading order and next-to-leading order calculations. As expected the 
contribution rises with the energy available in the centre-of-mass system.
 It is felt that if our
argument about heavy quarks not changing direction of their motion due to
soft collisions with partons is valid, then drag (or energy loss) alone will
not drastically alter this feature. It is needless to repeat that at LO all the
heavy quarks will be produced back-to-back resulting in a peak at $\Delta \phi=\pi$.
However, {\it {if}} the heavy quarks thermalize and flow with the
medium, this picture may undergo change. We shall come back to this.

In Fig.~\ref{mqq} we  show our results 
for the transverse momentum, rapidity, and invariant mass  
distribution of  charm and bottom quark pairs produced in $pp$ collisions at  
$\sqrt{s}$=2.76 and 5.5 TeV. Recall that the pair momentum will be balanced by
the momentum of the recoiling parton. Thus tagging on a high transverse momentum
recoiling parton in the case of heavy ion
collisions can give interesting details of how heavy quarks and (mostly)
gluons behave in the medium produced in the collision. These results also
contain a very interesting situation. Consider a heavy-quark produced in LO pQCD 
in a nucleus-nucleus collision.
They will be produced back-to-back and are most likely to cover different part and
length of the system, before they fragment (or coalesce with a light quark) to
form a D-meson. Thus they would lose a differing amount of energy and acquire
a net-transverse momentum which was initially identically zero. At least the
co-linear heavy-quarks produced during splitting of a off-shell gluon would,
on the other hand, cover similar distances under similar conditions in the plasma,
and thus their net transverse momentum will remain largely unaltered. It would be 
interesting to study such cases in future more detailed experiments.

We show our results for rapidity correlation where,
$\Delta y \,= \, y_1-y_2$,
of heavy quarks produced in such collisions in Fig.~\ref{Dy}.
We note that this correlation peaks at vanishing rapidity 
difference. We have also given the LO results for this along with 
a scaling of the LO results with a factor $\sigma_{\textrm{NLO}}/\sigma_{\textrm{LO}}$ 
to demonstrate that the NLO results can not, in general, be approximated
by a $K$ factor multiplying the LO results, and the inadequacy of this shows
up most strongly near $\Delta y$ equal to zero. It is also likely that the
rapidity difference, specially when the two rapidities have opposite signs
may encode effects of longitudinal flow in case of nucleus-nucleus collisions.

Fig.~\ref{R} shows the results of our
calculation for the jet-radius, $R$, correlation, where $R=\sqrt{\Delta\eta^2+\Delta\phi^2}$. 
It brings out the interesting differences between 
results for the leading order and next-to-leading 
calculations. Thus,  while at leading order
we do not have any contribution for R $< \pi$, there is a substantial 
contribution coming from next-to-leading processes for 0 $<$ R $<$ $\pi$.

\subsection{Lead Lead Collisions}

Now we proceed to our results for collision of lead nuclei 
at 2.76 ATeV and 5.5 ATeV.
In Fig.~\ref{azipbini} we show our results
for azimuthal distribution of heavy quarks produced from initial (prompt)
collision of partons, having transverse momenta of 1--4 GeV and rapidities
close to zero. The results for LO calculations, having
a peak at $\Delta \phi= \pi$ are given, to demonstrate the importance
of using NLO results as a base line for these studies. We see a sharpening of
the collinear and back-to-back correlations as the momenta of the quarks increases,
while the correlation, with the exception of the peak at $\Delta \phi= \pi$, gets
more flat, as NLO processes have a larger role, as the available energy increases.
We also find less production of pairs of bottom quarks with smaller $\Delta \phi$
at the same energy, compared to charm quarks, as expected.

We show our results for production of heavy quarks from
multiple scattering of jets in  Fig.~\ref{azipbjet}. We have limited our 
calculations to contributions from quarks and gluons having 
$p_T >$ 2 GeV. Let us first consider
our results for charm quarks. At both the energies, we see that while charm
quarks having transverse momenta around 1.5 GeV, show a correlation which
rises smoothly as we go from collinear to back-to-back correlations, the
charm-quarks having larger transverse momenta give rise to a flat distribution
for larger $\Delta \phi$. We also note that the contribution of multiple 
scattering of the jets, even though smaller at $\Delta \phi \approx \pi$ compared to
the contribution of initial production, is rather comparable at smaller angular
separations. We note that as the initial and the final times $\tau_i$ and $\tau_f$
appear only as a multiplicative factor $\ln(\tau_f/\tau_i)$, the shape the correlation
will remain unaffected by any change in  their value.  

The bottom quarks show a very interesting trend. For the lowest momentum considered,
the bottom-quarks are seen to be produced with a flat azimuthal correlation, while as 
their momenta increase, the distribution becomes more and more collinear. The observation
about comparable contributions of multiple scattering of jets and initial production
at $\Delta \phi < \pi$, seen earlier for charm quarks, applies to them as well.

The results for the angular correlations of heavy quarks produced 
from the passage of jets through QGP are shown in Fig.~\ref{azipbjetth}.
A very interesting and distinct picture emerges for these heavy quarks.
We see that these productions are dominated by collinear contributions,
confirming the nomenclature "jet-conversion"
 (see Ref.~\cite{liufries1,rainer}) 
for them. At small $\Delta\phi$ their contribution is similar to that
from initial production. The corresponding results for bottom quark-pairs
show similar trends, but those are an order of magnitude smaller than the
contribution of initial production.

And finally the results for the angular correlation of heavy quarks 
produced from scattering of thermalized partons is shown in Fig.~\ref{azipbth}.
Firstly, these contributions are smaller by more an
order of magnitude than the contributions
discussed above. 
However, we still discuss their features as these are quite interesting.
The azimuthal correlation of charm as well as bottom quark pairs
is rather flat for low transverse momenta but changes steadily to
back-to-back at the transverse momentum increases. This, we feel,
happens as heavy quarks having large transverse momenta can only
come from collisions of partons having large $\sqrt{s}$. This would be
possible for partons having almost equal and opposite momenta, thus leading
to heavy quarks which will be predominantly back-to-back.

In order to get a feeling of the relative contributions of
different mechanisms, we have shown the  correlation of charm
quarks for $p_T \approx$ 2 GeV/$c$ and 4 GeV/$c$ in
Fig.~\ref{compare}. The first thing we note is that
the shapes and relative contributions of the processes under
consideration remain unchanged with the change in the
momentum.  Next we note that the multiple scattering among the
high energy gluons and quarks, termed jet-jet interaction,
givesi rise to a rather flat azimuthal correlation between the
charm quarks, which is comparable to the results for prompt production
for $\Delta \phi \ne \pi$. The next large contribution is due to the
jet-thermal contribution, which is rather flat for
 $\Delta \phi \, \varepsilon \, [0:2]$
and then falls rapidly. Over this region it is again
comparable to the prompt and the jet-thermal contributions.
The thermal contributions are peaked toward $\Delta \phi = \pi$ and are 
rather small.

Recall that we have used a formation time of the plasma as
0.1 fm/$c$, inspired by the parton saturation model. A larger
value for $\tau_i$ will leave the jet-jet contribution
essentially unchanged, as we discussed earlier.
However the jet-thermal and thermal contributions are
expected to drop if the initial time is increased. Thus
recalling our results from Ref.~\cite{mdy1}, we estimate that
raising the $\tau_i$ to 0.5 fm/$c$ the jet-thermal contribution
may decrease by a factor of 2, while the thermal
contribution will come down by a factor of about 4.

\subsection{Effect of Flow}

We have suggested earlier that the effect of drag or energy loss
of heavy quarks alone may not be enough to change their direction
of motion, and thus the $p_T$-integrated azimuthal correlations discussed in
this work may not be affected by the energy loss. It may change for a given $p_T$ 
due to migration of quarks to the regions of lower $p_T$ and the $p_T$ dependence of
the heavy quark production. 
The flow of the medium can, however, affect the angular correlation
considerably,
if it is large and if the heavy quarks are thermalized. 
In order to estimate the effect of the flow on the correlation of the
heavy quarks, we use a toy model used earlier by Cuautle and Paic~\cite{paic}, 
and more recently in Ref.~\cite{tsiledakis}, for
studying correlations. 

In order to do this, we proceed as follows. We first give a
random orientation to the quark-pairs from the NLO pQCD calculations
(the NLO MNR code, e.g., at LO gives pairs with $p_{x_1}=p_{x_2}$=0). Then we place
them at $(x,y)$, randomly chosen according to the probability:
\begin{equation}
P=\frac{\int \int \, dx \, dy \, T_A(x, y, b=0) T_B(x,y,b=0)}
{T_{\textrm{AB}}(b=0)}~,
\end{equation}
where $T_i$ is the transverse density profile of the nucleus $i$ assumed to have
a uniform density of radius $R$, and
$T_{\textrm{AB}}(b=0)$ is the nuclear thickness for impact parameter, $b=0$. 
Assuming a flow, directed away from the centre, we add the flow
momentum $\bf{p}_{Tf}$= $p_{Tf}$ ($\bf{r}$/$r$) to the momentum of the heavy 
quark $\bf{p_T}$.

We use the blast-model ~\cite{schnedermann} to write $p_{Tf}$ as
\begin{equation}
p_{Tf}=\gamma \beta_r m_Q~,
\end{equation}
where 
\begin{equation}
\beta_r =\beta_s \times \left(\frac{r}{R_T}\right)^2~.
\end{equation}
and $r=\sqrt{x^2+y^2}$. 
We give results for $\beta_s=$ 0,
 0.3, and 0.6.
We show our results Fig.~\ref{flow1} for two ranges of $p_T$ of the
charm quarks, $p_T <$ 4 GeV and $p_T >$ 4 GeV. We see that even
though the 
azimuthal correlation is more strongly affected for charm quarks having
lower transverse momenta for reasonable values of the flow, the basic
nature of the correlation function remains unchanged. It is
likely that if the charm quarks are not completely thermalized, the
effective flow velocity for them could be smaller, and then the
above observation becomes even more relevant. Note that large values
of $\beta_s$ are normally reached only in the hadronic phase.

\section{Summary}

We have calculated azimuthal, rapidity difference, and 
transverse momentum correlations of heavy quark pairs produced 
in $pp$ collisions at several energies relevant for experiments being done at 
the Large Hadron Collider, using NLO pQCD.  Where-ever possible,
we have discussed how these could change due to final state 
effects in nucleus-nucleus collisions. These results will act as a base-line
for similar studies in the case of $Pb+Pb$ collisions at the corresponding
centre of mass energies/nucleon, to determine medium modifications.
We have noted that this picture is enriched (or complicated) by multiple
collisions among the partons having high energy, which can give very different
correlations of a magnitude comparable to that of initial productions
considered above. We have argued, but it remains to be verified,
that these correlations may not be drastically altered due to the energy
loss suffered by heavy quarks, as they may not change the direction
of their motion substantially, due to soft scatterings. These
may however, be affected by a strong flow of the medium, if the heavy quarks
are thermalized. 

In a forth-coming publication we shall address the issue of consequences of 
energy loss on these correlations. 

\section*{Acknowledgments}
One of us (MY) acknowledges financial support of the 
Department of Atomic Energy, Government of India
during the course of these studies. (UJ) acknowledges hospitality at VECC where 
part of this work was done.


\section*{References}
\begin{thebibliography}{110}
\bibitem{bass}C.~Shen, S.  ~A.~Bass, T.~Hirano, P.~Huovinen, Z.~Qiu, H.~Song and U.~W.~Heinz,
    Phys. \ Rev. \ Lett. {\bf 106}, 042301 (2011)
    arXiv:1106.6350 [nucl-th], H.~Song, S.~A.~Bass, U.~Heinz Phys. \ Rev. \ C 
{\bf 83}, 054912 (2011), 
B.~Schenke, S.~Jeon, C.~Gale Phys. \ Rev. \ Lett. {\bf 106}, 042301 (2011).
  
\bibitem{deadcone1} Y.~L.~Dokhshitzer, V.~A.~Khoze, S.~I.~Troian, J. \ Phys. \ G {\bf 17}, 1602 (1991);~ 
Yu.~L.~Dokshitzer, D.~E.~Kharzeev, Phys. \ Lett. \ B {\bf 519}, 199 (2001).

\bibitem{deadcone2} R.~Thomas, B.~Kampfer, G.~Soff, Acta Phys. \ Hung. \ A {\bf 22}, 83 (2005);
N.~Armesto, C.~A.~Salgado, U.~A.~Wiedemann, Phys. \ Rev. \ D {\bf 69}, 114003 (2004).

\bibitem{deadcone3} W.~C.~Xiang, H.~Ding, D.~Zhou, Chin. \ Phys. \ Lett. {\bf 22}, 72 (2005).

\bibitem{hvqtherm1} H.~van Hees and R.~Rapp,
    Phys.\ Rev.\  C {\bf 71}, 034907 (2005)
  [arXiv:nucl-th/0412015], P.~B.~Gossiaux, V.~Guiho and J.~Aichelin,
    J.\ Phys.\ G {\bf 32}, S359 (2006).


\bibitem{drag} B.~Svetitsky, Phys. \ Rev. \ D {\bf 37}, 2484 (1988),
M.~G.~Mustafa, D.~Pal and D.~K.~Srivastava,
  Phys.\ Rev.\  C {\bf 57}, 889 (1998)
  [Erratum-ibid.\  C {\bf 57}, 3499 (1998)],
  Santosh ~K.~Das, Jan-e Alam, P.~Mohanty, 
Phys. \ Rev. \ C {\bf 82}, 014908 (2010).

\bibitem{diffusion} R.~Rapp, H.~van ~Hees, arXiv:0803.0901v2 [hep-ph], 
H.~van Hees, M.~Mannarelli, V.~Greco, R.~Rapp Phys. \ Rev. \ Lett. {\bf 100}, 192301 (2008), 
R.~Rapp, H.~van Hees, J. \ Phys. \ G {\bf G32}, S351 (2006).


\bibitem{nuxu1} N.~Xu, X.~Zhu, P.~Zhuang, Phys. \ Rev. \ Lett.
 {\bf 100}, 152301 (2008), N.~Xu, X.~Zhu, P.~Zhuang, J.  \ Phys. \ G {\bf 36},
 064025 (2009).


\bibitem{lingyu} Z.~W.~Lin and M.~Gyulassy,
Phys.\ Rev.\  C {\bf 51}, 2177 (1995)
[Erratum-ibid.\  C {\bf 52}, 440 (1995)].

\bibitem{mdy1} Md.~Younus, D.~K.~Srivastava, 
J. \ Phys. \ G \ Nucl. \ Part. \ Phys. {\bf 37}, 115006 (2010).

\bibitem{lmw}P.~Levai, B.~M\"{u}ller, X.~N.~Wang,  Phys. \ Rev. \ C
 {\bf 51}, 6 (1995).


\bibitem{liufries1} W.~Liu, R.~J.~Fries, Phys. \ Rev. \ C 
{\bf 77}, 054902 (2008), 
W.~Liu, R. J.~Fries, Phys. \ Rev. \ C {\bf 78}, 037902 (2008).


\bibitem{shor} A.~Shor, Phys. \ Lett. {\bf B 215}, 375 (1988).

\bibitem{energyloss} E.~Batten, M.~H.~Thoma, Phys. \ Rev. \ D {\bf 44}, R2625 (1991), 
M.~G.~Mustafa, D.~Pal, D.~K.~Srivastava and M.~Thoma,
  Phys.\ Lett.\  B {\bf 428}, 234 (1998),
M.~Djordjevic and M.~Gyulassy 
Nucl. \ Phys. \ A {\bf 733}, 265 (2004), 
 S.~Wicks, W.~Harowitz, 
M.~Djordjevic, M.~Gyulassy, Nucl. \ Phys. \ A {\bf 784}, 426 (2007), 
W.~-C.~Xiang, H.~T.~Ding, D.~C.~Zhou, D.~Rohrick 
Eur. \ Phys. \ J {\bf A25}, 75 (2005),
 N.~Armesto, C.~A.~Salgado, U.~A.~Wiedemann, Phys. \ Rev. \ D 
{\bf 69}, 114003 (2004),
 M.~G.~Mustafa, Phys. \ Rev. \ C {\bf 72}, 014905 (2005), 
(erratum) Phys. \ Rev. \ C {\bf 74}, 019902 (2006),
S.~Peigne, A.~Peshier, Phys. \ Rev. \ D {\bf 77}, 114017 (2008).

\bibitem{renk} T.~Renk, Phys. \ Rev. \ C {\bf 74}, 034906 (2006).

\bibitem{wang} X.~-N.~Wang, Phys. \ Lett. \ B {\bf B595}, 165 (2004), 
A.~Mischke, [nucl-ex/1107.5138v1].

\bibitem{jamiln} E.~Eichten, I.~Hincliffe, K.~Lane, C.~Quigg, Rev. \ Mod. \ Phys. {\bf 56}, 4 (1984)
U.~Jamil, D.~K.~Srivastava, J. \ Phys. \ G \ Nucl. \ Part. \ Phys. {\bf 37}, 085106 (2010).

\bibitem{werner} S.~Vogel, P.~B.~Gossiaux, K.~Werner, J.~Aichelin, 
Phys. \ Rev. \ Lett. {\bf 107}, 032302 (2011),
 F.~-M.~Liu, K.~Werner,
  Phys.\ Rev.\ Lett.\  {\bf 106}, 242301 (2011).

\bibitem{combridge} B.~L.~Combridge Nucl. \ Phys. \ B {\bf 151}, 429 (1979).

\bibitem{Aamodt} K.~Aamodt et al(ALICE Collaboration), Phys. \ Rev. \ Lett. {\bf 105}, 252301 (2010).

\bibitem{rainer} Rainer ~J.~Fries, B.~M\"{u}ller, D. ~K. ~Srivastava, Phys. \ ReV. \ Lett., 
{\bf 90}, 132301 (2003).

\bibitem{bjorken} J.~D.~Bjorken, Phys. \ Rev. \ D {\bf 27}, 140 (1983).

\bibitem{kari}
  K.~J.~Eskola, H.~Honkanen, H.~Niemi, P.~V.~Ruuskanen, S.~S.~Rasanen,
  Phys.\ Rev.\  {\bf C72}, 044904 (2005).
  [hep-ph/0506049].

\bibitem{mangano} M.~L.~Mangano, P.~Nason, G.~Ridolfi, Nucl. \ Phys. \ B {\bf 373},
 295 (1992).

\bibitem{mangano2}  S.~Frixione, M.~L.~Mangano, P.~Nason, G.~Ridolfi,
  Adv.\ Ser.\ Direct.\ High Energy Phys.\  {\bf 15}, 609-706 (1998).
  [arXiv:hep-ph/9702287 [hep-ph]].


\bibitem{dainese} A.~Dainese (ALICE Collaboration), 
Quark Matter 2011, Annecy, France.

\bibitem{vogt} R.~Vogt, Acta Phys. \ Polon. \ Supp. \ {\bf 1}, 695 (2008), 
N.~Carrer, A. Dainese [hep-ph/0311225].

\bibitem{pete} 
  C.~Peterson, D.~Schlatter, I.~Schmitt, P.~M.~Zerwas,
  Phys.\ Rev.\  {\bf D27}, 105 (1983).

\bibitem{H1}
  A.~Aktas {\it et al.} [ H1 Collaboration ],
  Eur.\ Phys.\ J.\  {\bf C38}, 447-459 (2005).
  [hep-ex/0408149].

\bibitem{ZEUS} ZEUS Collaboration, JHEP {\bf 07}, 074 (2007).

\bibitem{alice-d}A.~Dainese, [ ALICE Collaboration], Talk given at QM 2011.

\bibitem{alta} G.~Altarelli, N.~Cabibbo, G.~Corbo, L.~Maiani,  G.~Martinelli,
Nucl. \ Phys. {\bf B 208}, 365 (1982).

\bibitem{cnvprl95} M.~Cacciari, P.~Nason, R.~Vogt, 
Phys. \ Rev. \ Lett. {\bf 95}, 122001 (2005).

\bibitem{electron} A.~Mischke, [ ALICE Collaboration], arXive:1106.1011.

\bibitem{braaten} G.~T.~Bodwin, E.~Braaten, J.~Lee, Phys. \ Rev. \ D {\bf 72}, 014004 (2005).

\bibitem{jpsi} R.~Aaij et al., [ LHCb Collaboration ], Eur. \ Phys. \ J. {bf C 71}, 1645 (2011).

\bibitem{jpsi_alice} K.~Aamodt {\it et al.} [ALICE Collaboration ],
  Phys.\ Lett.\  {\bf B704}, 442-455 (2011).
  [arXiv:1105.0380 [hep-ex]].

\bibitem{paic}   E.~Cuautle, G.~Paic,
    AIP Conf.\ Proc.\  {\bf 857}, 175-178 (2006).
  [hep-ph/0604246]. 

\bibitem{tsiledakis} G.~Tsiledakis, H.~Appelsh$\ddot{a}$user, K. ~Schweda, J. ~Stachel, 
Nucl. \ Phys. \ A {\bf 858}, 86 (2011), G.~Tsiledakis, K.~Schweda, Proc. of the ISMD08 Conf., DESY-PROC
-2009-001, 2009, p. 214, G.~Tsiledakis, Proc. of the 417$^{th}$ WE-Heraeus-Seminar, 2008, (Bad Honnef).

\bibitem{schnedermann}   E.~Schnedermann, J.~Sollfrank, U.~W.~Heinz,
    Phys.\ Rev.\  {\bf C48}, 2462-2475 (1993).


\end{thebibliography}
\end{document}